\documentclass[]{jfm}

\usepackage{graphicx}
\usepackage{newtxtext}
\usepackage{newtxmath}
\usepackage{natbib}
\usepackage{hyperref}
\hypersetup{
    colorlinks = true,
    urlcolor   = blue,
    citecolor  = black,
}

\newcommand{\RomanNumeralCaps}[1]
\linenumbers

\usepackage{graphicx}
\usepackage{comment}

\usepackage{amsmath}
\usepackage{mathtools}
\usepackage{amsfonts}
\usepackage{amssymb}
\usepackage{float}
\usepackage{caption}
\usepackage{subcaption}
\usepackage{physics}
\usepackage{url}
\usepackage{epstopdf}
\usepackage{lipsum}
\usepackage{titlesec}
\usepackage{wrapfig}
\usepackage{array}
\usepackage{cancel}
\usepackage{setspace}
\usepackage{multirow}
\usepackage{xcolor} 
\usepackage{algorithm,algpseudocode}
\usepackage{soul}
\definecolor{PrlBlue}{HTML}{2e3092}
\definecolor{CTI_ORANGE_SRG}{rgb}{0.996, 0.421, 0.046}

\newcommand{\tb}[1]{\mathsfbi{#1}}
\newcommand{\what}[1]{\widehat{#1}}
\newcommand{\oline}[1]{\overline{#1}}

\newcommand{\mc}[1]{\mathcal{#1}}

\newcommand{\vbar}[1]{\overline{\mathbf{#1}}}

\newcommand{\bre}[1]{\breve{#1}}

\newcommand{\wtil}[1]{\widetilde{#1}}
\newcommand{\vtil}[1]{\widetilde{\mathbf{#1}}}
\renewcommand{\vb}[1]{\boldsymbol{#1}}
\renewcommand{\vtil}[1]{\widetilde{\boldsymbol{#1}}}

\renewcommand{\vbar}[1]{\overline{\boldsymbol{#1}}}

\definecolor{PrlBlue}{HTML}{2e3092}
\definecolor{CTI_ORANGE_SRG}{rgb}{0.996, 0.421, 0.046}
\definecolor{BLACK}{rgb}{0.0,    0.0000,    0.0000}  
\definecolor{RED}{rgb}{0.6914,    0.0156,    0.0547}
\definecolor{BLUE}{rgb}{0.0039,    0.4219,    0.7188}
\definecolor{GREEN}{rgb}{0.3020,    0.6863,    0.2902}
\definecolor{PURPLE}{rgb}{0.5961,    0.3059,    0.6392}
\definecolor{ORANGE}{rgb}{0.9961    0.4219    0.0469}
\definecolor{YELLOW}{rgb}{1.0000,    1.0000,    0.2000}
\definecolor{BROWN}{rgb}{0.6510,    0.3373,    0.1569}
\definecolor{PINK}{rgb}{0.9686    0.5059    0.7490}
\definecolor{CYAN}{rgb}{0,    0.9961,    0.9961}
\definecolor{NEON}{rgb}{0.2227,    0.9961,    0.0781}
\definecolor{GREY}{rgb}{0.4500,    0.4500,    0.3500}

\title{Direct numerical simulations of supersonic three-dimensional turbulent boundary layers}

\author{Salvador R. Gomez\aff{1}
  \corresp{\email{gomezsr@stanford.edu}}}

\affiliation{\aff{1}Center for Turbulence Research, Stanford University, Stanford, CA 94305, USA}

\begin{document}


\maketitle
\begin{abstract}
    Supersonic turbulent channels subjected to sudden spanwise acceleration at initial friction Reynolds numbers of approximately 500 and different Mach numbers are studied through direct numerical simulations. The response to the spanwise acceleration creates a transient period where the flow exhibits three-dimensionality in the mean statistics. This enables a detailed study of the thermal transport and development of velocity transformations and Reynolds analogies for compressible turbulent flows in swept-like conditions. Extensions of velocity transformations to three-dimensional flows demonstrate near-wall self-similarity of the velocity, providing evidence for Morkovin's hypothesis in nonequilibrium conditions. A similarity solution for the spanwise velocity, valid during the initial transient, is also presented. During the transient, both the thermal fluctuations and turbulent kinetic energy decrease, consistent with previous observations in incompressible flows (Lozano-Duran, \textit{et al.} 2019, Moin, \textit{et al.} 1990). For sufficiently strong spanwise acceleration, $Q_{3}$ $(+T',+v')$ and $Q_{1}$ $(-T',-v')$ events become more significant than sweep and ejections across the channel, creating changes in sign in the velocity-temperature covariances. The temporal evolution of the orientation and sizes of the turbulent kinetic energy and temperature carrying structures is quantified through structure identification and spectra. Finally, the generalized Reynolds analogy (Zhang, \textit{et al.} 2012) is derived for a transient three-dimensional flow, allowing predictions of the mean temperature from the velocity.  
\end{abstract}

\section{Introduction}\label{Intro}
    
    Swept wings are commonly used in transonic and supersonic aircraft to delay or reduce high-speed drag and introduce mean three-dimensionality to the velocity statistics~\citep{vos2015introduction}. Apart from swept wings, high speed applications also exhibit three-dimensional wall-bounded turbulent compressible flow when encountering surface-mounted obstacles~\citep{subbareddy2014direct}, serpentine diffusers~\citep{harrison2013active}, and rotating detonation engines~\citep{bennewitz2018characterization}, to name a few. Despite the engineering relevance of supersonic three-dimensional flows, the majority of detailed compressible turbulent flow studies focus on statistically planar flows where the mean velocity is in the streamwise and wall-normal directions. As a result, many of the models commonly used to predict velocity and thermal statistics are agnostic to effects stemming from spanwise accelerated flows. In this study, commonly used velocity-temperature relations and near-wall velocity transformations are extended to supersonic turbulent wall-bounded flows with swept-like conditions. These relationships are tested with direct numerical simulations (DNS) of fully-developed statistically stationary compressible channels at moderate friction Reynolds number, $Re_{\tau}$, and subsonic and supersonic bulk Mach numbers, $Ma$, subjected to a sudden spanwise acceleration. During the transient response, the flow exhibits swept-like conditions as the flow adjusts to the new forcing direction.


    Due to viscous heating, supersonic flows experience significant wall-normal mean temperature variation which in turn introduce wall-normal variation of the transport properties like the mean density and viscosity, as well as their fluctuations~\citep{bradshaw1977compressible,lele1994compressibility,anderson2006hypersonic}. These property variations affect the velocity statistics in compressible flows. In compressible laminar boundary layer similarity solutions, the property variations are accounted for with similarity variables~\citep{dorodnitsyn1942laminar,lees1956laminar,schlichting2016boundary}. In turbulent flows, \citet{morkovin1962effects}'s hypothesis states that for sufficiently small turbulent Mach numbers or sufficiently small density fluctuations relative to the mean density, the compressible wall bounded turbulent flow can be mapped to an equivalent incompressible turbulent flow by accounting for property variations. These observations have inspired various velocity transformations in the literature that are able to apply the near-wall incompressible viscous scaling to compressible flows~\citep{van1951turbulent,zhang2012Mach,trettel2016mean,griffin2021velocity,hasan2023incorporating}, some of which have been applied to flows with exotic property variations~\citep{bai2022compressibleGFM}. In addition to velocity transformations, near-wall mean temperature fields have been shown to exhibit near-wall self-similarity when normalized with a friction temperature~\citep{kader1981temperature,kong2000direct,pirozzoli2016passive,chen2022unified}. Further attempts to characterize and predict the temperature field began with the work of \citet{Reynolds1874Analogy} who argued that the temperature is quadratically related to the velocity, developing the Reynolds analogy. This was shown to be true from compressible laminar flows~\citep{busemann1931handbuch,crocco1932sulla}. Further developments generalized the Reynolds analogy to non-unit Prandtl number and turbulent flows, even allowing for relations between the temperature and velocity fluctuations~\citep{van1951turbulent,morkovin1962effects,walz1962compressible,gaviglio1987reynolds,huang1995compressible,duan2011direct,zhang2014generalized}. These velocity transformations and Reynolds analogies focus primarily on statistically stationary two-dimensional flows, and their extension to flows with temporal non-equilibrium or three-dimensionality has not received much attention. These studies can improve compressible wall-bounded turbulence modeling~\citep{zhang2014generalized,griffin2023near}, yet the lack of extensions to flows with three-dimensional effects can limit their predictive capability in realistic engineering applications~\citep{lozano2020non}.

    Statistically stationary two-dimensional flows have received much attention in turbulent studies as canonical flow configurations. Incompressible studies have developed and provided evidence of the mean velocity scaling and multiscale energetic motions in wall-bounded flows~\citep{von1934turbulence,millikan1938critical,coles1956law,lee2015direct} as well as various tools to uncover the turbulent structure of the flow~\citep{wallace1972wall,wallace2016quadrant,lozano2012three}. Low $Ma$ turbulent studies have been used to consider heat transfer and temperature transport as a passive scalar~\citep{perry1976experimental,kader1981temperature,nagano1988statistical,kong2000direct,pirozzoli2016passive}. The majority of supersonic wall-bounded turbulent simulations have focused on supersonic boundary layers~\citep{duan2010direct,duan2011direct,pirozzoli2011turbulence,cogo2022direct_spectra} and channel flows~\citep{coleman1995numerical,huang1995compressible,modesti2016Reynolds,yu2019genuine,hasan2025intrinsic}, to name a few. These studies have enabled detailed assessment and development of the velocity transformations and Reynolds analogies, while enabling studies of the turbulent structure and turbulent statistics.

    Much of the studies of three-dimensional wall-bounded turbulent flows stem from incompressible studies. Initial experimental studies of swept-like conditions demonstrated a reduction in turbulent kinetic energy (TKE) and misalignment between the mean velocity and Reynolds shear stress directions~\citep{bradshaw1985measurements}. These studies were then corroborated with direct numerical simulation (DNS) of fully developed turbulent channels subjected to a sudden spanwise pressure gradient~\citep{moin1990direct,coleman1996numerical,lozano2020non}. These studies revealed that the reduction in the TKE, despite the net acceleration, occurs because of a decrease in the pressure-strain reducing the wall-normal velocity fluctuations that subsequently reduces the production of the streamwise Reynolds shear stress. This then reduces the production of the streamwise turbulent fluctuations faster than the the spanwise turbulent fluctuations are generated. Structural studies have revealed that the reduction in the Reynolds shear stress can be attributed to a mismatch between the orientation of the near-wall small-scales and the larger structures further from the wall. Additional cases of spanwise flows include drag-reduction studies where the walls are oscillated in the spanwise direction~\citep{quadrio2000numerical,choi2002drag,ge2017response,ricco2021review,marusic2021energy,rouhi2023turbulent,chandran2023turbulent}, among others. The physical arguments that explain the reduction in the TKE in the spanwise accelerated channels are similar to those that explain the drag reduction in the spanwise oscillated channels. The spanwise oscillations have been used to study drag reduction in turbulent supersonic channel~\citep{yao2019supersonic} and turbulent boundary layer~\citep{ni2016direct} flows as well. However, compressible flows mimicking swept-like conditions through nonzero mean spanwise velocity are missing.

    This paper uses DNS of fully developed compressible turbulent channels at an initial $Re_{\tau} \approx 500$ and $Ma = 0.3,$, $1.5$, and $3.0$ that are suddenly accelerated in the spanwise direction through a spanwise body force to study the temporal evolution of the velocity and temperature statistics. The paper is organized as follows. The configuration and numerical details are described in section \ref{Methodology}. In section \ref{S_Scaling_Mean_Field}, the near-wall velocity transformations and temperature scaling are extended to transient three-dimensional flows. The mean velocity and temperature are then presented, along with a similarity solution for the spanwise flow, valid for initial times. The turbulent statistics and structural organization of the flow are discussed in section \ref{S_Turb_Fluc}. The generalized Reynolds analogy of \citet{zhang2014generalized} is extended to temporally varying, three-dimensional flows in section \ref{S_Reynolds_analogies}. Conclusions are presented in \ref{Conclusion}. 

\section{Methodology} \label{Methodology}

        \begin{figure}
            \centering
            \includegraphics[width=\linewidth]{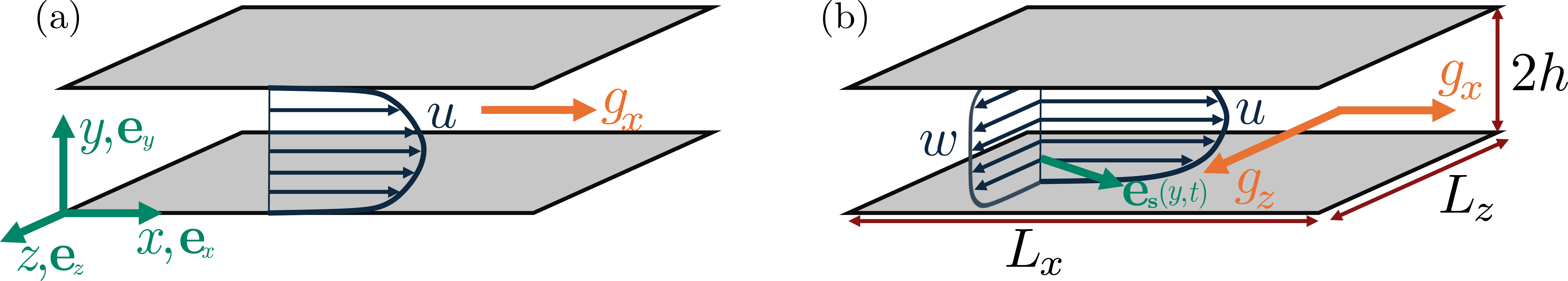}
            \caption{Schematic of the geometry, body force directions, and coordinates at time $t = 0$ (a) and $t>0$ (b)}
            \label{fig_Channel_Schematic}
        \end{figure}

    The compressible turbulent flow of a calorically perfect ideal gas within a channel is simulated with DNS using periodic streamwise and spanwise directions. The streamwise, wall-normal, and spanwise coordinates are $x \in \qty [0,L_{x}]$, $y\in \qty[0,2h]$, $z \in \qty [0,L_{z}]$, respectively, where $L_{x}$ and $L_{z}$ are the channel dimensions and $h$ the channel half-height. The unit vectors $\mathbf{e}_{x}$, $\mathbf{e}_{y}$, and $\mathbf{e}_{z}$ are along the streamwise, wall-normal, and spanwise directions, respectively. At the bottom and top walls ($y = 0, 2h)$, the flow satisfies the no-slip boundary condition and the walls are isothermal with wall temperature $T_{w} = 300 \textrm{K}$. At time $t \le 0$, the flow is a fully-developed canonical turbulent channel driven by a streamwise body force, $g_{x}$. For $t>0$, a spanwise body force, $g_{z} = \Pi g_{x}$ is applied which creates a transient period of three-dimensional mean flow. A schematic of these configurations is presented in figure \ref{fig_Channel_Schematic}. The choice of driving the flow with body forces is analogous to the applied pressure gradients used in incompressible studies of three-dimensional transient flow~\citep{moin1990direct,lozano2020non}. Previous literature has also shown that driving a compressible channel with a body force gives only slight differences in the statistics compared to a pressure-driven one~\citep{huang1995compressible}. In these simulations, the constant mass flux forcing that constrains the bulk mean velocity and is commonly applied in compressible simulations~\citep{coleman1995numerical,modesti2016Reynolds,yu2019genuine,hasan2025intrinsic} is not applied to allow the flow to respond solely to the imposed body forces. 

    The flow evolves under the compressible Navier--Stokes equations,
        \begin{align}
            \pdv{\rho}{t} + \div{\qty(\rho \vb{u})} & = 0, \label{Eq_Meth_Cont}  \\
            \pdv{\rho\vb{u}}{t} + \div{\qty(\rho \vb{u} \otimes\vb{u})} &= -\grad p + \div{\tb{\tau}} + \rho \vb{g}, \label{Eq_Meth_Mom}\\
            c_{v} \qty(\pdv{\rho T}{t} + \div{\qty(\rho T \vb{u})}) &= -p \div{\vb{u}} + \div{\vb{q}} - \tb{\tau}:\grad\vb{u},  \label{Eq_Meth_Temp}\\
            p &=\rho R T,\textrm{ and}  \label{Eq_Meth_EOS}\\
            \tb{\tau} &= \mu \qty( \grad \vb{u}^{T} + \grad \vb{u} - \frac{2}{3}\div{\vb{u}}\mc{I}), \label{Eq_Meth_Tau}
        \end{align}
    where $\rho$ denotes the density, $T$ the temperature, $p$ the pressure, $\vb{u} = u\vb{e}_{x} + v\vb{e}_{y} + w\vb{e}_{z}$ the velocity with its respective components, $R$ the universal gas constant, $c_{v}$ the specific heat capacity at constant volume, $\mu$ the dynamic viscosity, $\vb{q} = q_{x}\vb{e}_{x} + q_{y}\vb{e}_{y} + q_{z}\vb{e}_{z}$ the heat flux, and $\vb{g} = g_{x} \vb{e}_{x} + g_{z} \vb{e}_{z}$ the body force. For $t \le 0$, $g_{z} = 0$ and $t>0$, $g_{z} = \Pi g_{x}$. In equation \ref{Eq_Meth_Temp}, the cooling term that is commonly applied to the energy equation to enforce a constant bulk temperature~\citep{coleman1995numerical,yu2019genuine,hasan2025intrinsic} is omitted to study the transient response in the temperature as well. The dynamic viscosity follows Sutherland's law
        \begin{equation}
            \frac{\mu(T)}{\mu_{w}} = \qty(\frac{T}{T_{w}})^{3/2} \frac{T_{w} + S}{T + S},
        \end{equation}
    with $S = 110.4 \textrm{ K}$, and $\mu_{w}$ is $\mu$ evaluated at the wall. The heat flux follows Fourier's heating law, 
        \begin{equation}
            \vb{q} = - \frac{c_{p} \mu}{\Pr} \grad T,
        \end{equation} 
    where $c_{p}$ is the specific heat capacity at constant pressure and $\Pr=0.7$ is the Prandtl number. Along with the no-slip and isothermal wall boundary conditions, the viscous boundary condition for pressure, $\partial_{y} p = \partial_{y} \tau_{yy}$, is enforced at the channel walls.

    Because the flow is temporally evolving as a response to the sudden spanwise acceleration, the statistics are averaged across the streamwise and spanwise directions. A quantity, $f$, can be decomposed via a Reynolds decomposition, $f(x,y,z,t) = \oline{f}(y,t) + f'(x,y,z,t)$, or a Favre decomposition, $f(x,y,z,t) = \wtil{f}(y,t) + f''(x,y,z,t)$ where the Reynolds average, $\oline{f}$, is 
        \begin{equation}
            \oline{f}(y,t) = \frac{1}{L_{x}L_{z}} \int_{0}^{L_{x}}\int_{0}^{L_{z}} f(x,y,z,t) dz dx, \label{Meth_Eq_Reyn}
        \end{equation}
    and the Favre average, $\wtil{f}$, is
        \begin{equation}
            \oline{\rho}(y,t)\wtil{f}(y,t) = \frac{1}{L_{x}L_{z}} \int_{0}^{L_{x}}\int_{0}^{L_{z}} \rho(x,y,z,t)f(x,y,z,t) dz dx. \label{Meth_Eq_Favre}
        \end{equation}
    These averages are also averaged over the different ensembles and channel half-height, exploiting the symmetry or anti-symmetry of $f$ across $y=h$. Finally, the wall-normal average of a quantity $g(y,t)$ is defined as
        \begin{equation}
            \expval{g}(t) = \frac{1}{h}\int_{0}^{h} g(y,t) dy.
        \end{equation}
    This average is then used to define the bulk density, $\rho_{b} = \expval{\oline{\rho}}$, and bulk velocity, $u_{b} = \expval{\oline{\rho u}}/\rho_{b}$, where the former is constant in time from Equation \ref{Eq_Meth_Cont}. Finally, a quantity with a subscript $w$ is defined at the wall such that $f_{w}(t) = \oline{f}(0,t)$.

    Due to the temporal transient, the mean flow direction, $\vb{e}_{s}$, is a function of $y$ and $t$. Here, $\vb{e}_{s} = e_{1}\vb{e}_{x} + e_{3}\vb{e}_{z}$ where $e_{1} = \oline{u}/\norm{\vbar{u}}$ and $e_{3} = \oline{w}/\norm{\vbar{u}}$. At the wall, $\vb{e}_{s}$ is set to the mean shear direction such that $e_{1} = \oline{u}_{y}/\norm{\vbar{u}_{y}}$ and $e_{3} = \oline{w}_{y}/\norm{\vbar{u}_{y}}$ to avoid division by zero. Though different definitions can be prescribed for the instantaneous flow direction~\citep{bradshaw1985measurements,moin1990direct}, this choice is used because of its interpretability and success in defining temporally-local velocity transformations in section \ref{SS_Vel_Trans}. To avoid confusion, the streamwise direction does not denote $\vb{e}_{s}$. Rather, the streamwise and spanwise directions refer to $\vb{e}_{x}$ and $\vb{e}_{z}$, respectively. The wall shear stress along $\vb{e}_{s}$ is defined as $\vb{\tau}_{w} = \qty(\tau_{w,x}e_{w,1} + \tau_{w,z}e_{w,3})\vb{e}_{s,w} = \tau_{w}\vb{e}_{s,w}$. The friction velocities can be then defined as a streamwise friction velocity, $u_{\tau,x}(t) = \sqrt{\tau_{w,x}/\rho_{w}}$, spanwise friction velocity, $u_{\tau,z}(t) = \sqrt{\tau_{w,z}/\rho_{w}}$, and total friction velocity, $u_{\tau}(t) = \sqrt{\tau_{w}/\rho_{w}}$. The friction velocities then introduce $\ell_{\nu,x}(t) = \mu_{w}/\rho_{w}u_{\tau,x}$, $\ell_{\nu,z}(t) =\mu_{w}/\rho_{w}u_{\tau,z}$, and $\ell_{\nu}(t) =\mu_{w}/\rho_{w}u_{\tau}$ as the streamwise, spanwise, and total viscous length scales, respectively. The speed of sound at the wall is $a_{w} = \sqrt{\gamma R T_{w}}$, where $\gamma = 1.4$ is the ratio of specific heats. Henceforth, units of time with a $+$ superscript will be normalized with the initial viscous time unit such that $t^{+} = t u_{\tau}(0)/\ell_{\nu}(0)$. Finally, the use of Einstein summation convention will be used where applicable with the indeces $1$, $2$, and $3$ denoting streamwise, wall-normal, and spanwise components.

    \subsection{Simulation setup}

            \begin{table}
                \centering\scriptsize
                \vskip -0.1in
                \begin{tabular}{lccccccc}
                    case & $Re_{\tau}$ & $Ma$  & $\Pi$ & $L_{x}$  & $L_{z}$ & $\Delta y_{min}/\ell_{\nu}(0)$ & style                            \\
                    $1$ & $505$       & $0.3$ & $10$  & $4\upi h$ & $2\upi h$ & 0.34 & {\color{BLACK}\raisebox{1pt}{\rule{0.7em}{2pt}\hspace{.15em}\rule{0.2em}{2pt}\hspace{.15em}\rule{0.7em}{2pt}\hspace{.15em}\rule{0.2em}{2pt}}}  \\
                    $2$ & $560$       & $1.5$ & $10$  & $4\upi h$ & $2\upi h$ & 0.22 & {\color{RED}\raisebox{1pt}{\rule{0.7em}{2pt}\hspace{.15em}\rule{0.2em}{2pt}\hspace{.15em}\rule{0.7em}{2pt}\hspace{.15em}\rule{0.2em}{2pt}}}  \\
                    $3$ & $550$       & $3.0$ & $10$  & $6\upi h$ & $2\upi h$ & 0.25 & {\color{BLUE}\raisebox{1pt}{\rule{0.7em}{2pt}\hspace{.15em}\rule{0.2em}{2pt}\hspace{.15em}\rule{0.7em}{2pt}\hspace{.15em}\rule{0.2em}{2pt}}} \\
                    $4$ & $505$       & $0.3$ & $40$  & $4\upi h$ & $2\upi h$ & 0.34 & {\color{BLACK}\raisebox{1pt}{\rule{2.25em}{2pt}}} \\
                    $5$ & $560$       & $1.5$ & $40$  & $4\upi h$ & $2\upi h$ & 0.22 & {\color{RED}\raisebox{1pt}{\rule{2.25em}{2pt}}}    \\
                    $6$ & $550$       & $3.0$ & $40$  & $6\upi h$ & $2\upi h$ & 0.25 & {\color{BLUE}\raisebox{1pt}{\rule{2.25em}{2pt}}} 
                \end{tabular}
                \caption{Streamwise and spanwise domain lengths, $L_{x}$ and $L_{z}$, respectively, ratio of the driving body forces, $\Pi = g_{z}/g_{x}$, smallest wall-normal grid spacing, and the initial $Re_{\tau}$ and $Ma$ for each simulation studied. In some figures, the cases are distinguished by the colors and line styles shown above. \label{table_setup}}
            \end{table}

        The flow is studied with DNS using the Hypersonics Task-based Research (HTR) solver code~\citep{direnzo2020htr}.
        The flow is statistically stationary for $t\le0$ driven by $g_{x}\vb{e}_{x}$ to achieve a fixed friction Reynolds number, $Re_{\tau} = \rho_{w}u_{\tau,x}h/\mu_{w}$, and bulk Mach number, $Ma = u_{b}/a_{w}$. The simulations use $N_{x} = 768$, $N_{y} = 256$, and $N_{z} = 512$ grid points in $x$, $y$, and $z$, respectively, where the discretization uses a sixth-order hybrid Euler scheme and the time stepping uses the strong-stability-preserving third-order Runge–Kutta scheme keeping the Courant–Friedrichs–Lewy number below $0.5$. The points are uniform in $x$ and $z$, while the wall-normal grid points are stretched as $y = h \tanh{(s_{y}\wtil{y})}/\tanh{(s_{p})}$ where $\wtil{y} \in \qty[-1,1]$ are equispaced points. The stretching parameter $s_{y}$ is chosen such that the first grid point away from the wall, $\Delta y_{min},$ has $\Delta y_{min}/\ell_{\nu}(0) < 1$.

        While driven by $g_{x}\vb{e}_{x}$, the flow achieves a statistically stationary state at a fixed $Re_{\tau}$ and $Ma$. From this stationary period, $8$ statistically independent snapshots separated in time by at least an eddy-turnover with a time separation of $\Delta t \approx 550\ell_{\nu}(0)/u_{\tau}(0)$ are chosen to initialize the $8$ ensembles to average over. For $t> 0$, these initial conditions are integrated for a total time of $500 \ell_{\nu}(0)/u_{\tau}(0)$ and are driven by $g_{x}\vb{e}_{x} + \Pi g_{x}\vb{e}_{z}$. The $L_{x}$, $L_{z}$, $\Pi$, and initial $Re_{\tau}$ and $Ma$ are listed in Table \ref{table_setup}. The grid resolution is similar to that used in \citet{modesti2016Reynolds} at similar $Re_{\tau}$. The $Ma = 3.0$ cases use a longer streamwise domain because the larger $\oline{T}$ increases $\oline{\mu}$, making the local viscous length scales larger across the channel~\citep{modesti2016Reynolds}. Snapshots are saved in increments of $\Delta t = 10\ell_{\nu}(0)/u_{\tau}(0)$ for post-processing.

\section{Scaling of the mean flow field} \label{S_Scaling_Mean_Field}

        \begin{figure}
            \centering
            \includegraphics[width=\linewidth]{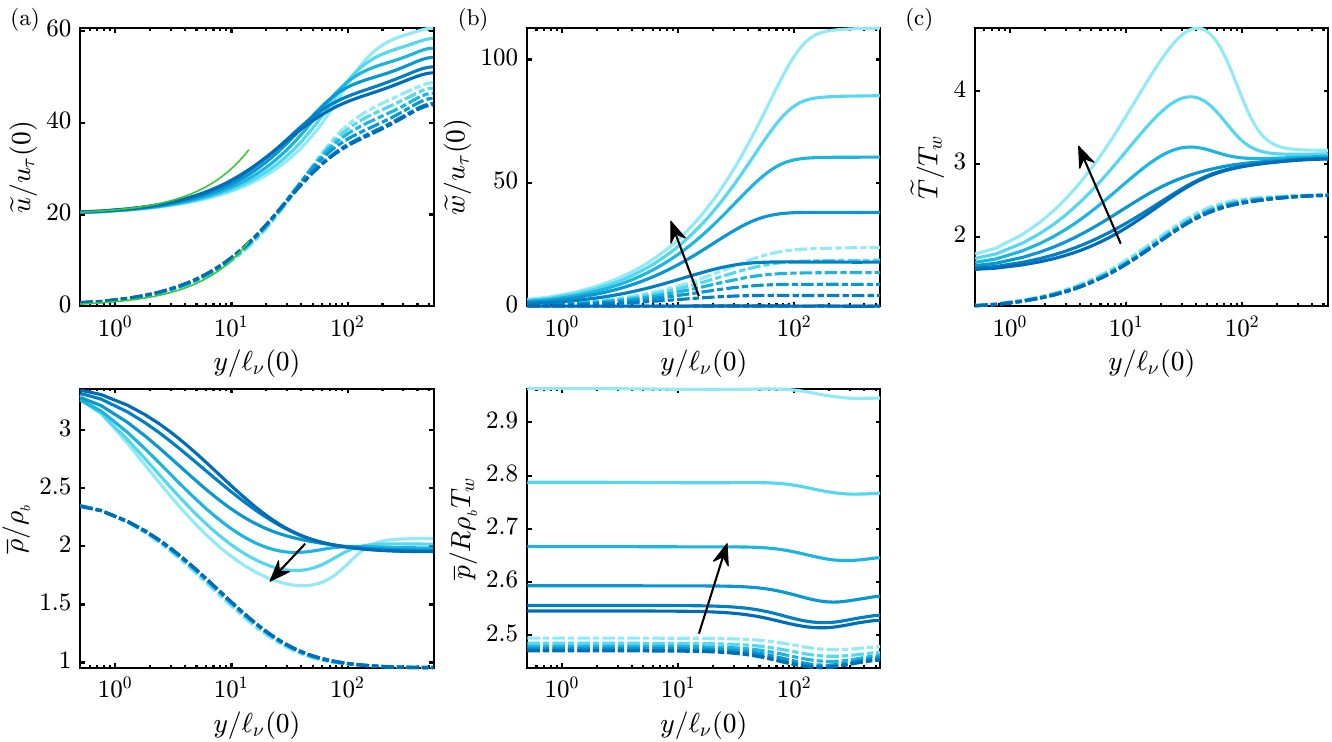}
            \caption{Temporal variation of $\widetilde{u}$ (a), $\widetilde{w}$ (b), normalized by the initial $u_{\tau}$, $\widetilde{T}$ normalized by $T_{w}$ (c), $\oline{\rho}$ normalized by $\rho_{b}$ (d), and $\oline{p}$ normalized by $R \rho_{b} T_{w}$ (e) for $Ma = 3$ and $\Pi = 10$ (dashed) and $\Pi = 40$ (solid). In (a,c,d,e), the plots of $\Pi = 40$ are offset vertically for visibility. The colors from dark to light indicate increasing time in increments of $\Delta t^{+} = 100$ and arrows indicate increasing $t$. The green lines in (a) plot $\oline{u}/u_{\tau}(0) = y/\ell_{\nu}(0)$, the viscous sublayer for a canonical incompressible flow. }
            \label{fig_Scaling_no_scaling}
        \end{figure}

    As a representative case, the mean flow fields during the transient period of the $\Pi = 10$ and $\Pi = 40$, $Ma = 3$ channels are presented in figure \ref{fig_Scaling_no_scaling} to highlight the temporal evolution of the mean profiles and highlight the $\oline{\rho}$ and $\wtil{T}$ variation. The other cases are presented in section \ref{SS_Vel_Trans} and are qualitatively similar to the $Ma = 3$ cases. In figures \ref{fig_Scaling_no_scaling}(a,b), $\widetilde{u}$ and $\widetilde{w}$ are plotted, normalized by the initial $u_{\tau}$. Due to the sudden application of $g_{z}$, $\widetilde{w}$ increases monotonically with time throughout the channel for both $\Pi$. On the other hand, the effect of $g_{z}$ on $\wtil{u}$ primarily affects the outer region and is less pronounced for $\Pi = 10$. Similarly, the effect of $g_{z}$ on $\wtil{T}$ is most appreciable for $\Pi = 40$ in figure \ref{fig_Scaling_no_scaling}(c), attaining a local near-wall temperature peak commonly observed in turbulent boundary layers~\citep{duan2010direct,cogo2022direct_spectra}. Due to the increase in viscous heating opposing the spanwise acceleration, $\wtil{T}$ increases with $t$. The density decreases in the near-wall region, and increases in the outer region to maintain a constant $\rho_{b}$, as shown in figure \ref{fig_Scaling_no_scaling}(d). The net increase in $\wtil{T}$ increases $\oline{p}$, despite the constant $\rho_{b}$. in the following sections, velocity transformations are presented for $\vbar{u}$ and $\wtil{T}$.

    \subsection{Extension of velocity-transformations to 3D nonequilibrium flows} \label{SS_Vel_Trans}

        Over the years, \citet{morkovin1962effects}'s hypothesis has inspired various velocity transformations for statistically stationary two-dimensional flows, like channels and boundary layers, that use the mean property variations to collapse the compressible mean velocity onto an equivalent incompressible mean velocity in the near-wall region~\citep{van1951turbulent,zhang2012Mach,trettel2016mean,griffin2021velocity,hasan2023incorporating}. However, extensions to statistically three-dimensional or temporally varying compressible flows have not yet been applied in the literature. Previous studies in transient three-dimensional non-equilibrium incompressible flows have shown that the initial $u_{\tau}$ and $\ell_{\nu}$ do not appropriately collapse the near-wall $\oline{u}$~\citep{lozano2020non}, while rescaling the three-dimensional velocity magnitude, $\norm{\vbar{u}}$, with the local $u_{\tau}(t)$ and $\ell_{\nu}(t)$ can collapse the near-wall statistics to the canonical counterpart~\citep{moin1990direct}. This suggest that an appropriate compressible velocity transformations for the near-wall region must account for the temporal variation of the mean flow field and the transport properties. 

            \begin{figure}
                \centering
                \includegraphics[width=\linewidth]{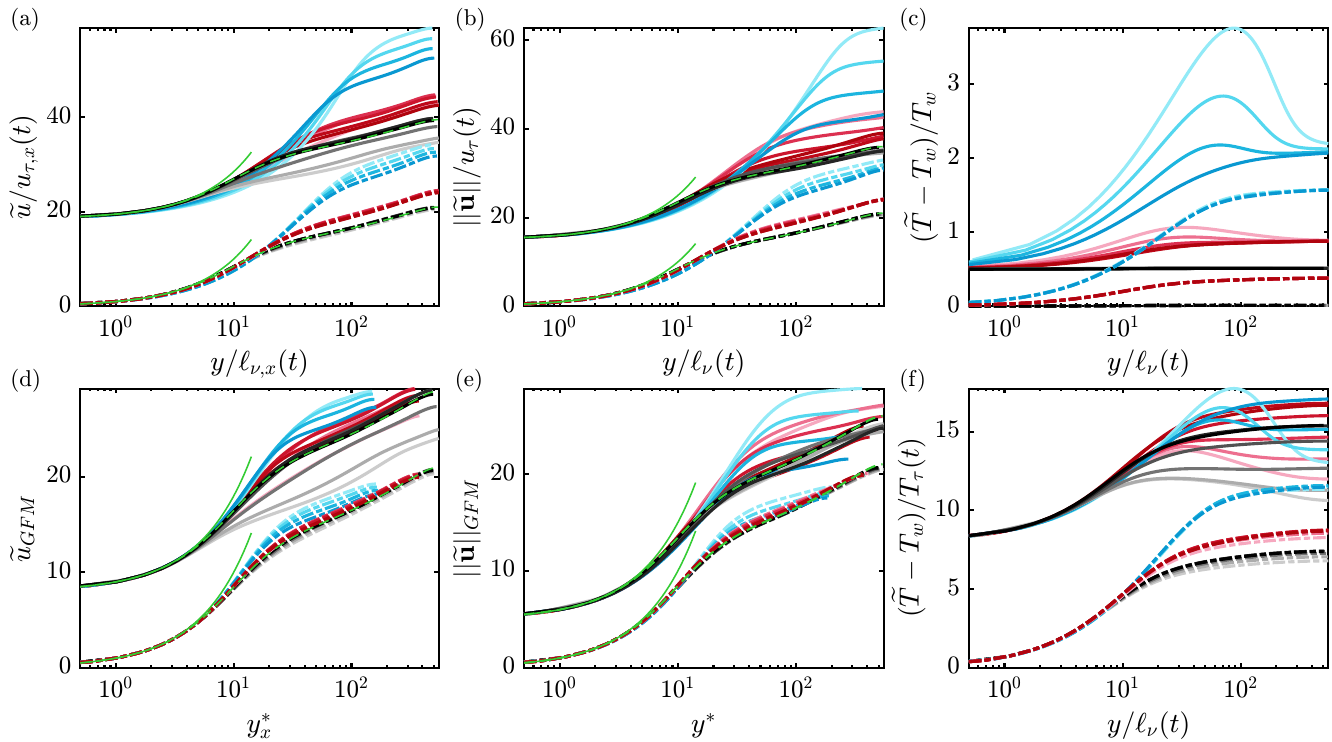}
                \caption{Quasi-steady friction scaling of $\wtil{u}$ (a) and $\norm{\vtil{u}}$ (b). $\wtil{T}$ normalized by $T_{w}$ with $y$ normalized by $\ell_{\nu}(t)$ (c). The GFM transformation of $\wtil{u}$ (d) and$\norm{\vtil{u}}$ (e) and friction scaling of $\wtil{T}$ by normalizing with $T_{\tau}(t)$ (f). The colors and line plots follow from table \ref{table_setup}. The colors from dark to light are offset by time increments of $\Delta t^{+} = 100$. In (a,b,d,e), the green line plots the viscous sublayer of an incompressible flow as $u^{+} = y^{+}$ and the green dashed line plots the mean velocity of an $Re_{\tau} = 550$ turbulent channel~\citep{lee2015direct}. The $\Pi = 40$ cases are offset vertically for visibility. } 
                \label{fig_Scaling_scaling_U_Umag_T}
            \end{figure}

        An extension of the \citet{griffin2021velocity} (GFM) velocity transformation is presented that provides near-wall self-similarity of $\wtil{u}$ and $\vtil{u}$ in a three-dimensional flow. The extension for the \citet{trettel2016mean} (TL) velocity transformation was presented in \citet{gomezcompressibilityARB} and are presented in section \ref{APP_TL}. The GFM transformation for a statistically stationary two-dimensional flow assumes that 
            \begin{equation}
                \frac{d \wtil{u}_{GFM}}{dy^{*}_{x}} = S_{t,x} = \frac{\tau^{+}_{x}S_{eq,x}}{\tau^{+}_{x} + S_{eq,x} - S_{TL,x}}, \label{Eq_Scaling_GFM_x}
            \end{equation}
        where $\tau^{+}_{x} = \qty(\oline{\tau}_{xy} + \oline{\rho} \wtil{u''v''})/\tau_{w,x}$, $S_{eq,x} = \qty(\mu_{w}/\oline{\mu})\qty(1/u_{\tau,x})\pdv*{\wtil{u}}{y^{*}_{x}}$, $y^{*}_{x} = u_{\tau,x}\qty(\oline{\rho}/\oline{\mu})y$, and $S_{TL,x} = \qty(\oline{\mu}/\mu_{w})\qty(\ell_{\nu,x}/u_{\tau,x})\pdv*{\wtil{u}}{y}$. The GFM transformation combines the near-wall scaling of \citet{trettel2016mean} with the quasi-equilibrium argument of \citet{zhang2012Mach} and has been shown to improve the collapse of a wide array of compressible flows to the incompressible counterpart in the inner regions~\citep{griffin2021velocity,bai2022compressibleGFM}. Since the original GFM transformation is formulated for a stationary flow, it can only apply at $t = 0$ when the flow is still under a canonical configuration. This restriction is relaxed in equation \ref{Eq_Scaling_GFM_x} by allowing for quasi-steady temporal variations in the averaged quantities, similar to how the streamwise evolution in a boundary layer is omitted for the GFM transformation~\citep{griffin2021velocity,bai2022compressibleGFM}. The GFM transformation of $\wtil{u}$ is contrasted against a quasi-steady friction scaling using $u_{\tau,x}$ and $\ell_{\nu,x}$ in figures \ref{fig_Scaling_scaling_U_Umag_T}(a,d). Using the quasi-steady friction scaling limits the near-wall self-similarity and introduces large wakes in the outer region. Through the GFM transformation, the self-similarity in the viscous sublayer improves while the wakes are mitigated in the $\Pi = 10$ case. With the velocity transformation, the supersonic $\wtil{u}_{GFM}$ approach the incompressible $\oline{u}$ of a canonical channel at $Re_{\tau} = 550$~\citep{lee2015direct} in the buffer region.

        In order to extend the GFM transformation to $\norm{\vtil{u}}$ the proposed scalings must also account for $\vb{e}_{s}$, the direction of $\vbar{u}$, and the temporal variation of the mean properties.
        First, the sum of the normalized viscous and Reynolds shear stresses along the direction $\vb{e}_{s}$ is $\tau^{+} = \qty[\qty(\tau_{xy} + \oline{\rho}\wtil{u''v''})e_{1} + \qty(\tau_{yz} + \oline{\rho}\wtil{w''v''})e_{3}]/\tau_{w}$. The semi-local coordinate along the mean flow direction is then $y^{*} = u_{\tau} \qty(\oline{\rho}/\oline{\mu}) y$ while the viscous length scale is $\ell_{\nu}$. These quantities are then used to define the normalized velocity gradient along $\vb{e}_{s}$ as $S_{eq} = \qty(\mu_{w}/\oline{\mu})\qty(1/u_{\tau})\pdv*{\norm{\vtil{u}}}{y^{*}}$ and $S_{TL} = \qty(\oline{\mu}/\mu_{w})\qty(\ell_{\nu}/u_{\tau})\pdv*{\norm{\vtil{u}}}{y}$. The extension of the GFM transformation to $\norm{\vtil{u}}$ is 
            \begin{equation}
                \frac{d \norm{\vtil{u}}_{GFM}}{dy^{*}} = S_{t} = \frac{\tau^{+}S_{eq}}{\tau^{+} + S_{eq} - S_{TL}}, \label{Eq_Scaling_GFM_mag}
            \end{equation}
        where the quantities are once again taken as quasi-steady in time. The friction scaling of $\norm{\vtil{u}}$ using $u_{\tau}$ and $\ell_{\nu}$ is shown in figure \ref{fig_Scaling_scaling_U_Umag_T}(b) as a contrast to its GFM velocity transformation in figure \ref{fig_Scaling_scaling_U_Umag_T}(e). Although the friction scaling in figures \ref{fig_Scaling_scaling_U_Umag_T}(b,d) is able to collapse the viscous sublayer, it can not account for the property variations further from the wall. The GFM velocity transformation offers an improvement in the collapse of $\wtil{u}$ and $\norm{\vtil{u}}$ for $y^{*}>10$ and suppresses the wakes in the larger $Ma$ flows by also incorporating information about the total stress across the channel into the transformation. For $\norm{\vtil{u}}$, the GFM velocity transformation is able to collapse the $\Pi = 10$ cases to a canonical channel~\citep{lee2015direct}, with slight deviations between the $Ma = 0.3$ and $Ma = 1.5$ cases. With the GFM velocity transformation, the near-wall collapse of $\norm{\vtil{u}}$ to a canonical $\oline{u}$ with viscous scales observed in \citet{moin1990direct} is reflected with these compressible cases. While this study focuses on the near-wall viscous scaling, the outer scaling of the velocity is discussed in \citet{gomezcompressibilityARB}.

        Following the success of the quasi-steady scaling in the velocities and the use of the friction temperature in the literature \citep{kader1981temperature,pirozzoli2016passive,kong2000direct,chen2022unified}, a quasi-steady friction temperature, $T_{\tau}$ is presented. The assumption here is that the near-wall temperature is governed by the total wall-shear stress such that $T_{\tau}(t) = q_{w}(t)/\qty(c_{p}\rho_{w}(t)u_{\tau}(t))$. This temperature scale is presented in figure \ref{fig_Scaling_scaling_U_Umag_T}(f) for $\wtil{T}$ where the wall-normal coordinate is normalized by $\ell_{\nu}(t)$ demonstrating near-wall collapse using temporally varying temperature and length scales that account for the mean flow direction. This near-wall collapse in $\wtil{T}$ is a marked improvement over the use of $T_{w}$ in figure \ref{fig_Scaling_scaling_U_Umag_T}(c) and suggests that the near-wall temperature field is also governed by its friction scales. It is worth also noting that the $Ma = 1.5$ and $Ma = 0.3$ cases also demonstrate a near-wall $\wtil{T}$ peak for $\Pi = 40$.

    \subsection{A self-similar velocity transformation for the spanwise velocity} \label{SS_Laminar_Solution}

        The $x$ and $z$ averaged spanwise momentum equation is 
            \begin{equation}
                \pdv{}{t}\qty(\oline{\rho}\wtil{w}) + \pdv{}{y}\qty(\oline{\rho}\wtil{w''v''}) + \pdv{}{y}\qty(\oline{\rho}\wtil{w}\wtil{v}) - \oline{\rho}g_{z} = \pdv{\oline{\tau}_{yz}}{y}= \pdv{}{y}\qty[\oline{\mu}\pdv{\oline{w}}{y} + \oline{\mu' \pdv{w'}{y}} + \oline{\mu' \pdv{v'}{z}}] .\label{eq_Scaling_MMB}
            \end{equation}
        Because $\oline{\rho}$ evolves in time and mass conservation, $\wtil{v} \ne 0$ for $t> 0$. However, the magnitude of $\wtil{v}$ remains negligible compared to $\wtil{w}$ and $\wtil{u}$ and its contribution is omitted for the rest of this section. For $t\le 0$, the spanwise Reynolds shear stress, $\oline{\rho}\wtil{w''v''}$, and spanwise turbulent viscous stresses, $\oline{\mu'\partial_{y}w'} + \oline{\mu'\partial_{z} v'}$, are initially zero due to the lack of net transport in the spanwise homogeneous direction and spanwise homogeneity. As time advances, the development of $\wtil{w}$ will lead to the generation of $\oline{\rho}\wtil{v''w''}$~\citep{moin1990direct,lozano2020non} and turbulent viscous stresses via the net transport of $\mu'$ in $\vb{e}_{z}$ through $\oline{\mu'\partial_{y}w'}$, though the latter are negligible~\citep{bradshaw1977compressible}.

            \begin{figure}
                \centering
                \includegraphics[width=\linewidth]{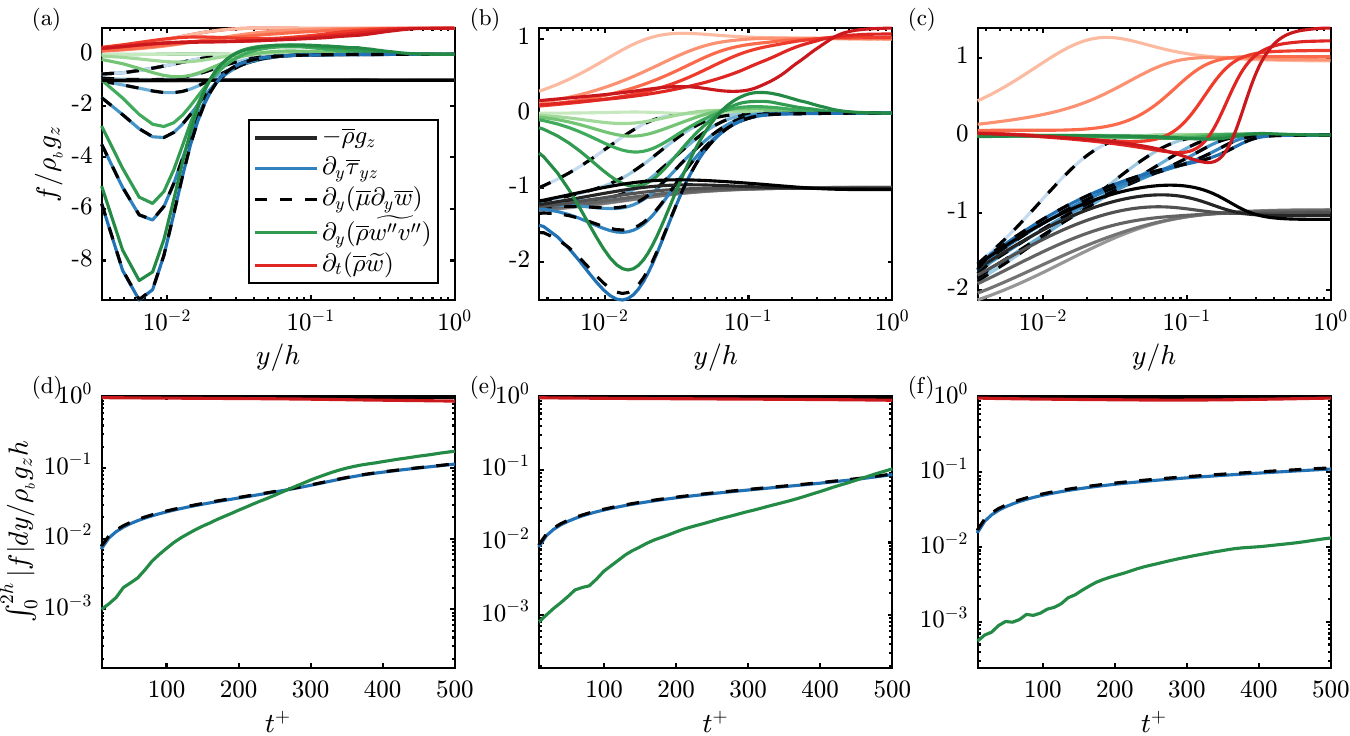}
                \caption{Spanwise mean momentum balance for $\Pi = 40$ and $Ma = 0.3$ (a,d), $1.5$ (b,e), $3.0$ (e,f). In (a--c), the colors from dark-light denote different time instances, in increments of $\Delta t^{+} = 100$. The quantity, $f$, is labeled in the legend of (a).}
                \label{fig_Scaling_MMB_PI_40}
            \end{figure}

    The time evolution of these terms is shown in figures (a--c) for $\Pi = 40$ and the three $Ma$. Note that the acceleration term, $\partial_{t}\qty(\oline{\rho}\wtil{w})$, is estimated from the saved snapshots using finite differencing. The acceleration term primarily balances $\oline{\rho}g_{z}$ in the outer region of the flow, while for short times, the viscous stress balances $\oline{\rho}g_{z}$ near the wall. As time advances, $\oline{\rho}\wtil{v''w''}$ intensifies, balancing the viscous stresses. This can be seen more clearly in figures (d--f) where the magnitudes of each of the terms is integrated across the wall as a measure of each term's effect across the channel. These plots show a period where the spanwise Reynolds stresses are negligible. An estimate for $\oline{\tau}_{yz}$ via $\oline{\mu}\partial_{y}\oline{w}$ is included, demonstrating that the turbulent viscous stresses are indeed negligible~\citep{bradshaw1977compressible}. Hence, for a short time interval, the effect of the Reynolds stresses and turbulent viscous stresses on the spanwise momentum equation may be neglected. For the supersonic cases, this interval extends much longer than the $Ma = 0.3$ case due to the increase in near-wall viscosity that dampens the production of the spanwise Reynolds shear stress. The time interval, $t_{c}$, is defined as the time where  $\int_{0}^{2h}\abs{\partial_{y}\oline{\tau}_{yz}}dy =  \int_{0}^{2h}\abs{\partial_{y}\qty(\oline{\rho \wtil{w''v''}})}dy$, that is the time where the viscous stress is no longer dominant over the Reynolds stress.

    To model the initial development of the spanwise mean flow during short times $t \ll t_{c}$, several assumptions are made. Following the observations from figure \ref{fig_Scaling_MMB_PI_40}, it is assumed that the turbulent viscous stresses can be neglected such that $\oline{\tau}_{yz} \approx \oline{\mu} \partial_{y}\wtil{w}$ and that $\oline{\rho}\wtil{v''w''}$ is negligible compared to $\oline{\tau}_{yz}$. The third assumption is that the mean properties are quasi-steady. This means that $\oline{\rho}$ and $\oline{\mu}$ are treated as constant in time with their values updated locally at each time and their temporal gradients are negligible. Finally, it is assumed that density fluctuations play a minor role such that $\oline{w} \approx \wtil{w}$. With these approximations, equation \ref{eq_Scaling_MMB} can be simplified as
        \begin{equation}
                \pdv{\wtil{w}}{t}  - g_{z} = \frac{1}{\oline{\rho}}\pdv{}{y}\qty[\oline{\mu}\pdv{\wtil{w}}{y} ] .\label{eq_Scaling_MMB_Simp}
        \end{equation}
    These assumptions essentially assume that a laminar prediction for $\wtil{w}$ holds. This has been shown to hold only for short times in incompressible simulations similar to the cases studied here~\citep{moin1990direct,lozano2020non} and for spanwise oscillating walls in drag reduction~\citep{quadrio2000numerical,choi2002drag}. The extent of the laminar prediction is expected to hold for a much longer time for the supersonic cases because of the larger $t_{c}$

    In \citet{gomezcompressibilityARB}, equation \ref{eq_Scaling_MMB_Simp} was solved using a series solution from the eigenmodes of $1/\oline{\rho}\partial_{y}\qty[\oline{\mu}\partial_{y} ] $ with Dirichelet boundary conditions at $y = 0$ and $2h$ after using a separation of variables. The series solution is an exact solution to equation \ref{eq_Scaling_MMB_Simp} under the assumptions described above. Here, a similarity solution is presented which provides a velocity transformation for $\wtil{w}$. First, it is assumed that 
        \begin{equation}
            \widetilde{w} = g_{z} t f(\eta) \label{Eq_Scaling_SimSol}
        \end{equation}
    where 
        \begin{equation}
            \eta(y,t) = \sqrt{\frac{\oline{\rho}(y) \oline{\mu}(y)}{t}} \int_{0}^{y}\frac{d\xi}{\oline{\mu}(\xi)}\label{Eq_Scaling_SimVar}
        \end{equation}
    is the similarity variable and the explicit time dependence of $\oline{\mu}$ and $\oline{\rho}$ is neglected. For an incompressible flow with constant transport properties, $\eta_{inc}(y,t) = y/\sqrt{\oline{\nu}t}$, where $\nu = \mu/\rho$ is the kinematic viscosity, is the same similarity variable as that used in Stoke's first problem~\citep{schlichting2016boundary}. The similarity variable $\eta$ is different to the one introduced by the Lees-Dorodnitsyn transformation~\citep{dorodnitsyn1942laminar,lees1956laminar,anderson2006hypersonic,schlichting2016boundary}, even when converting $x$ to $u_{e}t$, because the quantity in the square-root is wall-normally varying and the integrand is $\oline{\mu}^{-1}$ rather than $\oline{\rho}$. It can then be shown that using equation \ref{Eq_Scaling_SimSol}, equation \ref{eq_Scaling_MMB_Simp} becomes
        \begin{equation}
            f - \frac{\eta}{2}\dv{f}{\eta} -1 = \dv[2]{f}{\eta}.\label{Eq_Scaling_Eq_for_F}
        \end{equation}
    Note that equation \ref{Eq_Scaling_Eq_for_F} is the same for both incompressible and compressible flow. The difference in the solutions is the similarity variable, $\eta$. Reducing equation \ref{Eq_Scaling_SimVar} to the same incompressible similarity equation via $\eta$ demonstrates the applicability of \citet{morkovin1962effects}'s hypothesis where the same incompressible mechanisms govern the compressible spanwise response, provided that the mean property variations are accounted for. 

    For equation \ref{Eq_Scaling_SimSol} to be a valid solution, it must satisfy the no-slip boundary conditions at $y=0,2h$ and the initial condition $\wtil{w} = 0$ at $t = 0$. Since equation \ref{Eq_Scaling_Eq_for_F} is a second order differential equation, it can only satisfy two boundary conditions. To reconcile the boundary condition at $y = 2h$, the solution proposed in equation \ref{Eq_Scaling_SimSol} will only be valid for $y \in \qty[0,h]$ and $\wtil{w}$ will be assumed to be an even function in $y$ about $y = h$ so that its wall-normal derivative is continuous. The boundary condition at $y=0$ and the initial condition at $t = 0$ on $\wtil{w}$ translate to the boundary conditions that $f = 0$ at $\eta = 0$ and $f=1$ at $\eta \rightarrow \infty$, respectively. The latter condition ensures that $f$ is bounded as $t \rightarrow 0$. Due to the additional assumption of symmetry about $y = h$, there is an additional constraint that $\dv*{f}{\eta} = 0$ at $\eta(h,t)$. This presents an additional boundary condition that can not be reconciled analytically. However, it will be assumed that $\eta(h,t) \gg 1$ so that this symmetry constraint becomes equivalent to $\dv*{f}{\eta} = 0$ as $\eta \rightarrow \infty$. This is equivalent to assuming that the viscous length scales in the laminar solution are much smaller than $h$. An order of magnitude estimate for $\eta(h,t)$ can be found using viscous wall units and approximating $\eta$ with the incompressible counterpart, $\eta_{inc}$, such that  
        \begin{equation}
            \eta(h,t) \sim \frac{h}{\sqrt{\oline{\nu} t}} = \frac{Re_{\tau}}{\sqrt{t^+}}.\label{Eq_Scaling_estimate}
        \end{equation}
    The smallest $\eta(h,t)$ can be is at $t^{+} \approx 550$ during the duration of the simulation, which for the cases studied here is $\eta \approx Re_{\tau}/\sqrt{500} \approx 23$. It will be shown that for this estimate, $\dv*{f}{\eta} = 0$ can be satisfied to numerical round off error.

    The solution to equation \ref{Eq_Scaling_Eq_for_F} that satisfies the boundary conditions at $\eta = 0$ and $\eta \rightarrow \infty$ is 
        \begin{equation}
            f = 1 - \qty(\frac{\eta^2}{2} + 1)\textrm{erfc}\qty(\frac{\eta}{2}) + \frac{\eta}{\sqrt{\upi}}e^{-\frac{\eta^2}{4}}\label{Eq_Scaling_SimSolSol},
        \end{equation}
    where $\textrm{erfc}$ is the complimentary error function~\citep{boas2006mathematical}. The solution's derivative, $\dv*{f}{\eta} = 2/\sqrt{\upi}\exp\qty(-\eta^2/4) - \eta \textrm{erfc}(\eta/2)$, at the estimate for $\eta(h,t)$ from equation \ref{Eq_Scaling_estimate} evaluates to $\order{10^{-60}}$. The smallest value for all 6 compressible cases is $\eta(h,t) \approx 10$ at the end of the simulation, which evaluates $\dv*{f}{\eta} \approx 10^{-13}$. Though not exactly $0$, it is a sufficient approximation for symmetry about $y=h$. The value of $\eta(h,t)$ is much larger during the interval $t \in \qty[0,t_{c}]$ where the laminar assumptions are expected to hold which only improves the approximation regarding symmetry about $y=h$.

        \begin{figure}
            \centering
            \includegraphics[width=\linewidth]{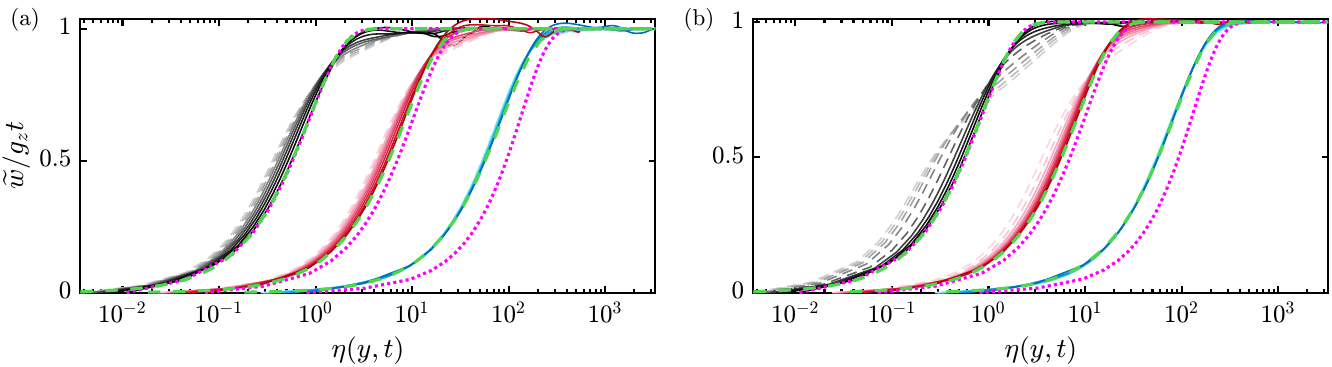}
            \caption{Similarity scaling of the spanwise flow for (a) $\Pi = 10$ and (b) $\Pi = 40$. The solid and dashed lines color coded with table \ref{table_setup} are DNS data, with the dashed lines plotted for $t > t_{c}$, and colors from dark to light indicate increasing time, plotted from $t^+ = 60$ in increments of $\Delta t^{+} = 60$. The green dashed line is the compressible similarity solution, $f\qty(\eta\qty(y,t))$. The magenta dotted line is the incompressible similarity solution, $f\qty(\eta_{inc}\qty(y,t))$, against $\eta\qty(y,t)$ at $t^{+} = 60$. The $Ma = 1.5$ and $Ma = 3.0$ cases are offset horizontally for visibility.  }
            \label{fig_Scaling_SimilaritySolution}
        \end{figure}

        The instantaneous $\wtil{w}$ are plotted in figures \ref{fig_Scaling_SimilaritySolution}(a,b) for all $6$ cases, normalized by $g_{z} t$, against $\eta(y,t)$. The similarity solution, $f$, is also included where $\eta$ is evaluated using equation \ref{Eq_Scaling_SimVar} for the compressible solution and $\eta_{inc}$ to illustrate an incompressible approximation. For short time intervals, $f(\eta)$ and $\wtil{w}/g_{z}t$ agree because of the absence of spanwise turbulent stresses for all 6 cases. As time advances, the laminar approximation breaks creating departures from the compressible-laminar similarity solution. This departure is most severe in the $Ma = 0.3$ cases where $t_{c}$, the time it takes $\oline{\rho}\wtil{v''w''}$ to dominate over $\oline{\tau}_{yz}$, is smallest. For the $Ma = 3$ cases, $\wtil{w}/g_{z}t$ and $f(\eta)$ agree during the full simulation time as $\oline{\rho}\wtil{v''w''}$ remains negligible compared to $\oline{\tau}_{yz}$. Although figure \ref{fig_Scaling_no_scaling}(a) demonstrates significant temporal variation in $\oline{\rho}$ for case 6, the agreement between the data and the similarity solution demonstrate that the quasisteady approximation is valid and any discrepancies in the smaller $Ma$ flows are indeed from the turbulent stresses. Finally, the incompressible solution, $f(\eta_{inc})$, agrees well with the $Ma = 0.3$ cases, but exhibits significant departures from $f(\eta)$ and $\wtil{w}/g_{z}t$ in the near-wall region as the $Ma$ increases. Properly accounting for the mean property variations in $\eta$ allows the similarity solution, $f$, to accurately predict $\wtil{w}$ for small times. 

    \subsection{Mean shear and heat transfer} \label{SS_Mean_Shear_Heat_Transfer}

            \begin{figure}
                \centering
                \includegraphics[width=\linewidth]{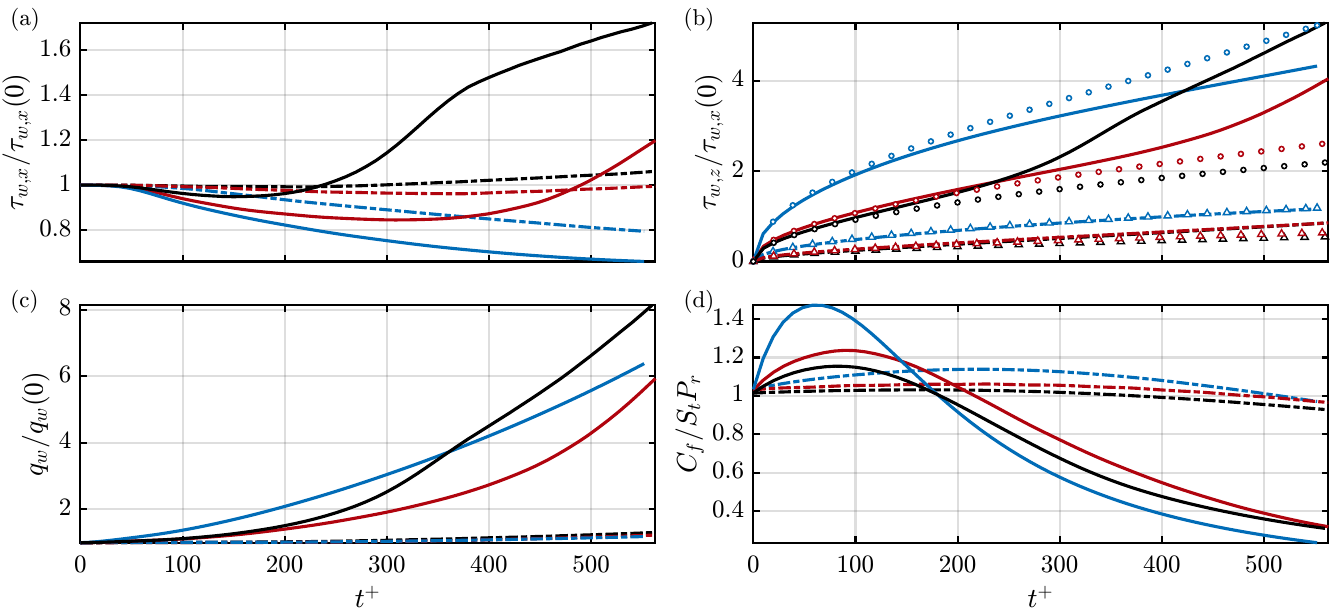}
                \caption{Temporal variation of (a) $\tau_{w,x}$, (b) $\tau_{w,z}$, (c) $q_{w}$, and (d) $C_{f}/(S_{t}\Pr)$. In (b), the symbols denote the laminar prediction, $2g_{z}\sqrt{\rho_{w}\mu_{w}t/\upi}$, with the circles and triangles denoting $\Pi = 40$ and $\Pi = 10$, respectively. The colors and line styles are defined in table \ref{table_setup}.}
                \label{fig_Scaling_WallShearStressHeatTransfer}
            \end{figure}

        The evolution of $\tau_{w,x}$ is shown in figure \ref{fig_Scaling_WallShearStressHeatTransfer}(a) for all $6$ cases. The temporal evolution of $\tau_{w,x}$ depends both on $Ma$ and $\Pi$ and is a direct result of the evolution of $u_{b}$, which is not here constrained as the streamwise flow is driven with a constant $g_{x}$. In incompressible studies of sudden spanwise acceleration~\citep{lozano2020non} and in the drag reduction observed in spanwise oscillations with constant $dp/dx$~\citep{ge2017response,ricco2021review}, $\tau_{w,x}$ is also affected by the spanwise acceleration.

        Owing to the success of the laminar prediction of $\wtil{w}$ for initial times, a laminar prediction of $\tau_{w,z}$ is also included via $\tau_{w,z} = \mu_{w}g_{z}\partial_{y}f(\eta)\vert_{y=0} = 2g_{z}\sqrt{\rho_{w}\mu_{w}t/\upi}$ in figure \ref{fig_Scaling_scaling_U_Umag_T}(b). This prediction agrees well for $\Pi = 10$ and for $\Pi = 40$ begins to deviate substantially for $t < t_{c}$ despite the success of the similarity solutions. The departure is present even for $Ma = 3$, indicating discrepancies in the near-wall prediction of $\wtil{w}$ from the similarity solution because of the production of turbulent stresses. Due to the net acceleration in the flow. $\tau_{w,z}$ increases monotonically during the transient. The increase in kinetic energy then results in additional viscous heating~\citep{lele1994compressibility}, raising both $\oline{T}$ in the channel and $q_{w}$, as shown in figure \ref{fig_Scaling_WallShearStressHeatTransfer}(c). The temporal evolution of $\tau_{w,z}$ and $q_{w}$ show more similarity for matched value of $\Pi$ rather than $Ma$.

        A common predictive tool in turbulent boundary layers is the Reynolds analogy factor, relating the friction coefficient, $C_{f}$ and the Stanton number, $S_{t}$~\citep{bradshaw1977compressible}. 
        While Fanning flow uses a friction coefficient based on the bulk velocity, here $C_{f}$ is related to the centerline quantities as an analogy for the freestream conditions. Here, $C_{f} = 2\norm{\vb{\tau}_{w}}/\qty(\rho_{c}\norm{\vtil{u}_{c}}^{2})$ and $S_{t} = q_{w}/\qty(\rho_{c}c_{p}\norm{\vtil{u}_{c}}\qty(T_{c} - T_{w}))$. Other definitions for these quantities have been used in the literature, primarily in the context of boundary layers. 
        The ratio, $C_{f}/S_{t}\Pr$, begins at $1$ and promptly evolves as $g_{z}$ is applied in figure \ref{fig_Scaling_scaling_U_Umag_T}(d). The Reynolds analogy factor, $f_{RA} = 2S_{t}/C_{f}$,~\citep{bradshaw1977compressible} is observed to be approximately $2\Pr$ for the canonical channels simulated herein at $t = 0$ and during the simulation time of the $\Pi = 10$ cases whereas $f_{RA}$ varies significantly for $\Pi = 40$.  
        The choice of $C_{f}$ and $S_{t}$ is non-unique. It is possible that different choices could lead to a constant ratio throughout the simulation time, though this is outside of the scope of this work. 

\section{Turbulent fluctuations} \label{S_Turb_Fluc}

            \begin{figure}
                \centering
                \includegraphics[width=\linewidth]{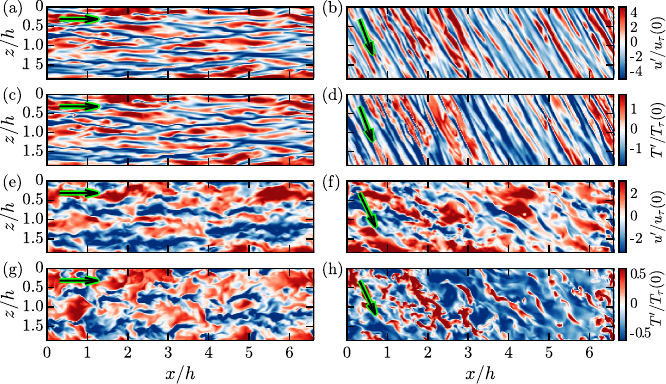}
                \caption{Instantaneous flow fields of (a,b,e,f) $u'/u_{\tau}(0)$ and (c,d,g,h) $T'/T_{\tau}(0)$ for $Ma = 1.5$ and $\Pi = 40$. The planes in (a--d) are at $y/\ell_{\nu}(0) = 15$ and (e--h) are at $y/\ell_{\nu}(0) = 100$ and (a,c,e,g) are at $t^{+} = 0$ and (b,d,f,h) are at $t^{+} = 415$. The black arrow denotes the instantaneous direction of $\vtil{U}$ at the wall-normal plane plotted. }
                \label{fig_turb_flucs_inst}
            \end{figure}

        Apart from changing the mean quantities, the spanwise acceleration also affects the turbulent structure throughout the channel. As a motivating picture, the instantaneous velocity and temperature fluctuation fields of $Ma = 1.5$ and $\Pi = 40$ at two different wall-normal heights are shown in figures \ref{fig_turb_flucs_inst}(a--h) for the canonical state at $t= 0$ and $t^{+} = 415$, where the spanwise flow has become significant. Qualitatively, the near-wall plane in figures \ref{fig_turb_flucs_inst}(b,d) demonstrate structures more uniformly aligned with the direction of the flow than the structures in the log-layer plane in figures \ref{fig_turb_flucs_inst}(f,h). In the near-wall planes, the $u'$ and $T'$ fields demonstrate a strong degree of correlation for both $t^{+} = 0$ and $t^{+} = 415$. The correlated $u'$ and $T'$ fields are also present at $t^{+} = 0$ for the log-layer plane. Due to the acceleration from the spanwise flow, these fields are anticorrelated at $t^{+} = 415$. 
        These qualitative observations from the instantaneous flows motivate the quantitative measures in the rest of this section.

    \subsection{Drop in turbulent kinetic energy and thermal fluctuations} \label{SS_Reynolds_Stresses}

        Spanwise acceleration has been shown to reduce the TKE via flow control strategies with active walls~\citep{quadrio2000numerical,choi2002drag,ge2017response,ricco2021review,marusic2021energy,rouhi2023turbulent,chandran2023turbulent}
        or the initial development of a spanwise flow~\citep{bradshaw1985measurements,moin1990direct,coleman1995numerical,lozano2020non}. From the sudden spanwise acceleration, incompressible studies~\citep{moin1990direct,lozano2020non} have concluded that due to the spanwise flow, the pressure-strain term, $\oline{p'\partial_{z}v'}$, drops, leading to a reduction in $\oline{v'v'}$. This leads to a reduction of the streamwise Reynolds shear stress production, $\oline{v'v'}\partial_{y}\oline{u}$, which reduces the magnitude of $\oline{u'v'}$. Finally, this causes a reduction in the production of the TKE, $\oline{u'v'}\partial_{y}\oline{u}$, leading to the reduction in the TKE. Once the spanwise flow is sufficiently developed, the additional spanwise Reynolds shear stresses lead to additional production in the TKE, ultimately increasing it. Mechanistically, the reduction in the Reynolds shear stresses occurs because of a misalignment between the near-wall structures and those further away from the wall leading to less efficient Reynold shear stress production~\citep{lozano2020non}. 
        The misalignment in the flow structures is described in more detail in section \ref{SS_Transport_of_Coherent_Structures}. In this section, the mechanisms described will be shown to be similar for the compressible Reynolds stresses and concludes with a description of the thermal turbulent transport.

            \begin{figure}
                \centering
                \includegraphics[width=\linewidth]{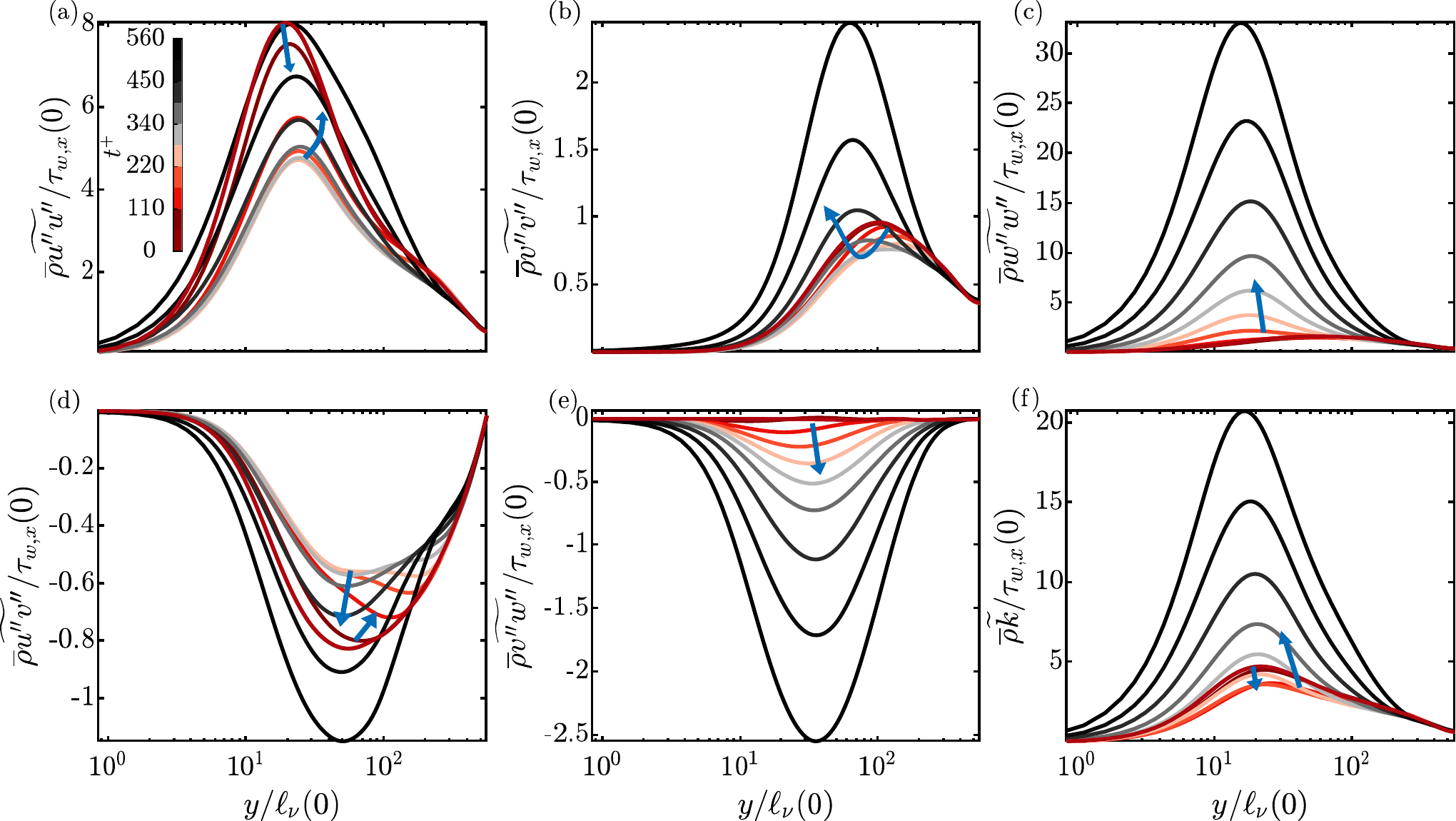}
                \caption{Temporal variation of (a--e) the Reynolds stresses and (f) TKE for $Ma = 1.5$ and $\Pi = 40$, normalized by the initial $\tau_{w,x}$. The lines are colored with the colorbar in (a) and arrows serve to illustrate the direction of the temporal variation in the statistics. }
                \label{figure_Turb_fluc_Pi_40}
            \end{figure}

        The TKE, $\oline{\rho}\wtil{k} = \oline{\rho}\qty(\wtil{u''u''} + \wtil{v''v''} + \wtil{w''w''})/2$ and some representative Reynolds stresses are plotted in figures \ref{figure_Turb_fluc_Pi_40}(a--f) for $Ma = 1.5$ and $\Pi = 40$. The evolution of the Reynolds stresses can be divided into two stages. The first stage is characterized by a reduction in $\oline{\rho}\wtil{u''u''}$ due to decreased production from the drop in $\oline{\rho}\wtil{u''v''}$. This coincides with a reduction in $\oline{\rho}\wtil{v''v''}$ due to the decrease in the production of $\oline{\rho}\wtil{u''v''}$~\citep{moin1990direct,lozano2020non}. While this is occurring, there is an increase in $\oline{\rho}\wtil{w''w''}$ due to the net transport in the spanwise direction's role in generating $\oline{\rho}\wtil{v''w''}$, leading to production of spanwise turbulent fluctuations. Despite the increase in kinetic energy and spanwise turbulent fluctuations, there is a reduction in the TKE driven by the reduction in production from $\oline{\rho}\wtil{u''v''}$ in agreement with various incompressible studies with imposed spanwise flows~\citep{bradshaw1977compressible,moin1990direct,lozano2020non}. During the second stage, the $\oline{\rho}\wtil{u''v''}$ intensifies, leading to an increase in $\oline{\rho}\wtil{u''u''}$ while $\oline{\rho}\wtil{w''w''}$ continues to increase. As a result, the TKE increases at this point as well. In \citet{gomezcompressibilityARB}, the $\Pi = 10$ and $Ma = 1.5$ case was shown to follow the same trends as the $\Pi = 40$ case, although the decrease in the magnitude of $\oline{\rho}\wtil{u''v''}$ was smaller, leading to less of a reduction in the TKE.

            \begin{figure}
                \centering
                \includegraphics[width=\linewidth]{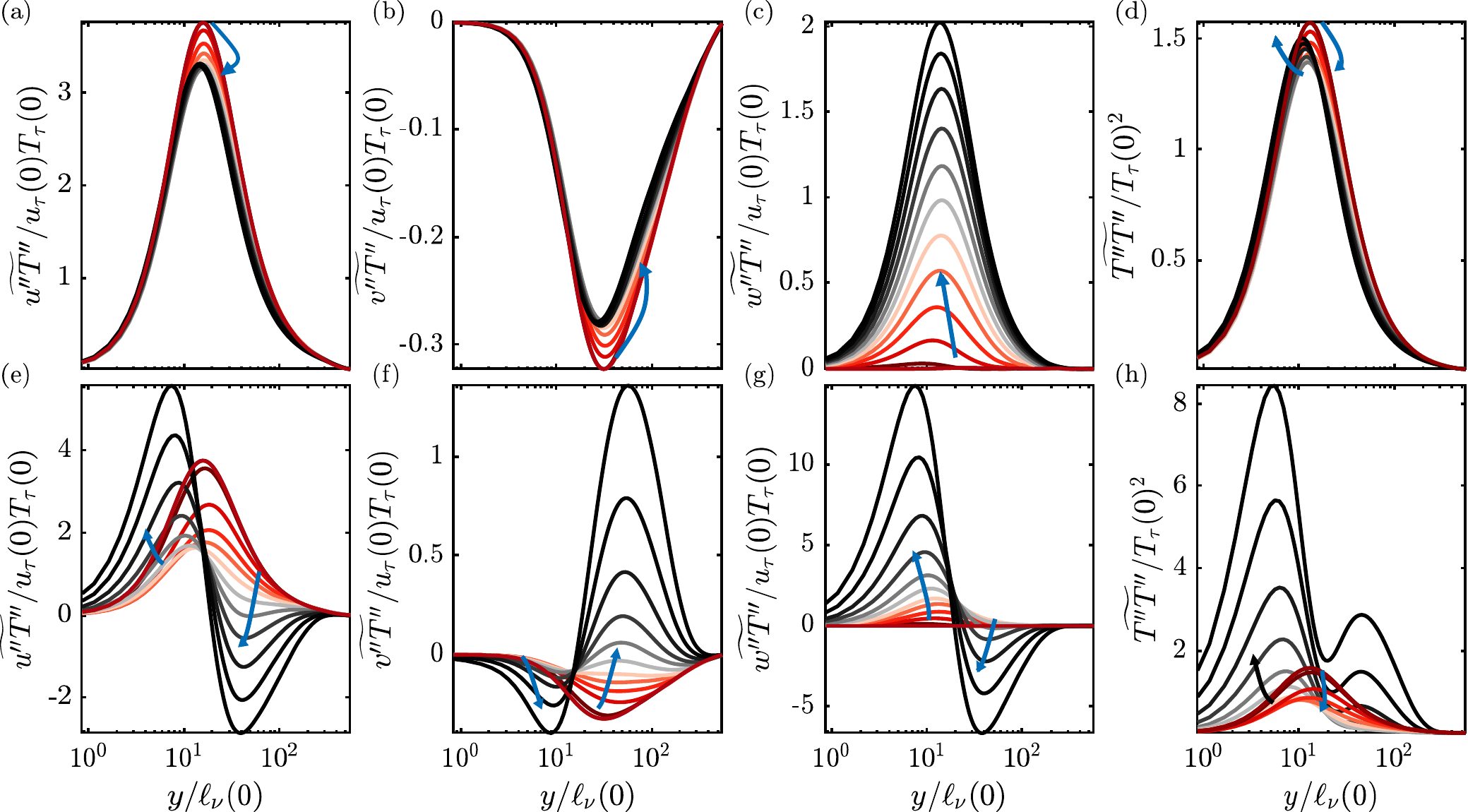}
                \caption{Temporal variation of (a--c, e--g) the velocity-temperature covariances normalized by $u_{\tau,x}(0)$ and $T_{\tau}(0)$ and (d,h) the thermal fluctuations normalized by $T_{\tau}(0)^{2}$ for $Ma = 1.5$ (a--d) $\Pi = 10$ and (e--h) $\Pi = 40$. The lines are colored with the colorbar in figure \ref{figure_Turb_fluc_Pi_40}(a) and arrows serve to illustrate the direction of the temporal variation in the statistics. }
                \label{figure_Temp_fluc_Pi_10}
            \end{figure}

        While the Reynolds stresses follow similar behavior between the $\Pi = 10$ and $\Pi = 40$ cases, the velocity-temperature covariances, $\wtil{\vb{u}''T''}$, have qualitatively different responses to the spanwise acceleration. Focusing first on the $\Pi = 10$ and $Ma = 1.5$ case in figures \ref{figure_Temp_fluc_Pi_10}(a--d), the covariances illustrate a slight reduction in $\wtil{u''T''}$ and $\wtil{v''T''}$ and a monotonic increase in $\wtil{w''T''}$. The latter reflects the increased transport in the spanwise direction. The slight changes in $\wtil{u''T''}$ reflect that the $u$ fluctuations are still correlated with the $T$ fluctuations. Additionally, the anticorrelation between $v''$ and $T''$ reflect the importance of sweep and ejection events in the compressible flow, despite the spanwise acceleration. Despite the change in the magnitudes, $\wtil{u''T''}$ and $\wtil{v''T''}$ behave similar to canonical compressible flows~\citep{coleman1995numerical}. Finally, $\wtil{T''T''}$ is shown to decrease in figure \ref{figure_Temp_fluc_Pi_10}(d), despite the net increase in $\wtil{T}$. This is similar to the net decrease in TKE.

        The $\Pi = 40$ and $Ma = 1.5$ case's velocity-temperature covariances and thermal fluctuations are plotted in figures \ref{figure_Temp_fluc_Pi_10}(e--h), illustrating a significantly different response than the $\Pi = 10$ case. For initial times, $\wtil{u''T''}$ and $\wtil{v''T''}$ decrease in magnitude while $\wtil{w''T''}$ increases, similar to the initial response in the $\Pi = 10$ case. Eventually, the velocity-temperature covariances change signs in the wall-normal direction at $t^{+} \approx 280$. This behavior is illustrated in the instantaneous visualizations of $u'$ and $T'$ in the near-wall and log-layer planes in figure \ref{fig_turb_flucs_inst}(b,d,f,h). This change in sign indicates a difference in the turbulent transport of the temperature fluctuations not present in the $\Pi = 10$ case. This can be seen by comparing $\wtil{T}$ between the $\Pi = 10$ and $\Pi = 40$ cases in figures \ref{fig_Scaling_scaling_U_Umag_T}(c,f). For $\Pi = 10$, $\wtil{T}$ remains monotonic in the wall-normal direction, whereas for $\Pi = 40$, a near-wall $\wtil{T}$ peak emerges for $t^{+} \approx 200$. Thus, for $t\gtrapprox200$, the lift-up mechanisms change the transport of $T'$ above this near-wall $\wtil{T}$ peak. Below this peak, $\wtil{u}$, $\wtil{w}$, and $\wtil{T}$ are all monotonically increasing like in the $\Pi = 10$ case, maintaining the same behavior in the velocity-temperature covariances. Away from the peak, $\wtil{T}$ is monotonically decreasing whereas $\wtil{u}$ and $\wtil{w}$ increase in $y$. This means that the lift-up mechanism advects low-speed, high temperature fluid up away from the temperature peak and high-speed, low temperature fluid down towards the peak. This then explains the change in sign in the velocity-temperature covariances above the near-wall $\wtil{T}$ peaks. This behavior is reflected in turbulent boundary layers~\citep{duan2010direct,pirozzoli2011turbulence}. In section \ref{SS_Organization_of_turb_flcus}, the lift-up mechanisms are discussed in more detail. Finally, $\wtil{T''T''}$ also decreases for initial times as shown in figure \ref{figure_Temp_fluc_Pi_10}(h), similar to what happened in the $\Pi = 10$ case. For later times, $\wtil{T''T''}$ increases and produces two peaks.

        The reduction in $\wtil{T''T''}$ can be explained by analyzing the terms in the budget of $\oline{\rho}\wtil{T''T''}$. The budgets are analyzed and plotted in appendix \ref{APP_Budgets} for more information. The term responsible for the decrease in $\oline{\rho}\wtil{T''T''}$ is the decrease in the thermal production stemming from $-2\oline{\rho}\wtil{v''T''}\partial_{y}\wtil{T}$ because of the reduction in the magnitude of $\wtil{v''T''}$. This is similar to how the reduction in the TKE is driven by a decrease in the production, $-\oline{\rho}\wtil{u''v''}\partial_{y}\wtil{u}$~\citep{moin1990direct,lozano2020non}.

    \subsection{Organization of turbulent and thermal transport}\label{SS_Organization_of_turb_flcus}

        In section \ref{SS_Reynolds_Stresses}, the presence of the near-wall peak in $\wtil{T}$ leads to a change in sign in the velocity-temperature covariances that is only observed in the $\Pi = 40$ case. Additionally, the development of $\wtil{w}$ leads to the production of $\oline{\rho}\wtil{w''w''}$ via the generation of $\oline{\rho}\wtil{v''w''}$. In this section, quadrant decompositions~\citep{wallace2016quadrant} will be used to study the organization of the turbulent and thermal transport. While the quadrant decomposition is most commonly used to quantify contributions of $u$ and $v$ based on their signs~\citep{wallace1972wall}, the quadrant decomposition has also been used to bin $T$ and $v$ fluctuations~\citep{perry1976experimental,nagano1988statistical,kong2000direct}. For a variable $a$ and wall-normal velocity $v$, the quadrants are organized into four quadrants, $Q_{i}$, where $Q_{1} = \qty{(a,v) : a''>0\textrm{ and } v''>0}$, $Q_{2} = \qty{(a,v) : a''<0\textrm{ and } v''>0}$, $Q_{3} = \qty{(a,v) : a''<0\textrm{ and } v''<0}$, and $Q_{4} = \qty{(a,v) : a''>0\textrm{ and } v''<0}$. For $u$, the $Q_{4}$ and $Q_{2}$ events are commonly denoted as sweep and ejection events, respectively. To avoid additional nomenclature, the $Q_{4}$ and $Q_{2}$ events for $w$ and $T$ will also be denoted as sweep and ejection events.

        The time evolution of the probability distributions of the $Q_{i}$ events was presented in \citet{gomezcompressibilityARB}. Here, the contributions of the $Q_{i}$ events to the Reynolds shear stresses and $\oline{\rho}\wtil{T''v''}$ are measured through
            \begin{equation}
                \expval{\oline{\rho}a''T''|Q_{i}} = \frac{1}{h}\int_{0}^{h}\oline{\rho}\qty(\wtil{a''T''}:(a,T)\in Q_{i})dy, \label{Eq_ThermalFluc_aT_Qi}
            \end{equation}
        for $a = u,w,T$. Equation \ref{Eq_ThermalFluc_aT_Qi} integrates $\oline{\rho}\wtil{a''T''}$ computed from the $Q_{i}$ across the channel. Note that $\sum_{i=1}^{4}\expval{\oline{\rho}a''T''|Q_{i}} = \expval{\oline{\rho}a''T''}$. 
        These measures are normalized with $W_{av}$, where
            \begin{equation}
                W_{av}(t)^{2} = \frac{1}{h^{2}}\int_{0}^{h}\oline{\rho}\wtil{a''a''}dy\int_{0}^{h}\oline{\rho}\wtil{v''v''}dy = \expval{\oline{\rho}a''a''}\expval{\oline{\rho}v''v''},
            \end{equation}
        for $a = u,w,T$, rather than $\expval{\oline{\rho}\wtil{a''v''}}$ since $W_{av}(t)$ is strictly positive. Normalizing with $\expval{\oline{\rho}\wtil{a''v''}}$ can be problematic since at $t = 0$, $\expval{\oline{\rho}\wtil{w''v''}} = 0$ and for the $\Pi = 40$ cases, due to the change in sign in $\wtil{v''T''}$, $\expval{\oline{\rho}\wtil{T''v''}}$ could be zero. Additionally, $\expval{\oline{\rho}a''T''|Q_{i}}/W_{av}(t) \in \qty[-1,1]$ via the Cauchy-Schwarz inequality.

            \begin{figure}
                \centering
                \includegraphics[width=\linewidth]{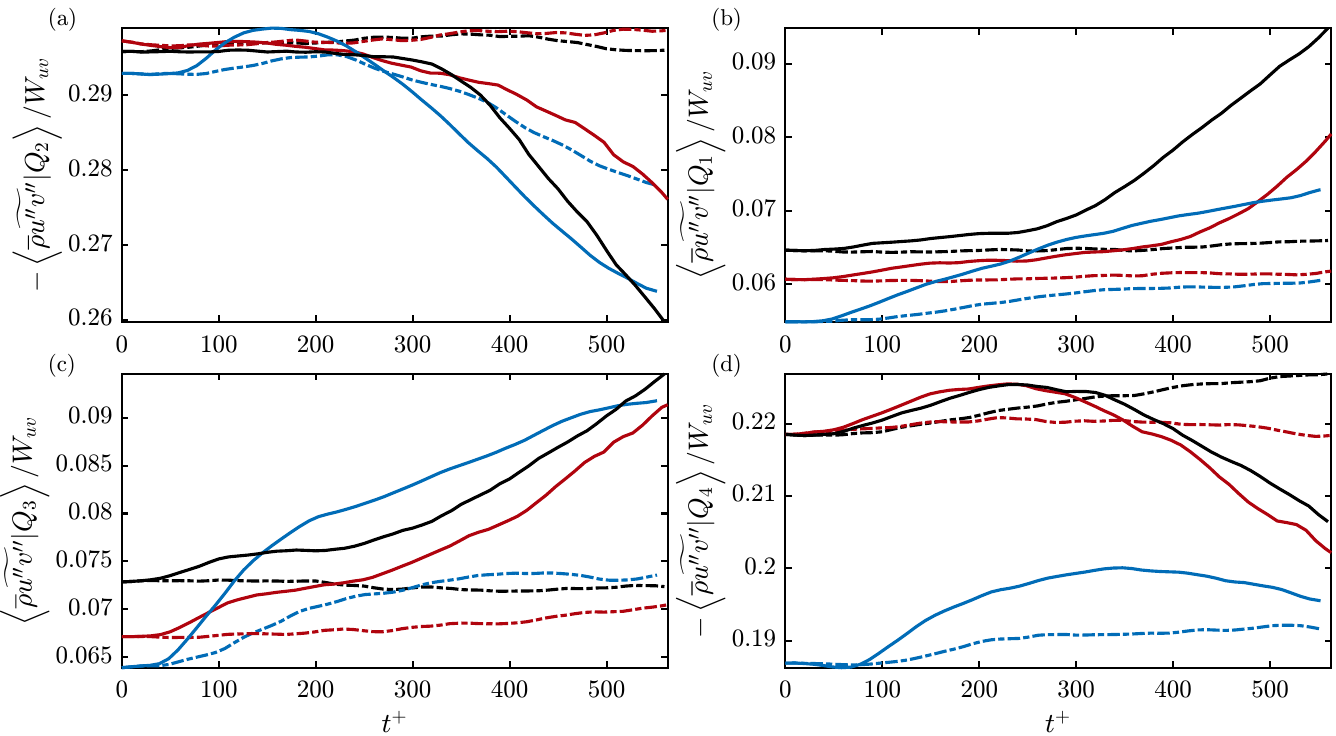}
                \caption{Temporal evolution of the (a) $Q_{2}$, (b) $Q_{1}$, (c) $Q_{3}$, and (d) $Q_{4}$ contributions to $\expval{\oline{\rho}\wtil{u''v''}}$, normalized by $W_{uv}(t)$. The colors and line styles are defined in table \ref{table_setup}. }
                \label{figure_Qi_uv}
            \end{figure}

        First, a classical quadrant decomposition is performed for $u''$ and $v''$ with the $Q_{i}$ contributions to $\expval{\oline{\rho}\wtil{u''v''}}$ presented in figure \ref{figure_Qi_uv}. Initially, the $Q_{2}$ and $Q_{4}$ events contribute the most to $\expval{\oline{\rho}u''v''}$, where the ejections play a bigger role in the contribution. The $Q_{1}$ and $Q_{3}$ contributions are small relative to the $Q_{2}$ and $Q_{4}$ events. This is in agreement with classical quadrant decomposition studies~\citep{wallace1972wall,perry1976experimental,wallace2016quadrant} since the flow is still in a canonical state. For later times, there is a slight increase in the $Q_{1}$ and $Q_{3}$ contributions for the $\Pi = 40$ cases, likely indicating a weakening of the streamwise alignment in the flow. However, the $Q_{2}$ and $Q_{4}$ events still remain the most dominant contribution as the flow evolves.

            \begin{figure}
                \centering
                \includegraphics[width=\linewidth]{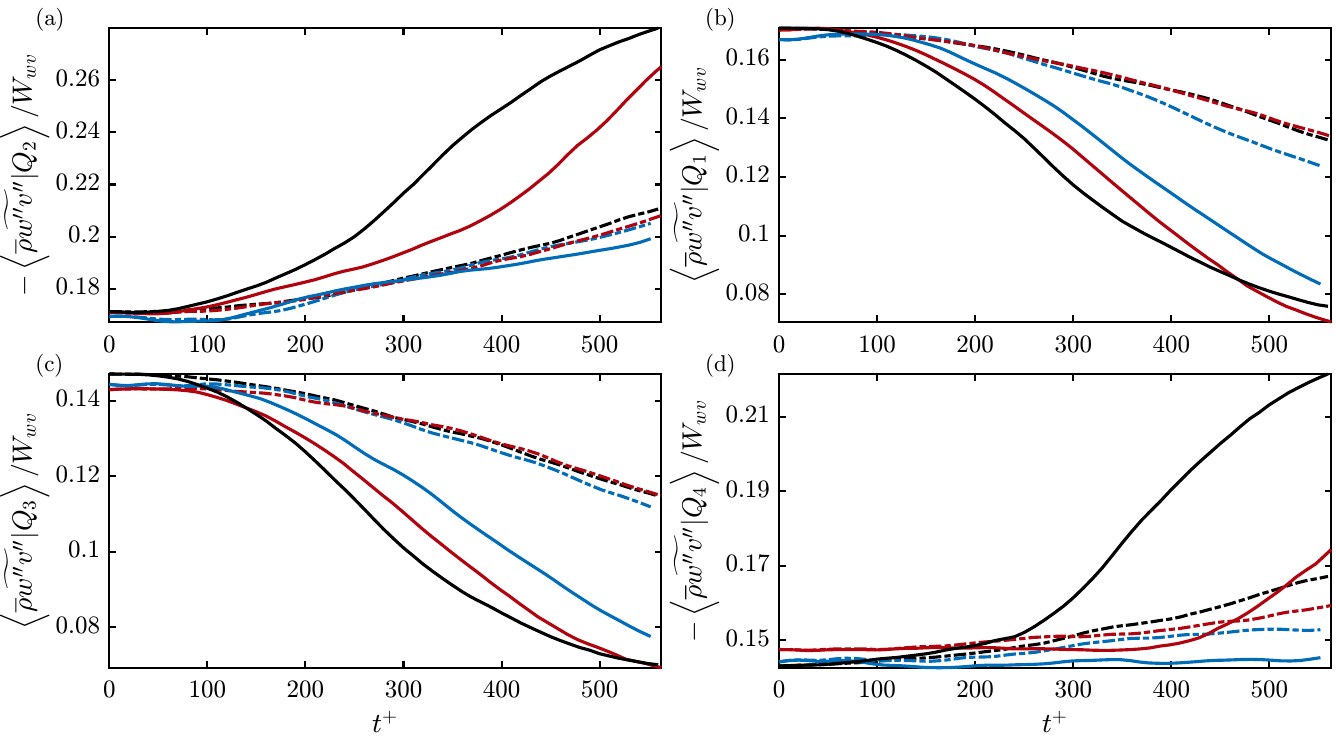}
                \caption{Temporal evolution of the (a) $Q_{2}$, (b) $Q_{1}$, (c) $Q_{3}$, and (d) $Q_{4}$ contributions to $\expval{\oline{\rho}\wtil{w''v''}}$, normalized by $W_{wv}(t)$. The colors and line styles are defined in table \ref{table_setup}. }
                \label{figure_Qi_wv}
            \end{figure}

        The $Q_{i}$ contributions to $\expval{\oline{\rho}\wtil{w''v''}}$ are shown in figure \ref{figure_Qi_wv}. At $t = 0$, they all contribute roughly equal amounts since $\expval{\oline{\rho}\wtil{w''v''}} = 0$ initially due to the lack of net transport in the spanwise direction. As the flow develops a net spanwise shear, $\wtil{w}_{y}$, the sweeps and ejection events are expected to play a role via lift-up effects.  Indeed, as time advances, $Q_{2}$ and $Q_{4}$ events play a larger role while the $Q_{1}$ and $Q_{3}$ become less relevant in contribution to $\expval{\oline{\rho}\wtil{w''v''}}$. There is a slight preference towards ejections rather than sweeps by comparing figures \ref{figure_Qi_wv}(a,d), especially as the $Ma$ increases. 

            \begin{figure}
                \centering
                \includegraphics[width=\linewidth]{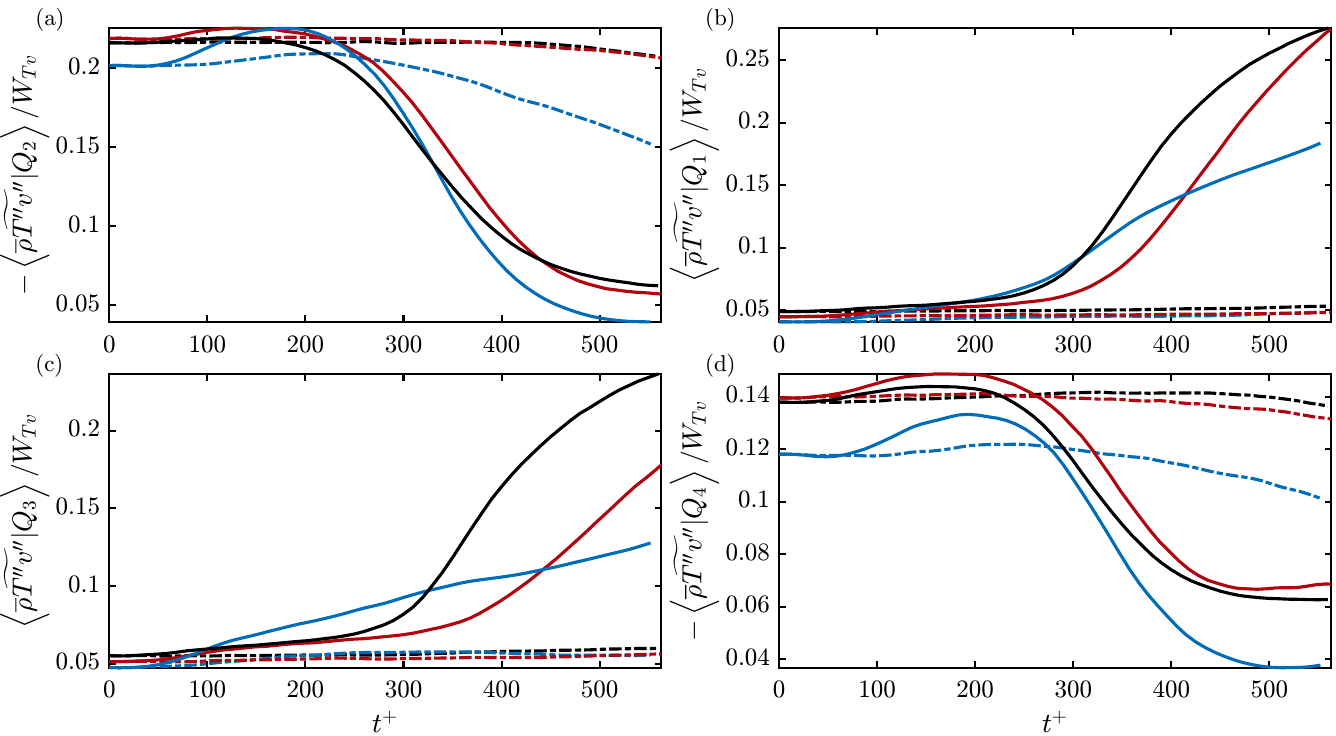}
                \caption{Temporal evolution of the (a) $Q_{2}$, (b) $Q_{1}$, (c) $Q_{3}$, and (d) $Q_{4}$ contributions to $\expval{\oline{\rho}\wtil{T''v''}}$, normalized by $W_{Tv}(t)$. The colors and line styles are defined in table \ref{table_setup}. }
                \label{figure_Qi_Tv}
            \end{figure}

        The contributions to $\expval{\oline{\rho}\wtil{T''v''}}$ from the $Q_{i}$ events are presented in figure \ref{figure_Qi_Tv}. At $t = 0$, the $Q_{2}$ and $Q_{4}$ contributions to $\expval{\oline{\rho}\wtil{T''v''}}$ are the most dominant, in agreement with the dominance of sweep and ejection events in the thermal transport of canonical shear flows~\citep{perry1976experimental,nagano1988statistical,kong2000direct}. As expected from the change in sign in the velocity-temperature covariances in figure \ref{figure_Temp_fluc_Pi_10}, the $Q_{i}$ contributions reveal distinct behavior for the $\Pi = 10$ and $\Pi = 40$ cases once the spanwise flow is sufficiently developed. For $\Pi = 10$, the $Q_{2}$ and $Q_{4}$ event contributions are dominant throughout the duration of the simulation, with only slight reductions in their magnitudes towards the end. The $\Pi = 40$ cases behave similar to the $\Pi = 10$ cases initially. For $t^{+} \approx 200$, coinciding with the time that the near-wall $\wtil{T}$ peak emerges, the $Q_{2}$ and $Q_{4}$ events begin to lose their dominance while the $Q_{1}$ and $Q_{3}$ contributions dominate. The change from $Q_{2}$ and $Q_{4}$ events to $Q_{1}$ and $Q_{3}$ events in $T$ support the explanation for the change in sign in the velocity-temperature covariances. That is, between the wall and the near-wall peak, the lift-up mechanism is dominated by sweeps and ejections such that $\mp v''$ coincide with $\pm T''$. Between the near-wall peak and $h$, the sweeps and ejections are no longer dominant as $\pm v''$ coincide with $\pm T''$ which give rise to $Q_{1}$ and $Q_{3}$ events. Since the near-wall peaks occur for $y/\ell_{\nu}(0) < 100$, the majority of the channel is no longer governed by sweeps and ejections of $T$. Because of the dominance in the $Q_{1}$ and $Q_{3}$ events, $\expval{\oline{\rho}\wtil{v''T''}}$ changes sign in the $\Pi = 40$ case. The trends in the $Q_{i}$ events for $T$ align with the qualitatively different behavior in the velocity-temperature covariances from figure \ref{figure_Temp_fluc_Pi_10} for the two $\Pi$'s. They further suggest that the $Ma$ is not responsible in changing the overall organization of the turbulent-thermal transport.

        The observations in figures \ref{figure_Qi_wv} and \ref{figure_Qi_Tv} agree with the results from \citet{gomezcompressibilityARB} where the contributions of the $Q_{i}$ events were measured through the time evolution of the probabilities of the $Q_{i}$ across the entire channel. Thus, whether the importance of the $Q_{i}$ events are computed as their contribution to the covariance or their probability, the conclusions are in agreement.

    \subsection{Transport of turbulent and thermal coherent structures} \label{SS_Transport_of_Coherent_Structures}

\begin{figure}
                \centering
                \includegraphics[width=\linewidth]{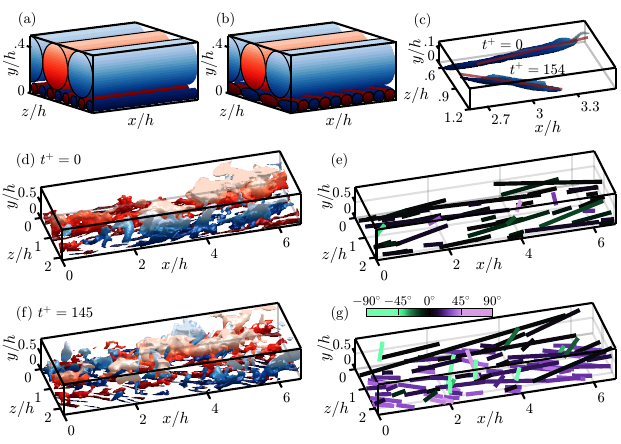}
                \caption{Cartoon of the near-wall small scale structures and large-scale structures at $t = 0$ (a) and $t>0$ (b) demonstrating the faster alignment of the near-wall structures than the larger structures further from the wall. Representative near-wall kinetic energy structures identified at $t^{+} = 0$ and $t^{+} = 145$ with their spines in red (c). Instantaneous realizations of identified kinetic energy structures (d,f) and their spines (e,g) at $t^{+} = 0$ (d,e) and $t^{+} = 145$ (f,g) for $Ma = 1.5$ and $\Pi = 40$. The shading of the isosurfaces reflects the wall-normal distance, darker colors are closer to the wall, and the blue and red denotes positive and negative fluctuations, respectively. The spines in (e,g) are colored based on their angle with respect to $\mathbf{e}_{x}$ as shown in the colorbar. Note tha the figures plot a subset of the full computational domain to highlight some representative structures and their spines.}
                \label{fig_3D_vizualizations}
            \end{figure}

        Having discussed the one-point statistics, the discussion now focuses on the structural properties of the three dimensional flow. 
        Previous work on the structural characteristics of three-dimensional boundary layers have reasoned that, in channels, the near-wall structures respond to the spanwise acceleration before the large scale structures. As a result, the large-scale structures and near-wall structures are misaligned~\citep{lozano2020non}. 
        This conceptual picture is illustrated in figures \ref{fig_3D_vizualizations}(a,b) for the canonical channel at $t=0$ and sometime after the spanwise body force has been imposed.
        Here, the structure identification proposed by \citet{lozano2012three} is used to identify the instantaneous kinetic energy structures and temperature structures to illustrate differences between hydrodynamic and thermal transport. In particular, the misalignment between the near-wall small-scale structures and large-scale structures will be quantified.

        To define the kinetic energy structures, the instantaneous kinetic energy is first defined as $K = \qty(u^2+v^2+w^2)/2$ and the fluctuating kinetic energy is then $K' = K - \oline{K}$. The kinetic energy structures and temperature structures are defined as connected regions, $\Omega$, where
            \begin{equation}
                \abs{K'} \ge 1.75\sqrt{\oline{K'K'}}
            \end{equation}
        and 
            \begin{equation}
                \abs{T'} \ge 1.75\sqrt{\oline{T'T'}},
            \end{equation}
        respectively. Note that because $\oline{K} = \qty(\oline{u_{i}'u_{i}'} + \oline{u}_{i}\oline{u}_{i})/2$, $\oline{K'K'}$ is different from the TKE--- it is the variance of the instantaneously fluctuating kinetic energy. The coefficient $1.75$ determines the threshold size and the results of the structure identification have been shown to be robust to changes in its value when used to identify Reynolds shear-stress structures~\citep{lozano2012three}.
        Connected regions whose volume is smaller than $\qty(30\ell_{\nu}(0))^{3}$ are rejected to avoid the accumulation of small disconnected structures. Each of the structure's sign can be computed based on the sign of $K'$ or $T'$ within each of the kinetic energy or temperature structures. To compute the angle of each structure, the center of mass of each structure, $\vb{x}_{m} = x_{m}\vb{e}_{x} + y_{m}\vb{e}_{y} + z_{m}\vb{e}_{z}$, is first computed as
            \begin{equation}
                \vb{x}_{m}V = \int_{\Omega}\vb{x}dxdydz,
            \end{equation}
        where $V$ is the volume of the region $\Omega$. Then, for each structure, the matrix, $\vb{X}$ is computed as 
            \begin{equation}
                X_{jk}V = \int_{\Omega} \qty(x_{j} - x_{j,m})\qty(x_{k} - x_{k,m})dxdydz.
            \end{equation}
        Since, $\vb{X}$ is symmetric, it can be written in its spectral representation as $\vb{X} = \sum_{j}^{3}\alpha_{j}\vb{p}_{j}\otimes\vb{p}_{j}$ where $\alpha_{1} \ge \alpha_{2} \ge \alpha_{3}$ and $\vb{p}_{j}$ are orthonormal eigenvectors often denoted as the principal axes. The principal direction of each structure is then $\vb{p}_{1}$. The direction along the wall-parallel plane is $\vb{e}_{p}$, the unit vector parallel to $\vb{p}_{1} - \qty(\vb{p}_{1}\cdot\vb{e}_{y})\vb{e}_{y}$. Finally, the length of each structure is defined as $r_{1} = \textrm{max}(d) - \textrm{min}(d)$, where
            \begin{equation}
                d = \qty[\qty(x-x_{m})\vb{e}_{x} + \qty(z-z_{m})\vb{e}_{z}]\cdot\vb{e}_{p} : (x,y,z) \in \Omega
            \end{equation}
        is the wall-parallel distance of $\Omega$ along $\vb{e}_{p}$. The angle of the structure is defined as $\theta$, the angle between $\vb{e}_{p}$ and $\vb{e}_{x}$, as $\cos(\theta) = \vb{e}_{p}\cdot\vb{e}_{x}$. Finally, the spines are defined as the line segment that intersects $\vb{x}_{m}$ parallel to $\vb{p}_{1}$ with a wall-parallel length of $r_{1}$.

        In figure \ref{fig_3D_vizualizations}(c), the structure identification is illustrated for two representative near-wall kinetic energy structures at $t^{+} = 0$ and $t^{+} = 145$ for the $Ma = 1.5$ and $\Pi = 40$ case. Their spines are included as well for visualization, illustrating that they can recover the statistical length of the structures and their orientation. They also highlight the change in the orientation of the near-wall structures as the flow is sufficiently accelerated. Two snapshots at $t = 0$ and $t^{+} = 145$ are presented in figures \ref{fig_3D_vizualizations}(d,f) with the surface contours of the identified structures within a subset of the channel domain. At $t^{+} = 0$, one can qualitatively note that the small-scale near-wall structures and large-scale structures are aligned along the streamwise direction, like in the cartoon presented in figure \ref{fig_3D_vizualizations}(a). At $t^{+} = 145$, the data reflects figure \ref{fig_3D_vizualizations}(b) as the near-wall small-scale structures are aligning towards the spanwise direction while the large-scales are mostly streamwise aligned. Figures \ref{fig_3D_vizualizations}(e,g) plot the spines of each of the identified structures in figures \ref{fig_3D_vizualizations}(d,f), respectively, color coded by their angle $\theta$, the angle between the spine and $\vb{e}_{x}$. At $t^+ = 0$, the spines all primarily have $\theta \approx 0^{\circ}$. There are some spines with large $\theta$ corresponding to smaller-scale structures that are longer in their spanwise extent than their streamwise extent. Thus, the spine detection algorithm treats these structures as spanwise aligned. At $t^{+} = 145$, the near-wall small-scale spines have $\theta \gtrsim 45^{\circ}$ while the longer spines away from the wall's $\theta$ reflect their streamwise alignment with smaller $\theta$. After a long enough time, the large-scale structures will eventually align with and equilibrate with the new direction of the flow. However, the focus of this section is on the initial period of the sudden spanwise acceleration.

        To further quantify the orientation of the spines, instantaneous weighted histograms of $y_{m}$ and $\theta$, $H(y_m,\theta)$, are computed for each time instance across all $8$ ensembles for each case and are weighted by the length $r_{1}(y_{m},\theta)$. Explicitly, the weighted histograms are computed as
            \begin{equation}
                R_{1}H(\theta,y_{m}) = \sum_{\bre{\theta} -\theta= \Delta\theta_{1}}^{\Delta \theta_{2}}\sum_{\bre{y}_{m} -y_{m} = \Delta y_{m,1}}^{\Delta y_{m,2}}r_{1}(\bre{y}_{m},\bre{\theta}), \label{Sec_Turb_Eq_Hist}
            \end{equation}
        where $R_{1} = \sum_{\theta}\sum_{y_{m}}r_{1}(y_{m},\theta)$ is the total length of all the structures and $\Delta\theta_{1},$ $\Delta \theta_{2}$, $\Delta y_{m,1}$, and $\Delta y_{m,2}$ denote the edges of the bins. The histograms use $22$ linearly spaced bins in $\theta \in \qty(-20^\circ,70^\circ)$ and $16$ logarithmically spaced bins in $y_{m}/\ell_{\nu}(0) \in \qty(2,450)$. The structures above the channel half-height are mapped to $y_{m} \rightarrow 2h - y_{m}$ in equation \ref{Sec_Turb_Eq_Hist} to take advantage of the statistical symmetry across the channel half-height. Weighing the histograms with $r_{1}$ removes the influence of small-scale structures away from the wall whose alignment is poorly defined, like those pictured in figures \ref{fig_3D_vizualizations}(e,g). In appendix \ref{APP_Unweighted_Histogram}, an unweighted histogram is shown that supports the conclusions from the weighted histogram. The orientation of the identified structures are also compared with the angle between two statistical quantities, $\wtil{a}$ and $\wtil{b}$, as 
            \begin{equation}
                \textrm{tan}\qty(\theta_{\wtil{a},\wtil{b}}) = \frac{\wtil{a}}{\wtil{b}},
            \end{equation}
        where $\wtil{a}$ and $\wtil{b}$ represent a streamwise and spanwise flow statistic, respectively. These angles are used to measure the orientation of the wall-parallel flow field, $\theta_{\wtil{u},\wtil{w}}$, the Reynolds shear stresses, $\theta_{\wtil{u''v''},\wtil{w''v''}}$, and velocity-temperature covariances, $\theta_{\wtil{u''T''},\wtil{w''T''}}$, with respect to $\vb{e}_{x}$. Both $\theta_{\wtil{u},\wtil{w}}$ and $\theta_{\wtil{u''v''},\wtil{w''v''}}$ have been used in incompressible studies, demonstrating a lag between the mean flow field and Reynolds shear stresses~\citep{bradshaw1985measurements,moin1990direct,lozano2020non}.
        The addition of $\theta_{\wtil{u''T''},\wtil{w''T''}}$ serves to define a direction based on the temperature-velocity covariances.

            \begin{figure}
                \centering
                \includegraphics[width=\linewidth]{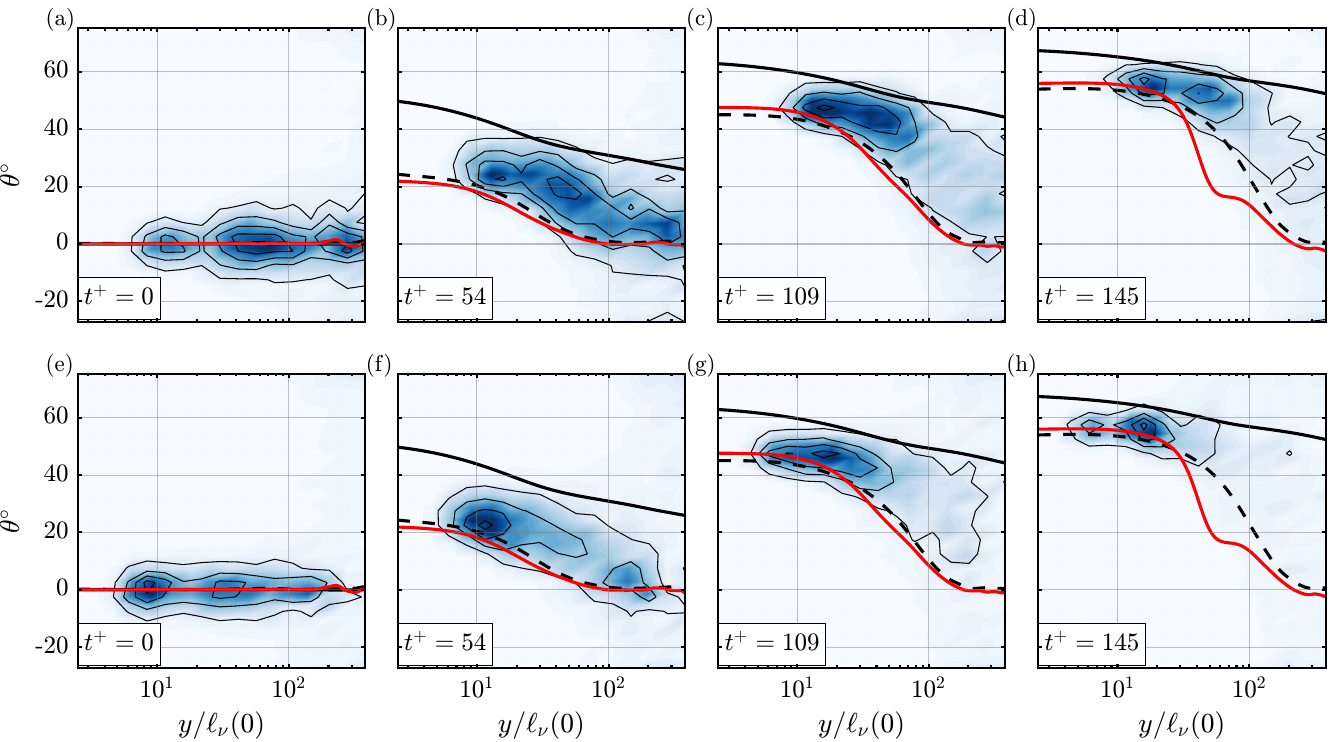}
                \caption{Weighted histogram of identified (a--d) kinetic energy structures and (e--h) temperature structures as a function of $y_{m}$ and $\theta$ for Ma = 1.5 and $\Pi = 40$. The solid black, dashed black, and solid red lines denote $\theta_{\wtil{u},\wtil{w}}$, $\theta_{\wtil{u''v''},\wtil{w''v''}}$, and $\theta_{\wtil{u''T''},\wtil{w''T''}}$, respectively. }
                \label{fig_y_vs_theta_k_T_Ma_1_5}
            \end{figure}

        The $H(\theta,y_{m})$ are computed for the temperature structures and kinetic energy structures of the $Ma = 1.5$ and $\Pi = 40$ case for each time snapshot. The $H(\theta,y_{m})$ of the kinetic energy structures in figures \ref{fig_y_vs_theta_k_T_Ma_1_5}(a--d) quantify the observations from figure \ref{fig_3D_vizualizations} across all ensembles. As expected, the structures begin statistically likely to be streamwise aligned at $t=0$ regardless of their wall-normal centroid, $y_{m}$. As evidenced by $H(\theta,y_{m})$, the kinetic energy structures closest to the wall are the first to respond to the spanwise acceleration while the structures further from the wall are more likely to be aligned closer to $\vb{e}_{x}$. The $H(\theta,y_{m})$ of the temperature structures in figures \ref{fig_y_vs_theta_k_T_Ma_1_5}(e--h) follow similar trends as the kinetic energy structures with near-wall structures more statistically likely to align in the spanwise direction. The kinetic energy and temperature structures' $H(\theta,y_{m})$ suggest that the temperature structures are more likely to be closer to the wall than the kinetic energy structures once $g_{z}$ has been applied. 
        Similar to incompressible channels, for $t>0$, $\theta_{\wtil{u},\wtil{w}}$ is greater than $\theta_{\wtil{u''v''},\wtil{w''v''}}$ such that the direction of the Reynolds shear stresses lag compared to the direction of the mean flow field. For initial times ($t^{+} \lessapprox 54$), both $\theta_{\wtil{u''v''},\wtil{w''v''}}$ and $\theta_{\wtil{u''T''},\wtil{w''T''}}$ are similar, albeit with a slight lag in the orientation of the velocity-temperature covariances' orientation with respect to the Reynolds stresses. For later times, near the wall, the $\theta_{\wtil{u''v''},\wtil{w''v''}} < \theta_{\wtil{u''T''},\wtil{w''T''}}$, coinciding the the increased presence of turbulent structures in that region. Further away from the wall, $\theta_{\wtil{u''v''},\wtil{w''v''}}$ and $\theta_{\wtil{u''T''},\wtil{w''T''}}$ differ for later times as the velocity-temperature covariances change sign, reflecting differences in the thermal fluctuation transport not reflected in the Reynolds stresses. For the times plotted here, the peaks of $H(\theta,y_{m})$ occur within $\theta_{\wtil{u''v''},\wtil{w''v''}}$ and $\theta_{\wtil{u},\wtil{w}}$.

            \begin{figure}
                \centering
                \includegraphics[width=\linewidth]{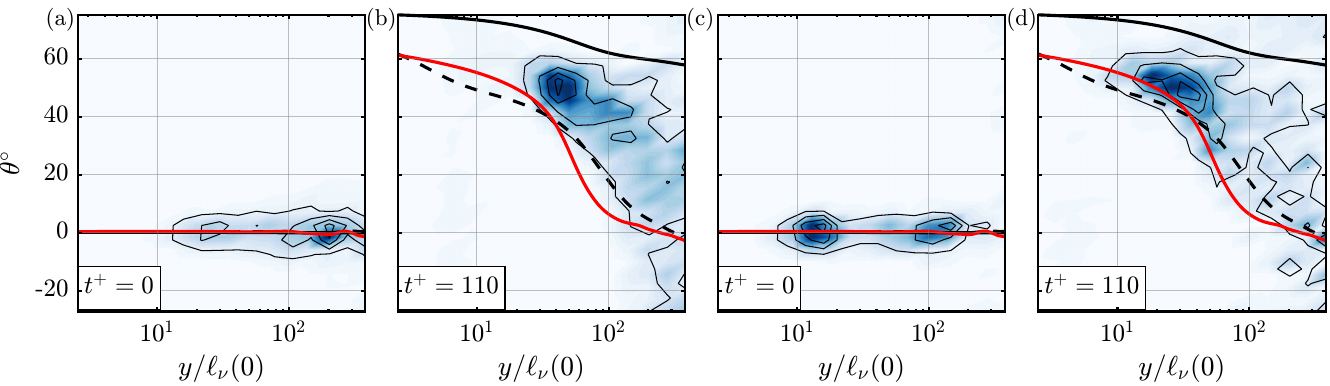}
                \caption{Weighted histogram of identified (a,b) kinetic energy structures and (c,d) temperature structures as a function of $y_{m}$ and $\theta$ for Ma = 3.0 and $\Pi = 40$. The solid black, dashed black, and solid red lines denote $\theta_{\wtil{u},\wtil{w}}$, $\theta_{\wtil{u''v''},\wtil{w''v''}}$, and $\theta_{\wtil{u''T''},\wtil{w''T''}}$, respectively. }
                \label{fig_y_vs_theta_k_T_Ma_3_0}
            \end{figure}

        In figure \ref{fig_y_vs_theta_k_T_Ma_3_0}, identified kinetic energy and temperature structures of the $Ma = 3.0$ and $\Pi = 40$ case are presented for two representative times. Just like the $Ma = 1.5$ case, the orientation of the near-wall kinetic energy and temperature structures align away from $\vb{e}_{x}$ before the structures further from the wall. Similar to the $Ma = 1.5$ case, the temperature structures are more likely to be closer to the wall than the kinetic energy structures based on the peak location of $H(\theta,y_{m})$. The discrepancy between the positions of the kinetic energy and temperature structures is larger for the $Ma = 3.0$ case than the $Ma = 1.5$ case. This discrepancy is also reflected in the difference in $\theta_{\wtil{u''v''},\wtil{w''v''}}$ and $\theta_{\wtil{u''T''},\wtil{w''T''}}$ in the near-wall region, where the latter is much more aligned with $\theta_{\wtil{u},\wtil{w}}$ than the former.

            \begin{figure}
                \centering
                \includegraphics[width=\linewidth]{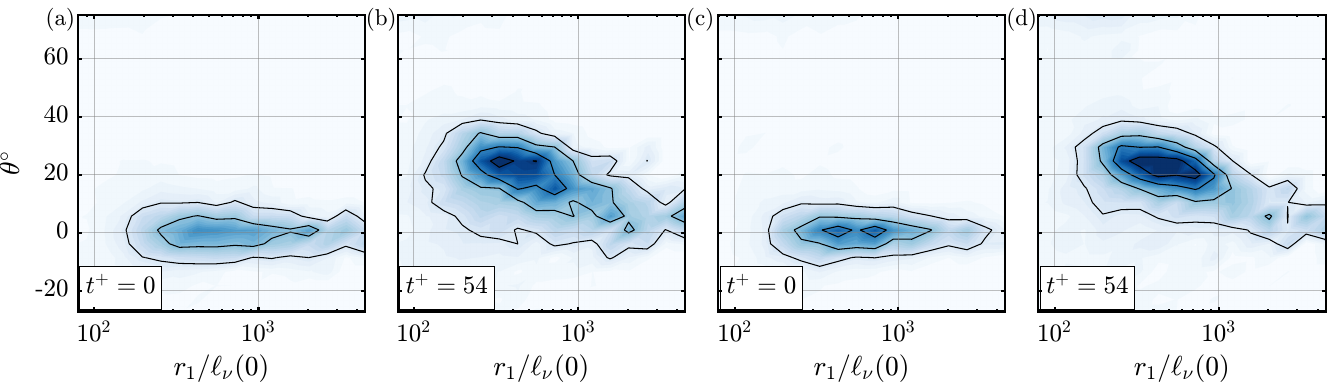}
                \caption{Weighted histogram of identified temperature structures (a,b) and kinetic energy structures (c,d) as a function of $r_{1}$ and $\theta$ for Ma = 1.5 and $\Pi = 40$.}
                \label{fig_r_1_vs_theta_k_T_Ma_1_5}
            \end{figure}

        Studying $H(\theta,y_{m})$ only reflects the alignment of the structures based on their wall-normal location. Here, the alignment of the structures is computed using a weighted histogram based on $\theta$ and $r_{1}$ via $H_{r}(\theta,r_{1})$ to determine the orientation based on a measure of the structure's size. This is computed as 
            \begin{equation}
                R_{1}H_{r}(\theta,r_{1}) = \sum_{\bre{\theta} -\theta= \Delta\theta_{1}}^{\Delta \theta_{2}}\sum_{\bre{r}_{1} -r_{1} = \Delta r_{1,1}}^{\Delta r_{1,2}}r_{1}(\bre{r}_{1},\bre{\theta}), \label{Sec_Turb_Eq_Hist_r_1}
            \end{equation}
        where $\Delta\theta_{1},$ $\Delta \theta_{2}$, $\Delta r_{1,1}$, and $\Delta r_{1,2}$ once denote the edges of the bins. This is similar to equation \ref{Sec_Turb_Eq_Hist}, except the binning and weighing are both done with respect to $r_{1}$. For $H_{r}(\theta,r_{1})$, there are $22$ linearly spaced bins in $\theta \in \qty(-20^\circ,70^\circ)$ and $16$ logarithmically spaced bins in $r_{1}/\ell_{\nu}(0) \in \qty(80,5000)$. These histograms are computed for the $Ma = 1.5$ and $\Pi = 40$ for both the kinetic energy and temperature structures in figures \ref{fig_r_1_vs_theta_k_T_Ma_1_5}(a--d) for $t^{+} = 0$ and $t^{+} = 54$. At $t^{+}=0$, the structures are statistically likely to be oriented along $\vb{e}_{x}$. For $t^{+} = 54$, $H_{r}(\theta,r_{1})$ suggest that the small-scale structures are more likely to be aligned with larger values of $\theta$ than the large-scale structures for both kinetic energy and temperature structures. This is expected, considering that larger structures tend to be found further from the wall. Together, $H(\theta,y_{m})$ and $H_{r}(\theta,r_{1})$ reveal that the small-scale near-wall structures align faster in the direction of the new body force than the large-scale structures away from the wall.

    \subsection{Temporal evolution of the wavenumber spectra} \label{SS_Spectra}

        The previous section illustrated differences in the turbulent structure as a result of the net spanwise acceleration through identified turbulent structures. Here, changes in the turbulent structure will be further studied by comparing the temporal evolution of the spectra. While this section considers Fourier transforms along the streamwise and spanwise directions, it could be possible to define the the Fourier transforms along $\vb{e}_{s}$ and its perpendicular, $\vb{e}_{s}\wedge
        \vb{e}_{y}$, or other directions defined by the flow statistics. However, the identified turbulent structures in the previous section were shown to be statistically likely to be aligned between $\theta_{\wtil{u},\wtil{w}}$ and $\theta_{\wtil{u''v''},\wtil{w''v''}}$, suggesting that axes based on the turbulent statistics do not capture the alignment of the structures. Thus, for ease of interpretation and avoiding wall-normally varying axes, the Fourier transforms here are taken along the fixed streamwise and spanwise directions.

        First, the streamwise Fourier modes of a variable $f$ are defined as
            \begin{equation}
                \what{f}(y,t;k_{x}) = \frac{1}{L_{z}L_{x}}\int_{0}^{L_{z}}\int_{0}^{L_{x}}f(x,y,z,t)e^{ik_{x}x}dxdz,
            \end{equation}
        averaged over the $8$ ensembles and over the channel half-height, where $k_{x} = 2\upi/\lambda_{x}$ is the streamwise wavenumber and $\lambda_{x}$ is the wavelength. The streamwise premultiplied power spectrum of the variable $f$ is then $E_{ff}(y,k_{x},t) = k_{x}\abs{\what{f}(y,t;k_{x})}^{2}$ for $k_{x}>0$. The spanwise premultiplied power spectrum, $E_{ff}(y,k_{z},t)$, is defined similarly by swapping $x$ and $z$ using the spanwise wavenumber, $k_{z}$, and spanwise wavelength, $\lambda_{z}$.

            \begin{figure}
                \centering
                \includegraphics[width=\linewidth]{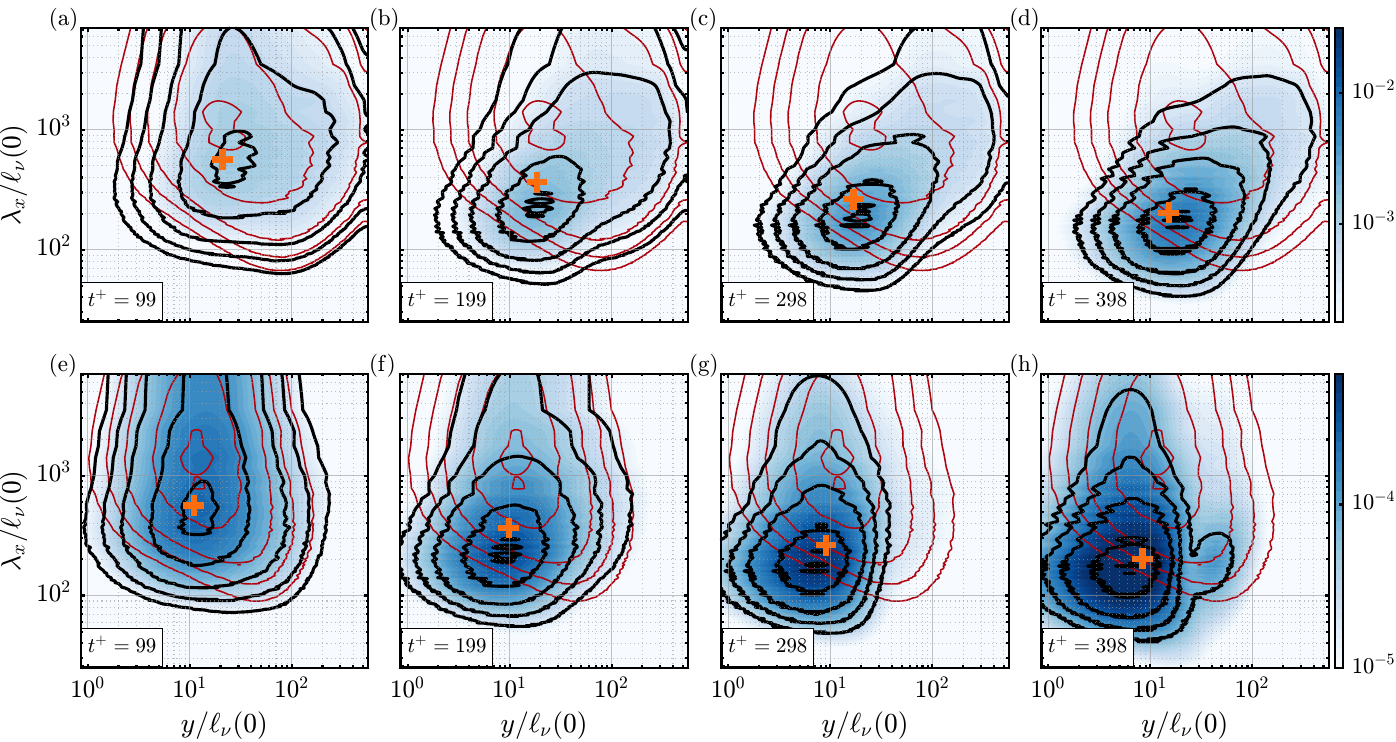}
                \caption{Instantaneous contours of the viscous-scaled streamwise premultiplied spectra for $Ma = 1.5$ and $\Pi = 40$ with $\ell_{\nu}(0)E_{kk}(y,k_{x},t)/u_{\tau}(0)^{2}$ (a--d) and $\ell_{\nu}(0)E_{TT}(y,k_{x},t)/T_{\tau}(0)^{2}$  (e--h). The solid lines are isocontours of $5\%$, $10\%$, $20\%$, $50\%$, and $95\%$ of the maximum instantaneous premultiplied spectra. The black solid lines and colored contours are at the time listed while the red solid lines are taken at $t^{+} = 0$ as a comparison. The solid orange crosses denote the predicted location of the near-wall peak via equation \ref{S_Turb_eq_PredLoc}.}
                \label{fig_y_vs_lambda_x_Ma_1_5_PI_40}
            \end{figure}

        The premultiplied kinetic energy spectra, $E_{kk} = \qty(E_{uu} + E_{vv} + E_{ww})/2$, and premultiplied temperature spectra, $E_{TT}$, are compared for the $Ma = 1.5$ and $\Pi = 40$ case in figure \ref{fig_y_vs_lambda_x_Ma_1_5_PI_40} for four representative times. At $t^{+} = 0$, $E_{kk}(y,k_{x},t)$ and $E_{TT}(y,k_{x},t)$ peak near $\qty(y/\ell_{\nu}(0),\lambda_{x}/\ell_{\nu}(0)) = \qty(20,1000)$ and $\qty(10,1000)$, respectively, as expected for the near-wall cycle in wall-bounded turbulent flows~\citep{lee2015direct}. 
        Figures \ref{fig_y_vs_lambda_x_Ma_1_5_PI_40}(a,e) reveal that the most significant initial temporal evolution in the spectra is seen in the near-wall small-scales while the spectra in the large-scale outer region mostly coincide with the canonical spectra. As time advances, the near-wall peaks move to smaller values of $\lambda_{x}/\ell_{\nu}(0)$ for both $E_{kk}$ and $E_{TT}$. For $t^{+} \le 200$, the contours for large $\lambda_{x}$ away from the wall still coincide with the contours from $t^{+} = 0$, reinforcing the observations from section \ref{SS_Transport_of_Coherent_Structures} that these large-scales in the outer region are affected less by the spanwise acceleration than the small-scales near the wall. Similar observations were made with structural observations of wall-attached Reynolds shear-stress carrying eddies in incompressible flow~\citep{lozano2020non}. Eventually, the large-scale structures become less energetic as they too orient away from $\vb{e}_{x}$.

        The movement of the near-wall peak to smaller $\lambda_{x}$ and $y$ is related to two mechanisms. The first is as $\norm{\vb{\tau}_{w}}$ increases, the viscous length-scale decreases causing the characteristic size of the near-wall structures to subsequently shrink. The second is that as the near-wall structures align with the direction of the net acceleration, their extent along the streamwise direction decreases as in the schematics of figures \ref{fig_3D_vizualizations}(a,b). As a result of these two mechanisms and assuming these near-wall structures have an orientation of $\theta$, the characteristic streamwise wavelength, $l$, and wall-normal location, $y_{l}$, of the structures in the near-wall cycle can be approximated as
            \begin{equation}
                \qty(l(t),y_{l}(t)) \approx \qty(1000 \cos(\theta(t)),y_{0})\ell_{\nu}(t),\label{S_Turb_eq_PredLoc}
            \end{equation} 
        where $y_{0} = 20\ell_{\nu}(0)$ and $10 \ell_{\nu}(0)$ for $E_{kk}(y,k_{x},t)$ and $E_{TT}(y,k_{x},t)$, respectively. These are shown to approximately fall near the location of the near-wall peaks, albeit in log-space.

            \begin{figure}
                \centering
                \includegraphics[width=\linewidth]{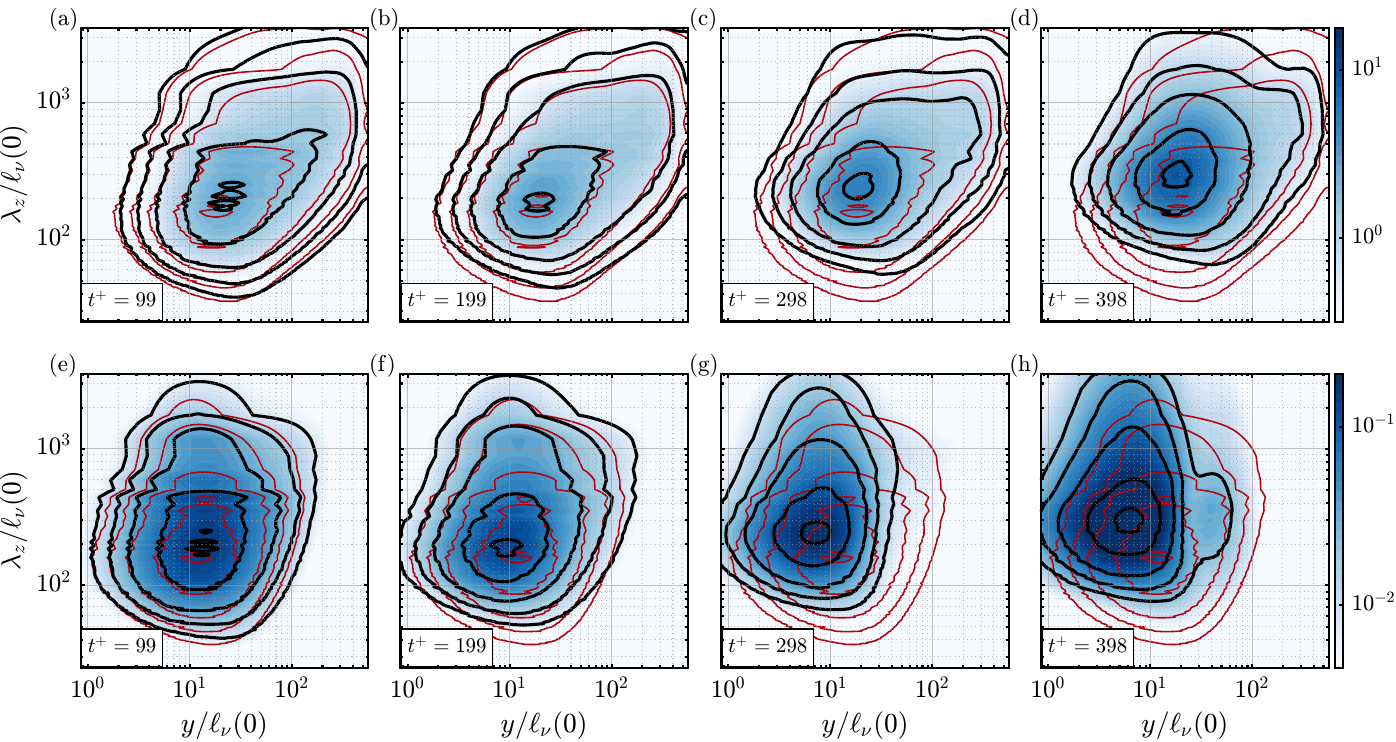}
                \caption{Same as Figure \ref{fig_y_vs_lambda_x_Ma_1_5_PI_40}, except for the premultiplied spanwise spectra.}
                \label{fig_y_vs_lambda_z_Ma_1_5_PI_40}
            \end{figure}

        Finally, the premultiplied spanwise spectra, $E_{kk}(y,k_{z},t)$ and $E_{TT}(y,k_{z},t)$ are plotted in figure \ref{fig_y_vs_lambda_z_Ma_1_5_PI_40}. At the initial time, both $E_{kk}$ and $E_{TT}$ peak around $y/\ell_{\nu} \approx 15$ and $\lambda_{z}/\ell_{\nu}(0) \approx 200$, which are characteristic of the near-wall cycle in both incompressible and compressible flows~\citep{lee2015direct,cogo2022direct_spectra}. Unlike the streamwise spectra, $E_{kk}(y,k_{z},t)$, reacts slower to the spanwise acceleration such that the isocontours for up to $t^{+} = 199$ approximately overlap with those at $t^{+} = 0$. As time advances, $E_{kk}$ moves to larger $\lambda_{z}$ due to the increased alignment of the kinetic energy structures towards $\vb{e}_{z}$. For $E_{TT}$, the isocontours begin to depart from those at $t^{+} = 0$ for earlier times. The near-wall peaks of $E_{TT}$ move significantly closer to the wall while also moving towards larger $\lambda_{z}$. Finally, in both figures \ref{fig_y_vs_lambda_x_Ma_1_5_PI_40}(h) and \ref{fig_y_vs_lambda_z_Ma_1_5_PI_40}(h), both temperature spectra reveal a secondary peak of small-scales in the buffer layer away from the primary near-wall peak. This secondary peak is responsible for the secondary peak in the plots of $\wtil{T''T''}$ from figure \ref{figure_Temp_fluc_Pi_10}(h). 

\section{Extension of the Generalized Reynolds analogy for nonequilibrium flows} \label{S_Reynolds_analogies}

    Of engineering relevance is the prediction of the temperature field and its fluctuations from velocity information. 
    There has been a long history in relating or predicting the temperature fields from velocities~\citep{Reynolds1874Analogy,busemann1931handbuch,crocco1932sulla,van1951turbulent,morkovin1962effects,walz1962compressible,gaviglio1987reynolds,huang1995compressible,duan2011direct,zhang2014generalized}.
    However, these Reynolds analogies have only been developed and tested in the context of mostly unidirectional flows, like channels and boundary layers. Here, the GRA of \citet{zhang2014generalized} is extended for three-dimensional, unsteady flows and tested for the $6$ cases studied herein.

    Generally, the GRA predicts $\wtil{T}$ and $\wtil{T''T''}$ from the velocity statistics, heat transfer at the wall, freestream temperature, wall temperature, and fluid properties. In the channel, the freestream temperature will be replaced with $T_{c}$, the centerline temperature. The initial assumptions for the GRA, and other Reynolds analogies, begin with an assumption of steady flow for a unidirectional flow. The Reynolds analogies then assume fluctuations of either the total enthalpy or recovery enthalpy are equal to $U_{w}u''$, where $U_{w}$ is a constant velocity factor. They then show that the difference between the mean recovery enthalpy and $U_{w}\wtil{u}$ is constant. From this relationship, the modeled temperature, $T_{m}$, can be recovered as a quadratic polynomial of $\wtil{u}$. In appendix \ref{APP_Derivation}, it is shown that a transient and three-dimensional flow with a similar Reynolds analogy between the enthalpy and $\vb{u}''$ results in a constant difference between a recovery enthalpy and $\vb{U}_{w}\cdot\vtil{u}$. Hence, the mean temperature can be modeled as a quadratic polynomial in $\norm{\vtil{u}}$ as well. The modeled mean temperature is then 
        \begin{equation}
            \frac{T_{m}(\norm{\vtil{u}},\tau_{w},\beta)}{T_{w}} = 1 + \frac{T_{rg} - T_{w}}{T_{w}}\qty(\frac{\norm{\vtil{u}}\cos(\beta)}{\norm{\vtil{u}}_{c}\cos(\beta_{c})}) + \frac{\oline{T}_{c} - T_{rg}}{T_{w}}\qty(\frac{\norm{\vtil{u}}}{\norm{\vtil{u}}_{c}})^{2}, \label{Eq_GRA_Temp}
        \end{equation}
    where the subscript $c$ is evaluated at the centerline, $T_{rg}$ is the recovery temperature, and $\beta$ is the angle between $\vb{u}$ and $\vb{\tau}_{w}$ such that $\cos(\beta) = \vb{e}_{s}\cdot\vb{e}_{s,w}$. The recovery temperature depends on thermal transport properties, heat transfer, centerline velocity, and wall shear stress as 
        \begin{equation}
            T_{rg}(\oline{u},\tau_{w},\beta_{c}) = T_{c} + \qty(\frac{T_{w} - T_{c}}{\oline{u}_{c}^{2}/2c_{p}} - \frac{2\Pr q_{w}\cos(\beta_{c})}{\oline{u}_{c}\tau_{w}} )\frac{\oline{u}_{c}^{2}}{2 c_{p}} = T_{w} - \frac{\Pr q_{w}\oline{u}_{c}\cos(\beta_{c})}{\tau_{w}c_{p}},
        \end{equation}
    where the quantity in parenthesis is often referred to as the recovery factor, $r_{g}$~\citep{zhang2014generalized}. If the flow is unidirectional, equation \ref{Eq_GRA_Temp} reduces to the GRA of \citet{zhang2014generalized} since $\beta = 0$, $\norm{\vtil{u}} = \wtil{u}$, and $\tau_{w} = \tau_{w,x}$. $T_{m}$ accounts for the three-dimensionality of the flow by incorporating the misalignment between $\vtil{u}$ and $\vb{\tau}_{w}$ as well as the velocity magnitude.

        \begin{figure}
            \centering
            \includegraphics[width=\linewidth]{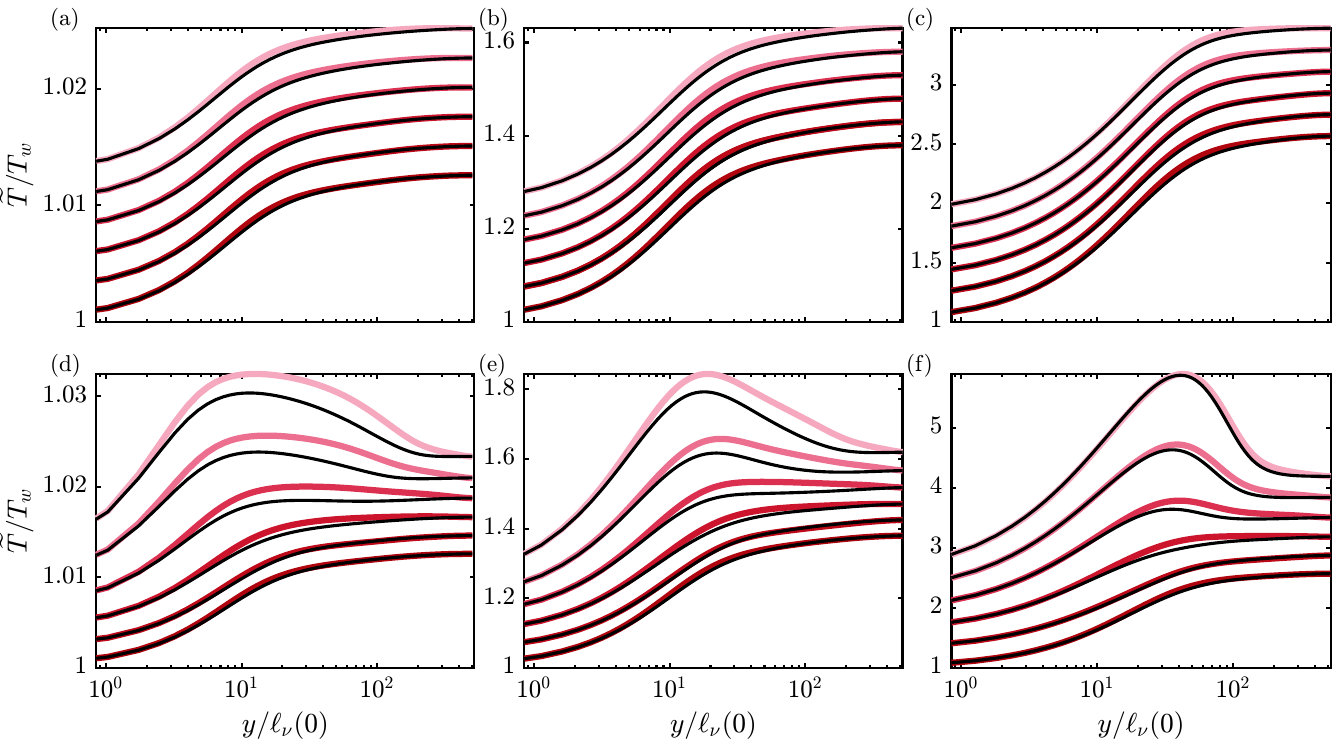}
            \caption{Prediction of $\oline{T}$ from the GRA, $T_{m}$, (red thick lines) and $\oline{T}$ (black) normalized by $T_{w}$ for (a) $Ma = 0.3$, $\Pi = 10$, (b) $Ma = 1.5$, $\Pi = 10$, (c) $Ma = 3.0$, $\Pi = 10$, (d) $Ma = 0.3$, $\Pi = 40$, (e) $Ma = 1.5$, $\Pi = 40$, and (f) $Ma = 3.0$, $\Pi = 40$. Each line is an increment of $\Delta t^{+} = 100$ and is offset vertically for visibility.}
            \label{figure_ReynoldsAnalogy_Mean}
        \end{figure}

    A comparison between $T_{m}$ and $\wtil{T}$ is shown in figure \ref{figure_ReynoldsAnalogy_Mean} for all $6$ cases. The $\Pi = 10$ cases in figures \ref{figure_ReynoldsAnalogy_Mean}(a--c) demonstrate the best agreement between the model and the DNS mean temperature. Although the temporal evolution of $\wtil{T}$ was small for $\Pi = 10$, the GRA is able to account for the significant spanwise acceleration in the prediction for $\wtil{T}$. For $\Pi = 40$ in figures \ref{figure_ReynoldsAnalogy_Mean}(d--f), there are discrepancies between $T_{m}$ and $\wtil{T}$, primarily for $Ma = 0.3$ and $Ma = 1.5$. The agreement between the prediction and $\wtil{T}$ is best for $Ma= 3$. Despite the discrepancies, $T_{m}$ is able to capture the near-wall temperature peak that emerges in $\Pi  =40$.

    In appendix \ref{APP_Derivation}, similar arguments from the GRA are used to create a model for the thermal fluctuations in the spanwise accelerating flow. The thermal fluctuations are modeled as 
        \begin{equation}
            \wtil{T''T''}_{m} = \qty(\frac{\wtil{v''T''}}{e_{i}\wtil{v''u_{i}''}} - \qty( 1 - \cos(\beta) )\frac{\Pr q_{w}}{c_{p}\tau_{w}})^{2} \wtil{\qty(e_{i}u_{i}'')^{2}}, \label{Eq_GRA_TT_Fluc}
        \end{equation}
    relating $\wtil{T''T''}$ to the fluctuating velocity along $\vb{e}_{s}$. In the case of a streamwise-aligned flow, the second term in the parenthesis is $0$ and $ \wtil{T''T''}_{m} = \qty(\wtil{v''T''}/\wtil{v''u''})^{2}\wtil{u''u''}$ as in the GRA. The first term inside the parenthesis is similar to what appears in unidirectional flows, except the Reynolds shear stresses are taken along $\vb{e}_{s}$. The second term accounts for a mismatch between the direction of $\vb{\tau}_{w}$ and $\vtil{u}$, though its magnitude is small compared to $\wtil{v''T''}/\qty(e_{i}\wtil{v''u_{i}''})$.

        \begin{figure}
            \centering
            \includegraphics[width=\linewidth]{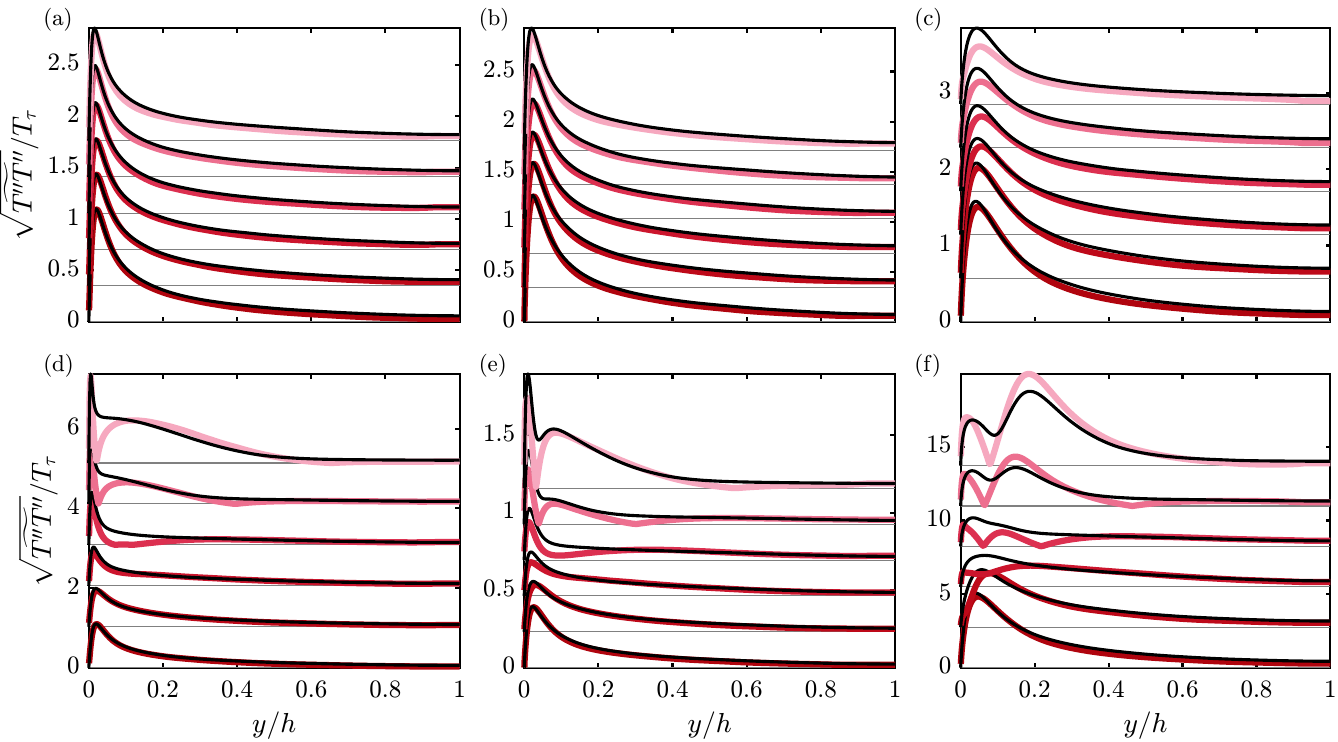}
            \caption{Prediction of $\oline{T'T'}$ from the GRA (red dotted lines) and $\oline{T'T'}$ (black) normalized by $T_{\tau}(0)$ for (a) $Ma = 0.3$, $\Pi = 10$, (b) $Ma = 1.5$, $\Pi = 10$, (c) $Ma = 3.0$, $\Pi = 10$, (d) $Ma = 0.3$, $\Pi = 40$, (e) $Ma = 1.5$, $\Pi = 40$, and (f) $Ma = 3.0$, $\Pi = 40$. Here, $y$ is normalized by $h$ to visualize the outer region. Each line is an increment of $\Delta t^{+} = 100$ with a vertical offset denoted by the horizontal grey lines for visibility.}
            \label{figure_ReynoldsAnalogy_Flucs}
        \end{figure}

    In figure \ref{figure_ReynoldsAnalogy_Flucs}, $\wtil{T''T''}_{m}$ and $\wtil{T''T''}$ are compared for all $6$ cases. There is agreement between the thermal fluctuations and their prediction for $\Pi = 10$ and $Ma = 0.3$ and $Ma = 1.5$ throughout the entire channel. Discrepancies emerge in the $Ma = 3.0$ case in predicting the near-wall $\wtil{T''T''}$ peak, though the outer region is well predicted. For $\Pi = 40$ in figures \ref{figure_ReynoldsAnalogy_Flucs}(d--f), significant disagreements emerge in the near-wall region. Namely, $\wtil{T''T''}_{m}\approx 0$ near the zero-crossing in $\wtil{v''T''}$. These discrepancies arise from a a lack of correlation between $\wtil{v''T''}$ and $e_{i}\wtil{v''u_{i}''}$ in the presence of strong spanwise acceleration. In the outer region of the flow, both $\wtil{T''T''}_{m}$ and $\wtil{T''T''}$ agree since $\wtil{v''T''}$ and $e_{i}\wtil{v''u_{i}''}$ are well correlated. The second term, $\qty( 1 - \cos(\beta) )\Pr q_{w}/c_{p}\tau_{w}$, can address the lack of correlation in the near-wall region, however its magnitude is too small to affect the result. Future work will need to address the lack of correlation between the Reynolds shear stresses and $\wtil{v''T''}$ as improvements to the GRA predictions. These are particularly important when changes in sign are present in the velocity-temperature covariances, even in unidirectional flows.

    \subsection{Discussion} \label{S_Discussion}
            \begin{figure}
                \centering
                \includegraphics[width=\linewidth]{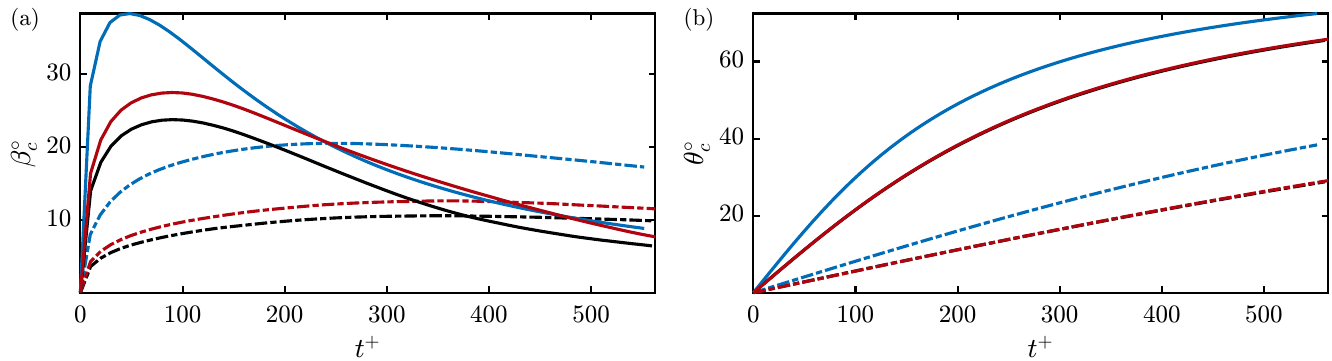}
                \caption{Measures of the sweep angle via $\beta_{c}$, the angle of the centerline velocity relative to the angle of the wall shear stress (a) and $\theta_{c}$, the angle of the centerline velocity (b) against $t^{+}$ in degrees. The colors and line symbols are denoted in table \ref{table_setup}. Note that in (b), the lines corresponding to $Ma = 0.3$ are just underneath the lines for $Ma = 1.5$.}
                \label{figure_SweepAngleMeasures}
            \end{figure}

        The two choices of $\Pi$ create two qualitatively distinct responses to the spanwise accelaration despite the $Ma$. These effects are primarily observed in the temperature field where a secondary peak in $\wtil{T}$ emerges for $\Pi = 40$ that is not observed in the $\Pi = 10$ cases. This is indicative of increased viscous heating due to the increase in mean kinetic energy. The near-wall temperature peak is also associated with an increased $q_{w}$ for the $\Pi = 40$ cases compared to the $\Pi = 10$ cases. Furthermore, the presence of the near-wall temperature peak changes the turbulent transport of temperature fluctuations by creating a change in sign across the peak in the velocity-temperature variances and an increased contribution of $Q_{1}$ and $Q_{3}$ in $\wtil{T''v''}$. This change in sign is not reflected in the Reynolds shear stresses, making $\wtil{\vb{u}''v''}$ and $\wtil{T''v''}$ uncorrelated, which ultimately limits the predictive capabilities of the GRA. 
        Despite their differences, both the $\Pi = 10$ and $\Pi = 40$ cases observed a decrease in $\wtil{T''T''}$ across the channel for early times despite the net increase in $\wtil{T}$. 

        If the angle of $\vb{u}_{c}$ relative to $\vb{\tau}_{w}$ ($\beta$) is taken as a measure for the sweep angle, the sweep angles are at their peak, above $20^\circ$ for $\Pi = 40$. Taking instead the relative velocity of $\vb{u}_{c}$ to $\vb{e}_{x}$ as the sweep angle gives, for the largest $\Pi$, sweep angles around $60^{\circ}$ at the end of the simulation time. These sweep angles are similar to the sweep angles of supersonic aircraft, such as $\Lambda = 55^\circ$ for a Concorde. However, the rate of acceleration is much different in the case of the aircraft as opposed to the ones shown herein. Using the Concorde as an example once again, its spanwise acceleration can be estimated as $\qty(U_{\infty}\sin(\Lambda))^2/L_{c}$~\citep{vos2015introduction}, where the cruising velocity is $U_{\infty} \approx 600 \textrm{ m/s}$ and the chord is $L_{c}\approx 20 \textrm{ m}$. This will be normalized with $a_{w}^{2}/\delta$, where the boundary layer thickness, $\delta$, is estimated as $\delta \approx .1\textrm{ m}$ using a $1/5$ power-law estimate~\citep{schlichting2016boundary}. 
        The normalized acceleration is then $A_{a} = U^{2}_{\infty}\sin(\Lambda)^{2}\delta/L_{c}a_{w}^{2} \approx 1.2\times 10^{-2}$ for the Concorde. The normalized acceleration can also be calculated for the channel as $A_{c} = g_{z}h/a_{w}^{2}$. For the $\Pi = 10$ cases used herein, the normalized acceleration is $A_{c} \approx 2.5\times 10^{-3}$, $6.9\times 10^{-2}$, and $3.3\times 10^{-1}$ for $Ma = 0.3$, $1.5$, and $3.0$, respectively. Thus for similar $Ma$ as the Concord, case 2 would be at a similar spanwise acceleration, relative to the speed of sound and boundary layer thickness, as those expected from real operations. Both $T_{m}$ and $\wtil{T''T''}_{m}$ showed the best predictive capability for $\Pi = 10$, suggesting that these approaches may have applicability in realistic conditions.

\section{Conclusion} \label{Conclusion}
    While the use of a compressible flow enables the study of the heat transfer and temperature responses to spanwise acceleration, the velocity response in the compressible flow is qualitatively similar to that of the incompressible spanwise response~\citep{moin1990direct,lozano2020non}. By properly accounting for the property variations and the new flow direction, the GFM velocity transformation is able to collapse the mean velocity in the viscous sublayer. Velocity transformations were also illustrated in the similarity variable for the initial spanwise flow, whereby introducing the similarity variable $\eta$ the spanwise compressible mean momentum equation could be transformed into the same similarity equation for the incompressible flow. The similarity solution for the mean spanwise velocity held for short times, until $\oline{\rho}\wtil{w''v''}$ became nonneglgible. From this similarity solution, $\tau_{w,z}$ could also be predicted for short times.

    The turbulent fluctuations were also qualitatively similar to the incompressible regime. For example, both the incompressible and compressible flows observe a net decrease in $\oline{\rho}\wtil{u''v''}$ leading to a reduction in $\oline{\rho}\wtil{u''u''}$ and thus a reduction in $\oline{\rho}\wtil{k''}$. While this is occurring, $\oline{\rho}\wtil{w''v''}$ increases in magnitude leading to an increase in $\oline{\rho}\wtil{w''w''}$ and eventually, a net increase in $\oline{\rho}\wtil{k''}$. The transport of the energy-containing eddies is also similar to the incompressible case, where the small-scale near-wall eddies orient themselves along the direction of the new spanwise flow before the large-scale eddies further from the wall. In addition, there is a mismatch between the direction of the Reynolds shear stresses and the mean flow. The lag between the mean flow direction and the Reynolds shear stresses has been observed in both incompressible simulations and experiments of swept wings.

    The temperature response depends on the strength of the spanwise acceleration. For the weaker acceleration, $\Pi = 10$, the mean temperature varies little remaining monotonically increasing, though there is a slight increase in overall temperature. For $\Pi = 40$, while the temperature continues to rise substantially, there is an emergence of a near-wall temperature peak. Despite the different responses in the mean temperature, there is near-wall collapse of the mean temperature when rescaled using the friction temperature, $T_{\tau}(t)$, and viscous length, $\ell_{\nu}(t)$, which both account for the new direction of $\tau_{w}$. Due to the presence of the near-wall peak, the turbulent transport of thermal fluctuations fundamentally changes. This manifests itself as a change in sign in the velocity-temperature covariances for $\Pi = 40$ while for $\Pi = 10$, their signs remain unchanged. Additionally, for $\Pi = 40$, the non-monotonic mean temperature presents an increase in $Q_{1}$ and $Q_{3}$ and decrease in $Q_{2}$ and $Q_{4}$ events in the thermal transport not seen in $\Pi = 10$. These observations hint at different transport mechanisms between the thermal and momentum fluctuations away from the wall. Similar to the drop in kinetic energy, the thermal fluctuation intensity was shown to also drop before eventually increasing. This initial reduction stems from an initial decrease in the thermal production of the $\oline{\rho}\wtil{T''T''}$ budgets.

    As the flow is accelerated, the energetic near-wall structures move closer to the wall and their streamwise extent shrinks due to their change in orientation. The spanwise size of these structures increases as the flow moves primarily to the spanwise direction. Between the kinetic energy-carrying structures and the temperature-carrying structures, the latter move closer to the wall as $\tau_{w}$ increases.

    The GRA was shown to extend to statistically three-dimensional transient flows, provided that the mean velocity's three-dimensionality is accounted for via the velocity magnitude and angle between $\vb{\tau}_{w}$ and $\vtil{u}$. The modeled temperature agreed well for both the $\Pi =10$ and $\Pi = 40$ cases, though discrepancies were observed in the latter. For $\Pi = 10$, the modeled $\wtil{T''T''}$ agree well for $Ma = 0.3$ and $1.5$. For $\Pi = 40$, the change in sign in $\wtil{v''T''}$ that is not present in $\wtil{u_{i}''v''}e_{i}$ limits the predictions of $\wtil{T''T''}$ in the near-wall region since the GRA assumes $\wtil{v''T''}$ and $\wtil{u_{i}''v''}e_{i}$ are correlated. Future work will need to account for this lack of correlation to account for the zero-crossings in $\wtil{v''T''}$ to improve the correction of $\wtil{T''T''}$.

    This flow introduces spanwise flow via a transient response to a strong, suddenly applied spanwise body force. In the case of swept wings, the spanwise flow occurs because of changes in geometry where spanwise pressure gradients cause the flow to accelerate. From estimates of real supersonic aircraft with swept wings, it can be argued that the spanwise pressure gradients are on a similar order of magnitude as the acceleration from the $\Pi = 10$, $Ma = 1.5$ case, when normalized with the speed of sound and boundary layer thickness. Thus, one may expect that the observations from the $\Pi = 10$, $Ma = 1.5$ case may extend to realistic aircraft design, provided that these observations hold for large $Re_{\tau}$. 

\backsection[Acknowledgements]{S.R.G thanks Christopher T. Williams and Professor Parviz Moin for insightful discussions and acknowledge the use of computational resources from Lawrence Livermore National Laboratory, which were instrumental in this work.}

\backsection[Funding]{Funding from the Center for Turbulence Research postdoctoral fellowship is gratefully acknowledged.}

\backsection[Declaration of interests]{The author reports no conflict of interest.}

\backsection[Author ORCIDs]{S. Gomez, https://orcid.org/0000-0002-7568-721X}

\appendix
\section{Trettle and Larson Velocity Transformation}\label{APP_TL}
        \begin{figure}
            \centering
            \includegraphics[width=0.75\linewidth]{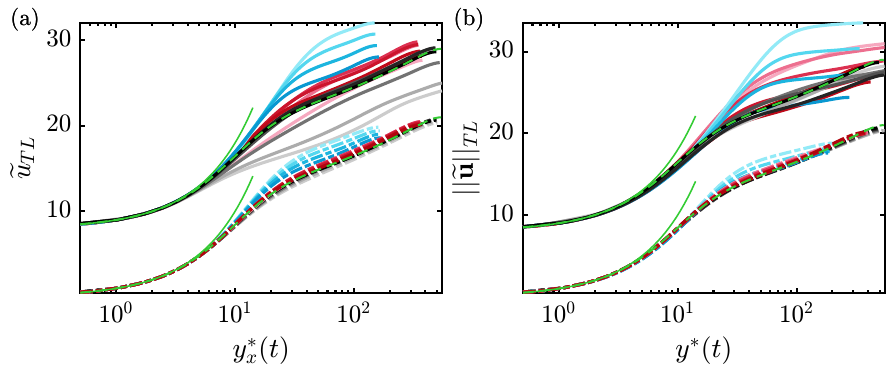}
            \caption{TL scaling for $\oline{u}$ (a) and $\norm{\vbar{u}}$ (b). The colors and linestyles are the same as Figure \ref{fig_Scaling_scaling_U_Umag_T}. The $\Pi = 40$ cases are offset vertically for visibility.}
            \label{fig_APP_TL_Scaling}
        \end{figure}

    The \citet{trettel2016mean} velocity transformation (TL) is applied to both the streamwise velocity and the velocity magnitude in figure \ref{fig_APP_TL_Scaling}. Similar to equations \ref{Eq_Scaling_GFM_x} and \ref{Eq_Scaling_GFM_mag}, the velocity transformation are $\pdv*{\oline{u}_{TL}}{y^{*}_{x}} = S_{TL,x}$ and $\pdv*{\vbar{u}_{TL}}{y^{*}} = S_{TL}$. These velocity transformations are equivalent in the near-wall region since the GFM and TL scalings are, by construction, similar~\citep{griffin2021velocity}. 

\section{Budgets of thermal fluctuations} \label{APP_Budgets}

    The temporal evolution of $\oline{\rho}\wtil{T''T''}$ can be found by 1) subtracting the spatially averaged equation \ref{Eq_Meth_Temp} from itself, 2) multiplying the difference by $T''$, and 3) spatially averaging. This is similar to how the TKE or Reynolds stress budget is found~\citep{moin1990direct,lozano2012three,pope2000turbulent}. The temporal evolution is then
        \begin{equation}
            \partial_{t}\qty(\oline{\rho}\wtil{T''T''}) = C_{o,TT} + P_{TT} + \Pi_{TT} + \mc{E}_{TT} + D_{TT} + C_{TT},
        \end{equation}
    where $C_{o,TT} = \partial_{y}\qty(\wtil{v}\oline{\rho}\wtil{T''T''})$ are wall-normal convective terms that are negligible, $P_{TT} = 2\qty(-\oline{\rho}\wtil{v''T''}\partial_{y}\wtil{T} -\qty(\gamma -1)\oline{\rho}\wtil{T''T''}\partial_{y}\wtil{v} + c_{v}^{-1}\oline{T''\tau_{yi}}\partial_{y}\wtil{u}_{i})$ is the thermal production, $\Pi_{TT} = 2c_{v}^{-1}\qty( \oline{ T''S_{ij}''\tau'_{ij} } - \oline{ T'' \partial_{x_{i}}u''_{i}p' } )$ is the thermal pressure-stress strain, $\mc{E}_{TT} = 2c_{v}^{-1}\oline{\partial_{x_{i}}T''q_{i} }$ is the heat dissipation, $D_{TT} =  -\partial_{y}\qty(\oline{\rho}\oline{T''T''v''} ) -2 c_{v}^{-1}\partial_{y}\oline{T''q_{y}}$ is the turbulent and heat transport contribution, and $C_{TT} = 2\qty( c_{v}^{-1} \oline{T''}\oline{\tau}_{yi}\partial_{y}\oline{u}_{i} + c_{v}^{-1} \oline{S_{ij}''T''}\qty(\oline{\tau}_{ij} - \oline{p}\delta_{ij}) ) $ are the compressible contributions. Here, $S_{ij} = \qty(\partial_{x_{j}}u_{i} + \partial_{x_{i}}u_{j} )/2$ is the strain rate tensor and $\delta_{ij}$ is the Kroenecker delta. An increase or decrease in each of these terms will provide insight into the gain or reduction of $\oline{\rho}\wtil{T''T''}$.

        \begin{figure}
            \centering
            \includegraphics[width=\textwidth]{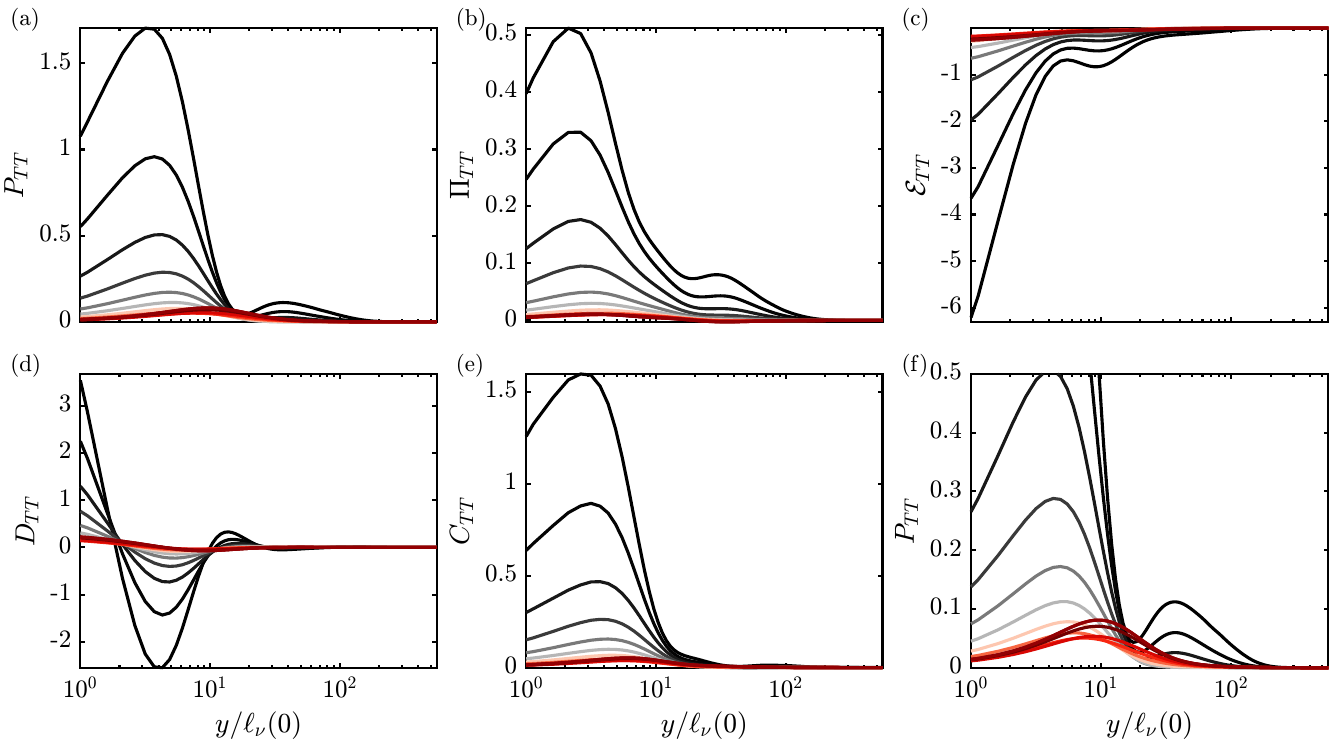}
            \caption{The individual budget terms of $\partial_{t}\qty(\oline{\rho}\wtil{T''T''})$, normalized by $u_{\tau}(0)\rho_{W}(0)T_{\tau}(0)^{2}/\ell_{\nu}(0)$ for $\Pi = 40$ and $Ma = 1.5$. A zoom in view of $P_{TT}$ is included in (f) for visibility. The colors are the same as those in figure \ref{figure_Turb_fluc_Pi_40}(a).}
            \label{fig_App_Budgets}
        \end{figure}

    In figure \ref{fig_App_Budgets}, the budget terms for $\Pi = 40$ and $Ma = 1.5$ are plotted, with $C_{o,TT}$ omitted since it is negligible compared to the rest. The only term that appreciably drops near $y/\ell_{\nu}(0)$ during the time that $\wtil{T''T''}$ is reduced is the thermal production, $P_{TT}$. $\Pi_{TT}$, $D_{TT}$, and $C_{TT}$ all increase in magnitude even when $\wtil{T''T''}$ initially decreases. There is a slight change in $\mc{E}_{TT}$ during this initial time, however, it also initially increases before eventually decreasing after $\wtil{T''T''}$ has reached its smallest value. Once $\wtil{T''T''}$ begins to increase again, there is the presence of a secondary peak in the buffer region above the near-wall $\wtil{T}$ peak. These are accompanied by secondary peaks in $P_{TT}$ and $\Pi_{TT}$.

        \begin{figure}
            \centering
            \includegraphics[width=\textwidth]{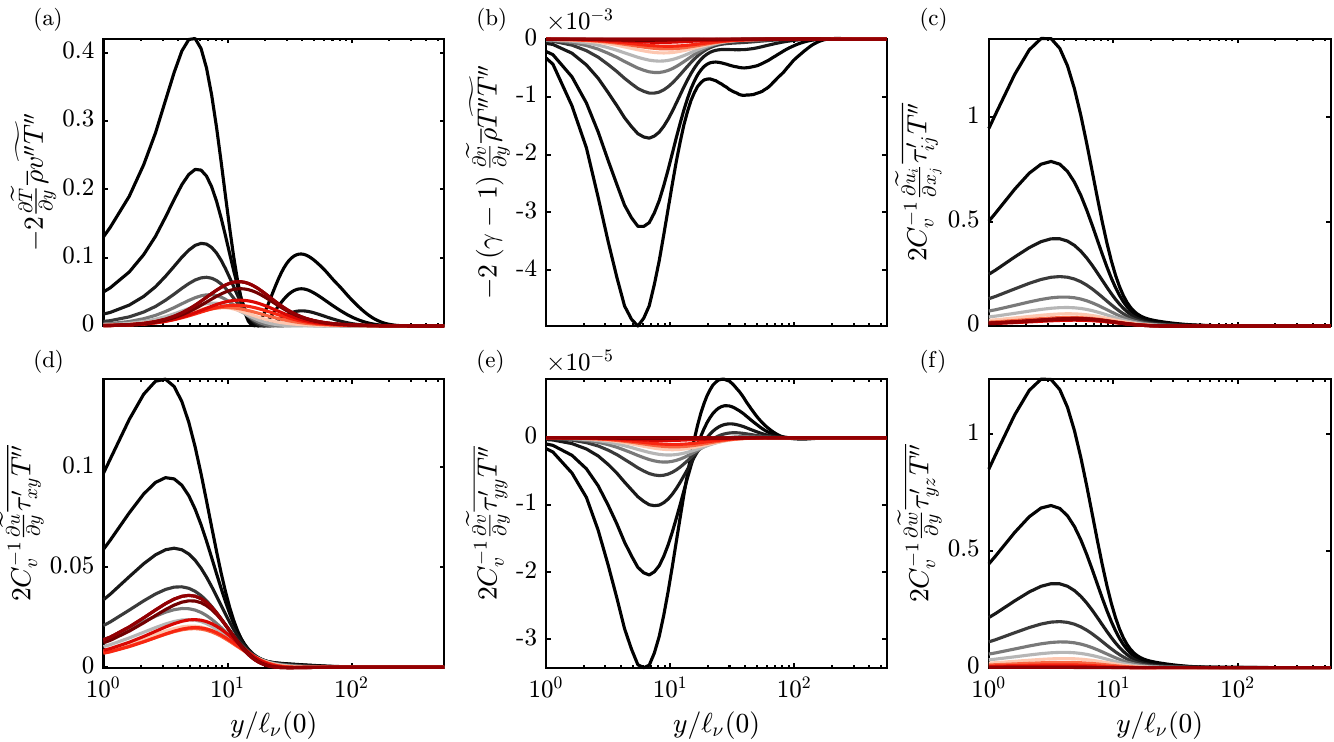}
            \caption{The individual terms of the production, $P_{TT}$, normalized by $u_{\tau}(0)\rho_{W}(0)T_{\tau}(0)^{2}/\ell_{\nu}(0)$ for $\Pi = 40$ and $Ma = 1.5$. The colors are the same as those in figure \ref{figure_Turb_fluc_Pi_40}(a). }
            \label{fig_App_Budgets_Detailed}
        \end{figure}

    The individual components of $P_{TT}$ are then plotted in figure \ref{fig_App_Budgets_Detailed} to isolate the term that contributes the most to the reduction in $\wtil{T''T''}$. The terms that are negligible are those that are related to the mean compressibility, $\partial_{y}\wtil{v}$, in figures \ref{fig_App_Budgets_Detailed}(b,e), though they do decrease near $y/\ell_{\nu}(0)$. The most substantial decrease in the thermal production stems from the drop in $-2\oline{\rho}\wtil{v''T''}\partial_{y}\wtil{T}$. This term is similar to the production in the budget for $\wtil{u''u''}$, $-\oline{\rho}\wtil{u''v''}\partial_{y}\wtil{u}$, whose reduction is responsible for the decrease in $\wtil{u''u''}$ and the TKE~\citep{moin1990direct,lozano2020non}. Just like the reduction in $-\oline{\rho}\wtil{u''v''}\partial_{y}\wtil{u}$, the reduction in $-2\oline{\rho}\wtil{v''T''}\partial_{y}\wtil{T}$ is driven by a reduction in $\wtil{v''T''}$, as seen in figure \ref{figure_Temp_fluc_Pi_10}(f). A secondary reduction is also seen in the slight reduction in $2c_{v}^{-1}\oline{\tau_{xy}'T''}\partial_{y}\wtil{u}$, though this reduction is most significant below $y/\ell_{\nu}(0)= 10$. Thus, the reduction in $\wtil{T''T''}$ is driven by a decrease in the thermal production, specifically by the term $-2\oline{\rho}\wtil{v''T''}\partial_{y}\wtil{T}$. This is also reflected in the other cases as well. 

\section{Unweighted histogram}\label{APP_Unweighted_Histogram}

    The results in section \ref{SS_Transport_of_Coherent_Structures} relied on the weighted histogram, $H(\theta,y_{m})$ to reduce the impact of the small-scale structures in the outer region whose alignment was spurious. To show that the weighting does not affect the results, an unweighted historgram is computed similar to equation \ref{Sec_Turb_Eq_Hist} as
        \begin{equation}
            N_{T}H_{u}(\theta,y_{m}) = \sum_{\bre{\theta} -\theta= \Delta\theta_{1}}^{\Delta \theta_{2}}\sum_{\bre{y}_{m} -y_{m} = \Delta y_{m,1}}^{\Delta y_{m,2}} 1,
        \end{equation}
    where $N_{T} = \sum_{\theta}\sum_{y_{m}}1$ is the total number of identified structures. Both $H$ and $H_{u}$ use the same bins.

        \begin{figure}
            \centering
            \includegraphics[width=0.99\linewidth]{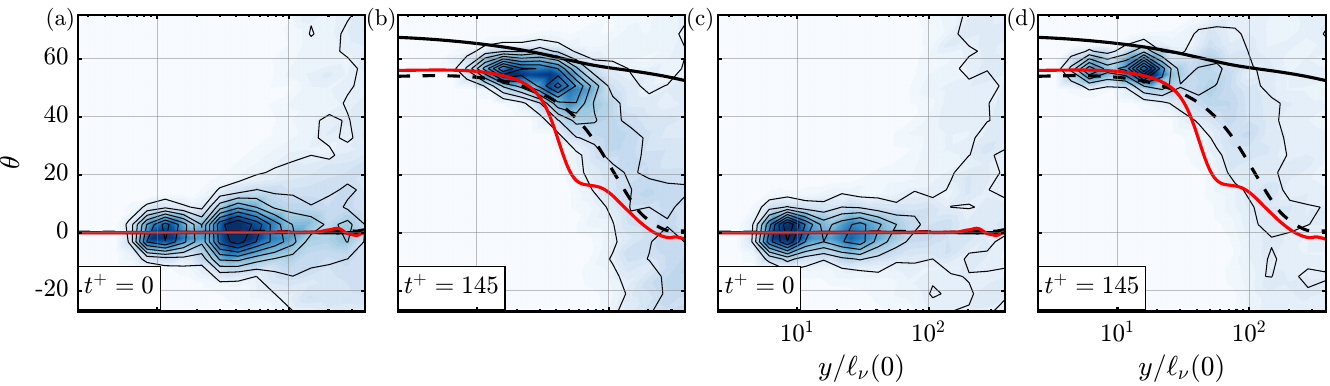}
            \caption{Unweighted histogram for the kinetic energy structures (a,b) and temperature structures (c,d). The solid black, dashed black, and solid red lines denote $\theta_{\wtil{u},\wtil{w}}$, $\theta_{\wtil{u''v''},\wtil{w''v''}}$, and $\theta_{\wtil{u''T''},\wtil{w''T''}}$, respectively.}
            \label{fig_App_NoWeighting}
        \end{figure}

    Figure \ref{fig_App_NoWeighting} shows $H_{u}$ for both kinetic energy and temperature structures. Similar to figure \ref{fig_y_vs_theta_k_T_Ma_1_5}, the unweighted histograms show that the near-wall structures orient themselves in the spanwise direction before the structures away from the wall. However, the unweighted histogram does not identify a clear orientation angle in the outer region. This likely occurs because of small-scale structures that are identified in the outer region that are not filtered away. These small-scale structures do not represent typical energy-containing eddies and are likely spherical in shape, rather than long oblong structures that are typically observed in wall-bounded turbulence, like elongated streaks. Due to their small size, a proper inclination angle can not be determined from the method described in section \ref{SS_Transport_of_Coherent_Structures}. These small-scale outer-region identified structures can be seen in figures \ref{fig_3D_vizualizations}(e,g) as the green spines that are primarily vertically aligned. 

\section{Derivation of Generalized Reynolds analogy for transient, 3D flows} \label{APP_Derivation}

    The derivation follows similar assumptions to 
    except, the averaged quantities are functions of $x$ and $y$ rather than $x$ and $y$. This derivation assumes negligible density and viscosity fluctuations such that $\wtil{f}\approx \bar{f}$, $\bar{\tau}_{ij} \approx \oline{\mu}\qty(\partial_{x_{i}}\wtil{u}_{j} + \partial_{x_{j}}\wtil{u}_{i})$, and $\oline{q}_{y} \approx -c_{p}\partial_{y} \wtil{h}_{e}/\Pr$, where $h_{e} = c_{p}T$ is the enthalpy and $\wtil{v}\ll \wtil{u}$. Finally, the kinetic energy is approximated as $u_{i}u_{i}/2\approx \wtil{u}_{i}\wtil{u}_{i}/2 + \wtil{u}_{i}u_{i}''$ because mean kinetic energy is large relative to the turbulent kinetic energy. It is convenient to work with the total enthalpy, $H = h_{e} + u^{2}/2$ and total stress, $\bar{\tau}_{T,ij} = \bar{\tau}_{ij} - \bar{\rho}\wtil{u_{i}''u_{j}''}$. The Reynolds averaged momentum and total enthalpy equations are 
        \begin{align}
            \oline{\rho}\mc{D}_{t}\wtil{u}_{i} &= \partial_{y}\qty(\oline{\mu}\partial_{y}\wtil{u}_{i} - \oline{\rho}\wtil{u_{i}''v''}) + \oline{\rho}g_{i} \label{Eq_App_Mom}\\
            \oline{\rho}\mc{D}_{t}\wtil{H} &= \partial_{y}\qty(\oline{\mu}\wtil{u}_{i}\partial_{y}\wtil{u}_{i} - \oline{\rho}\wtil{H''v''}) + \oline{\rho}g_{i}\wtil{u}_{i} + \partial_{t} \wtil{p},\label{Eq_App_Enth}
        \end{align}
    where $\mc{D}_{t} =\partial_{t} + \wtil{v}\partial_{y}$.

    The GRA of \citet{zhang2014generalized} is based on extensions of the Crocco-Buseman relationship~\citep{busemann1931handbuch,crocco1932sulla} 
    and strong Reynolds analogies~\citep{morkovin1962effects} to $\Pr\ne 1$ through a recovery enthalpy, $H_{g} = h_{e}  + r_{g}u_{i}u_{i}/2$ where $r_{g}$ is a recovery factor. Similar to the GRA, it is assumed that 
        \begin{align}
            \wtil{H}_{g} + \wtil{H}_{w} &= U_{w,i}\wtil{u}_{i}, \label{Eq_H_G_Bar}\\
            H''_{g} + c_{p}\phi'' &= U_{w,i}u_{i}''. \label{Eq_H_G_prime}
        \end{align}
    for some to be determined wall velocity $U_{w,i}$, temperature difference $\phi$, and constant wall enthalpy, $H_{w}$. It is assumed that $U_{w,i}$ is quasisteady in time, similar to how it is considered quasiparallel in a turbulent boundary layer.

    Manipulation of $H_{g}$ and equations \ref{Eq_H_G_Bar} and \ref{Eq_H_G_prime} identifies the following relationships
        \begin{align}
            \wtil{H} &= \wtil{H}_{g} + \qty(1-r_{g})\wtil{u}_{i}\wtil{u}_{i}/2, \label{Eq_B1}\\
            \oline{\mu}\wtil{u}_{i}\wtil{u}_{i}\partial_{y}r_{g} &= \qty(U_{w,i} - r_{g}\wtil{u}_{i})\oline{\tau}_{i,y} + \Pr \oline{q}_{y}, \label{Eq_B5}\\
            \oline{\rho}\wtil{v''H''} &= \oline{\rho}\wtil{v''H''_{g}} +\qty(1-r_{g})\wtil{u}_{i}\oline{\rho}\wtil{u_{i}''v''}, \label{Eq_B3}\\
            \oline{\rho}\wtil{v''h_{e}''} &= \qty(U_{w,i}-r_{g}\wtil{u}_{i})\oline{\rho}\wtil{u_{i}''v''} - \oline{\rho}c_{p}\wtil{\phi''v''}, \label{Eq_B4}\\
            \oline{\mu}\wtil{u}_{i}\partial_{y}\wtil{u}_{i} &= \oline{\mu}\partial_{y}\wtil{H}_{g} - \oline{\mu}\partial_{y}\wtil{h}_{e} + \oline{\mu}\partial_{y}\qty((1-r_{g})\wtil{u}_{i}\wtil{u}_{i}/2). \label{Eq_B2}
        \end{align}
    An equation for the evolution of $\wtil{H}_{g}$ can be found by combining equations \ref{Eq_B1}, \ref{Eq_B2}, and \ref{Eq_B3} into equation \ref{Eq_App_Enth} as
        \begin{align}
            \oline{\rho}\mc{D}_{t}\wtil{H}_{g} &= \partial_{y}\qty[\oline{\mu}\partial_{y}\wtil{H}_{g} - \oline{\rho}\wtil{v''H''_{g}}] + \oline{\rho}g_{i}\wtil{u}_{i} + \partial_{t} \oline{p} + R_{g} \label{Eq_App_Hg}\\
            R_{g} = \oline{\rho}\mc{D}_{t}\qty[(r_{g}-1)\wtil{u}_{i}\wtil{u}_{i}/2] &- \partial_{y}\qty[ ( r_{g} - 1 )\oline{\tau}_{T,i}\wtil{u}_{i} + (1-\Pr)\oline{q}_{y} + \oline{\mu}\wtil{u}_{i}\wtil{u}_{i}\partial_{y}r_{g}/2 ]. \label{Eq_APP_Rg}
        \end{align}
    By applying $\oline{\rho}\wtil(\partial_{t}+\wtil{v}\partial_{y})$ to equation \ref{Eq_H_G_Bar} and using equation \ref{Eq_App_Mom}, an expression for $\wtil{h}_{e}$ is found as
        \begin{equation}
            \oline{\rho}\mc{D}_{t}\wtil{h}_{e} = \oline{\rho}\mc{D}_{t}\qty((1-r_{g})\wtil{u}_{i}\wtil{u}_{i}/2) + \oline{\rho}g_{i}\qty(U_{w,i} - \oline{u}_{i}) + \qty(U_{w,i} - \wtil{u}_{i})\partial_{y}\oline{\tau}_{T,i}. \label{Eq_APP_Enth_1}
        \end{equation}
    Subtracting $\wtil{u}_{i}$ times equation \ref{Eq_App_Mom} from equation \ref{Eq_App_Enth} and using equation \ref{Eq_B4} provides an alternate expression for $\wtil{h}_{e}$ as
        \begin{equation}
            \oline{\rho}\mc{D}_{t}\wtil{h}_{e} = \partial_{t}\wtil{p} + \oline{\tau}_{T,i}\partial_{y}\wtil{u}_{i} - \oline{\rho}\wtil{v''u''_{i}}\partial_{y}\qty(U_{w,i} - r_{g}\wtil{u}_{i}) - c_{p}\oline{\rho}\wtil{v''\phi''} -c_{p}\oline{\mu}\partial_{y} \wtil{T}/\Pr. \label{Eq_APP_Enth_2}
        \end{equation}
    From these two expressions for $\oline{\rho}\mc{D}_{t}\wtil{h}_{e}$, an expression for the recovery kinetic energy emerges as 
        \begin{align}
            \begin{split}
                \oline{\rho}\mc{D}_{t}\qty((r_{g} - 1)\wtil{u}_{i}\wtil{u}_{i}/2) = &\partial_{y}\qty((U_{w,i} - r_{g}\wtil{u}_{i})\oline{\rho}\wtil{v''u_{i}''} - c_{p}\oline{\rho}\wtil{v''\phi''} + c_{p}\oline{\mu}\partial_{y}\wtil{T}/\Pr) \\
                &+ \oline{\rho}g_{i}\qty(U_{w,i} - \wtil{u}_{i}) + U_{w,i}\partial_{y}\oline{\tau}_{T,i} - \partial_{y}\qty(\wtil{u}_{i}\oline{\tau}_{T,i}) + \partial_{t}\oline{p}
            \end{split}\label{Eq_APP_B10}
        \end{align}
    Equations \ref{Eq_APP_B10} and \ref{Eq_B5} are then used to simplify equation \ref{Eq_APP_Rg} as
        \begin{equation}
            R_g = -c_{p}\partial_{y}\qty(\oline{\rho}\wtil{v''\phi''}) + \oline{\rho}g_{i}\qty(U_{w,i} - \wtil{u}_{i}) - \partial_{t}\wtil{p}.
        \end{equation}
    With the simplified expression of $R_{g}$ and equation \ref{Eq_H_G_prime}, equation \ref{Eq_App_Hg} can be simplified as 
        \begin{equation}
            \oline{\rho}\mc{D}_{t}\wtil{H}_{g} = \partial_{y}\qty(\oline{\mu}\partial_{y}\wtil{H}_{g} - \oline{\rho}U_{w,i}\wtil{v''u_{i}''}).
        \end{equation}
    Subtracting this expression from the product of $U_{w,i}$ and equation \ref{Eq_App_Mom}, reveals that
        \begin{equation}
            \qty(\oline{\rho}\mc{D}_{t} - \partial_{y}\qty(\oline{\mu}\partial_{y}))\qty(\wtil{H}_{g} - U_{w,i}\wtil{u}_{i}) = 0
        \end{equation}
    which confirms equation \ref{Eq_App_Enth}.

    To determine $U_{w,i}$, the dot product $U_{w,i}\wtil{u}_{i}$ is rewritten as $\norm{\vb{U}_{w}}\norm{\vtil{u}}\cos(\beta)$, where $\beta$ is the angle between $\vb{U}_{w}$ and $\vtil{u}$ and $\wtil{u}_{i}\wtil{u}_{i}$ as $\norm{\vtil{u}}^{2}$. Using these expressions, equation \ref{Eq_App_Enth} is differentiated with respect to $y$ and evaluated at the wall such that
        \begin{equation}
            c_{p}\partial_{y}\wtil{T}\vert_{w} = \norm{\vb{U}_{w}}\qty(\partial_{y}\norm{\vtil{u}})\vert_{w}\cos(\beta_{w}).
        \end{equation}
    By introducing $\tau_{w}$ and $q_{w}$, the expression for $\norm{\vb{U}_{w}}$ is then $\norm{\vb{U}_{w}} = q_{w}\Pr/\tau_{w}\cos(\beta_{w})$. Since $\norm{\vb{U}_{w}}$ is defined with respect to wall quantities, $\beta_{w}$ is chosen to be $0$ such that $\vb{U}_{w}$ is parallel to $\vb{\tau}_{w}$ and $\vb{e}_{s,w}$. Hence, combining the expression fro $\vb{U}_{w}$ and equation \ref{Eq_App_Enth} gives the modeled expression for $\wtil{T}$ from equation \ref{Eq_GRA_Temp}.

    In \citet{zhang2014generalized}, the expression for $\phi$ gives $T'' = \Pr_{t}^{-1}\pdv*{\wtil{T}}{\wtil{u}}u''$ where $\Pr_{t}$ is a turbulent Prandtl number. This linear relationship between $T''$ and $u''$ is also reflected in the strong Reynolds analogy with stronger assumptions~\citep{morkovin1962effects}. 
    Motivated by these observations, $\phi''$ is chosen such that a linear relationship emerges between $T''$ and $u_{i}''$. By taking the derivative of $\wtil{H}_{g}$ with respect to $\norm{\vtil{u}}$, it can be shown that 
        \begin{equation}
            r_{g} = \frac{\norm{\vb{U}_{w}}\cos(\beta)}{\norm{\vtil{u}}} - \frac{c_{p}}{\norm{\vtil{u}}}\pdv{\wtil{T}}{\norm{\vtil{u}}}.
        \end{equation}
    By using this expression in equation \ref{Eq_H_G_prime}, it can be shown that
        \begin{equation}
            c_{p}T'' - c_{p}\pdv{\wtil{T}}{\norm{\vtil{u}}}\frac{\wtil{u}_{i}}{\norm{\vtil{u}}} u_{i}'' + c_{p}\phi'' = 0. \label{eq_app_temp}
        \end{equation}
    In line with the results from the literature, $\phi''$ is chosen to achieve a linear relationship between $T''$ and $u_{i}''$ as
        \begin{equation}
            T'' = \qty[\qty( 1 - \cos(\beta) )\frac{\Pr q_{w}}{c_{p}\tau_{w}} + \pdv{\wtil{T}}{\norm{\vtil{u}}}]e_{s,i}u_{i}''. \label{eq_app_temp1}
        \end{equation} 
    By then multiplying equation \ref{eq_app_temp1} with $\rho v''$ and averaging, it can be shown that 
        \begin{equation}
            \frac{\wtil{v''T''}}{e_{s,i}\wtil{v''u_{i}''}} - \qty( 1 - \cos(\beta) )\frac{\Pr q_{w}}{c_{p}\tau_{w}} = \pdv{\wtil{T}}{\norm{\vtil{u}}}.\label{eq_app_temp2}
        \end{equation}
    It is important to note that because the second term on the left is small, equation \ref{eq_app_temp2} suggests that $\wtil{v''T''}$ and $e_{s,i}\wtil{v''u_{i}''}$ are correlated. This is not entirely true, especially in the near-wall region of the $\Pi = 40$ cases. However, it is shown that this agreement holds in the outer region. Finally, by squaring both sides of equation \ref{eq_app_temp}, multiplying by $\rho$, and averaging, provides the relationship in equation \ref{Eq_GRA_TT_Fluc}. Corrections for the lack of correlation between $\wtil{v''T''}$ and $e_{s,i}\wtil{v''u_{i}''}$ will in the future motivate better choices of $\phi''$.

\bibliographystyle{jfm}
\bibliography{biblio}

\begin{thebibliography}{63}
\expandafter\ifx\csname natexlab\endcsname\relax\def\natexlab#1{#1}\fi
\def\au#1{#1} \def\ed#1{#1} \def\yr#1{#1}\def\at#1{#1}\def\jt#1{\textit{#1}} \def\bt#1{#1}\def\bvol#1{\textbf{#1}} \def\vol#1{#1} \def\pg#1{#1} \def\publ#1{#1}\def\arxiv#1{#1}\def\org#1{#1}\def\st#1{\textit{#1}}

\bibitem[Anderson~Jr.(2006)]{anderson2006hypersonic}
{\sc \au{Anderson~Jr., John~David}} \yr{2006} {\em Hypersonic and High-Temperature Gas Dynamics\/}, 2nd edn.  \publ{Reston, Virginia: American Institute of Aeronautics and Astronautics}.

\bibitem[Bai {\em et~al.\/}(2022)Bai, Griffin \& Fu]{bai2022compressibleGFM}
{\sc \au{Bai, Tianyi}, \au{Griffin, Kevin~P} \& \au{Fu, Lin}} \yr{2022}  \at{Compressible velocity transformations for various noncanonical wall-bounded turbulent flows}.  \jt{AIAA journal}  \bvol{60}~(7),  \pg{4325--4337}.

\bibitem[Bennewitz {\em et~al.\/}(2018)Bennewitz, Bigler, Hargus, Danczyk \& Smith]{bennewitz2018characterization}
{\sc \au{Bennewitz, John~W}, \au{Bigler, Blaine~R}, \au{Hargus, William~A}, \au{Danczyk, Stephen~A} \& \au{Smith, Richard~D}} \yr{2018} Characterization of detonation wave propagation in a rotating detonation rocket engine using direct high-speed imaging.  \bt{In {\em 2018 Joint Propulsion Conference\/}},  \pg{p. 4688}.

\bibitem[Boas(2006)]{boas2006mathematical}
{\sc \au{Boas, Mary~L}} \yr{2006} {\em Mathematical methods in the physical sciences\/}.  \publ{John Wiley \& Sons}.

\bibitem[Bradshaw(1977)]{bradshaw1977compressible}
{\sc \au{Bradshaw, Peter}} \yr{1977}  \at{Compressible turbulent shear layers}.  \jt{Annual Review of Fluid Mechanics}  \bvol{9},  \pg{33--52}.

\bibitem[Bradshaw \& Pontikos(1985)]{bradshaw1985measurements}
{\sc \au{Bradshaw, P} \& \au{Pontikos, NS}} \yr{1985}  \at{Measurements in the turbulent boundary layer on an ‘infinite’swept wing}.  \jt{Journal of Fluid Mechanics}  \bvol{159},  \pg{105--130}.

\bibitem[Busemann(1931)]{busemann1931handbuch}
{\sc \au{Busemann, Adolph}} \yr{1931}  \at{Handbuch der experimentalphysik, vol. 4}.  \jt{Geest und Portig} .

\bibitem[Chandran {\em et~al.\/}(2023)Chandran, Zampiron, Rouhi, Fu, Wine, Holloway, Smits \& Marusic]{chandran2023turbulent}
{\sc \au{Chandran, Dileep}, \au{Zampiron, Andrea}, \au{Rouhi, Amirreza}, \au{Fu, Matt~K}, \au{Wine, David}, \au{Holloway, Brian}, \au{Smits, Alexander~J} \& \au{Marusic, Ivan}} \yr{2023}  \at{Turbulent drag reduction by spanwise wall forcing. part 2. high-reynolds-number experiments}.  \jt{Journal of Fluid Mechanics}  \bvol{968},  \pg{A7}.

\bibitem[Chen {\em et~al.\/}(2022)Chen, Huang, Shi, Yang \& Lv]{chen2022unified}
{\sc \au{Chen, Peng~ES}, \au{Huang, George~P}, \au{Shi, Yipeng}, \au{Yang, Xiang~IA} \& \au{Lv, Yu}} \yr{2022}  \at{A unified temperature transformation for high-mach-number flows above adiabatic and isothermal walls}.  \jt{Journal of Fluid Mechanics}  \bvol{951},  \pg{A38}.

\bibitem[Choi {\em et~al.\/}(2002)Choi, Xu \& Sung]{choi2002drag}
{\sc \au{Choi, Jung-Il}, \au{Xu, Chun-Xiao} \& \au{Sung, Hyung~Jin}} \yr{2002}  \at{Drag reduction by spanwise wall oscillation in wall-bounded turbulent flows}.  \jt{AIAA journal}  \bvol{40}~(5),  \pg{842--850}.

\bibitem[Cogo {\em et~al.\/}(2022)Cogo, Salvadore, Picano \& Bernardini]{cogo2022direct_spectra}
{\sc \au{Cogo, Michele}, \au{Salvadore, Francesco}, \au{Picano, Francesco} \& \au{Bernardini, Matteo}} \yr{2022}  \at{Direct numerical simulation of supersonic and hypersonic turbulent boundary layers at moderate-high reynolds numbers and isothermal wall condition}.  \jt{Journal of Fluid Mechanics}  \bvol{945},  \pg{A30}.

\bibitem[Coleman {\em et~al.\/}(1996)Coleman, Kim \& Le]{coleman1996numerical}
{\sc \au{Coleman, Gary~N}, \au{Kim, John} \& \au{Le, Anh-Tuan}} \yr{1996}  \at{A numerical study of three-dimensional wall-bounded flows}.  \jt{International journal of heat and fluid flow}  \bvol{17}~(3),  \pg{333--342}.

\bibitem[Coleman {\em et~al.\/}(1995)Coleman, Kim \& Moser]{coleman1995numerical}
{\sc \au{Coleman, Gary~N}, \au{Kim, John} \& \au{Moser, Robert~D}} \yr{1995}  \at{A numerical study of turbulent supersonic isothermal-wall channel flow}.  \jt{Journal of Fluid Mechanics}  \bvol{305},  \pg{159--183}.

\bibitem[Coles(1956)]{coles1956law}
{\sc \au{Coles, Donald}} \yr{1956}  \at{The law of the wake in the turbulent boundary layer}.  \jt{Journal of Fluid Mechanics}  \bvol{1}~(2),  \pg{191--226}.

\bibitem[Crocco(1932)]{crocco1932sulla}
{\sc \au{Crocco, Luigi}} \yr{1932}  \at{Sulla trasmissione del calore da una lamina piana a un fluido scorrente ad alta velocita}.  \jt{L’Aerotecnica}  \bvol{12}~(181-197),  \pg{126}.

\bibitem[Di~Renzo {\em et~al.\/}(2020)Di~Renzo, Fu \& Urzay]{direnzo2020htr}
{\sc \au{Di~Renzo, Mario}, \au{Fu, Lin} \& \au{Urzay, Javier}} \yr{2020}  \at{{HTR} solver: An open-source exascale-oriented task-based multi-gpu high-order code for hypersonic aerothermodynamics}.  \jt{Computer Physics Communications}  \bvol{255},  \pg{107262}.

\bibitem[Dorodnitsyn(1942)]{dorodnitsyn1942laminar}
{\sc \au{Dorodnitsyn, A}} \yr{1942} Laminar boundary layer in compressible fluid.  \bt{In {\em Dokl. Akad. Nauk SSSR\/}}, ,  \vol{vol.~34}.

\bibitem[Duan {\em et~al.\/}(2010)Duan, Beekman \& Martin]{duan2010direct}
{\sc \au{Duan, Lian}, \au{Beekman, I} \& \au{Martin, MP}} \yr{2010}  \at{Direct numerical simulation of hypersonic turbulent boundary layers. part 2. effect of wall temperature}.  \jt{Journal of Fluid Mechanics}  \bvol{655},  \pg{419--445}.

\bibitem[Duan \& Martin(2011)]{duan2011direct}
{\sc \au{Duan, Lian} \& \au{Martin, MP}} \yr{2011}  \at{Direct numerical simulation of hypersonic turbulent boundary layers. part 4. effect of high enthalpy}.  \jt{Journal of Fluid Mechanics}  \bvol{684},  \pg{25--59}.

\bibitem[Gaviglio(1987)]{gaviglio1987reynolds}
{\sc \au{Gaviglio, J}} \yr{1987}  \at{Reynolds analogies and experimental study of heat transfer in the supersonic boundary layer}.  \jt{International journal of heat and mass transfer}  \bvol{30}~(5),  \pg{911--926}.

\bibitem[Ge \& Jin(2017)]{ge2017response}
{\sc \au{Ge, Mingwei} \& \au{Jin, Guodong}} \yr{2017}  \at{Response of turbulent enstrophy to sudden implementation of spanwise wall oscillation in channel flow}.  \jt{Applied Mathematics and Mechanics}  \bvol{38},  \pg{1159--1170}.

\bibitem[Gomez(2025)]{gomezcompressibilityARB}
{\sc \au{Gomez, Salvador~Rey}} \yr{2025}  \at{Compressibility effects in a non-equilibrium three-dimensional supersonic turbulent boundary layer}.  \jt{CTR Annual Research Briefs}  \pg{pp. 65--78}.

\bibitem[Griffin {\em et~al.\/}(2021)Griffin, Fu \& Moin]{griffin2021velocity}
{\sc \au{Griffin, Kevin~Patrick}, \au{Fu, Lin} \& \au{Moin, Parviz}} \yr{2021}  \at{Velocity transformation for compressible wall-bounded turbulent flows with and without heat transfer}.  \jt{Proceedings of the National Academy of Sciences}  \bvol{118}~(34),  \pg{e2111144118}.

\bibitem[Griffin {\em et~al.\/}(2023)Griffin, Fu \& Moin]{griffin2023near}
{\sc \au{Griffin, Kevin~P}, \au{Fu, Lin} \& \au{Moin, Parviz}} \yr{2023}  \at{Near-wall model for compressible turbulent boundary layers based on an inverse velocity transformation}.  \jt{Journal of Fluid Mechanics}  \bvol{970},  \pg{A36}.

\bibitem[Harrison {\em et~al.\/}(2013)Harrison, Anderson, Fleming \& Ng]{harrison2013active}
{\sc \au{Harrison, Neal~A}, \au{Anderson, Jason}, \au{Fleming, Jon~L} \& \au{Ng, Wing~F}} \yr{2013}  \at{Active flow control of a boundary layer-ingesting serpentine inlet diffuser}.  \jt{Journal of Aircraft}  \bvol{50}~(1),  \pg{262--271}.

\bibitem[Hasan {\em et~al.\/}(2025)Hasan, Costa, Larsson, Pirozzoli \& Pecnik]{hasan2025intrinsic}
{\sc \au{Hasan, Asif~Manzoor}, \au{Costa, Pedro}, \au{Larsson, Johan}, \au{Pirozzoli, Sergio} \& \au{Pecnik, Rene}} \yr{2025}  \at{Intrinsic compressibility effects in near-wall turbulence}.  \jt{Journal of Fluid Mechanics}  \bvol{1006},  \pg{A14}.

\bibitem[Hasan {\em et~al.\/}(2023)Hasan, Larsson, Pirozzoli \& Pecnik]{hasan2023incorporating}
{\sc \au{Hasan, Asif~Manzoor}, \au{Larsson, Johan}, \au{Pirozzoli, Sergio} \& \au{Pecnik, Rene}} \yr{2023}  \at{Incorporating intrinsic compressibility effects in velocity transformations for wall-bounded turbulent flows}.  \jt{Physical Review Fluids}  \bvol{8}~(11),  \pg{L112601}.

\bibitem[Huang {\em et~al.\/}(1995)Huang, Coleman \& Bradshaw]{huang1995compressible}
{\sc \au{Huang, PG}, \au{Coleman, Gary~N} \& \au{Bradshaw, P}} \yr{1995}  \at{Compressible turbulent channel flows: {DNS} results and modelling}.  \jt{Journal of Fluid Mechanics}  \bvol{305},  \pg{185--218}.

\bibitem[Kader(1981)]{kader1981temperature}
{\sc \au{Kader, BA}} \yr{1981}  \at{Temperature and concentration profiles in fully turbulent boundary layers}.  \jt{International journal of heat and mass transfer}  \bvol{24}~(9),  \pg{1541--1544}.

\bibitem[Kong {\em et~al.\/}(2000)Kong, Choi \& Lee]{kong2000direct}
{\sc \au{Kong, Hojin}, \au{Choi, Haecheon} \& \au{Lee, Joon~Sik}} \yr{2000}  \at{Direct numerical simulation of turbulent thermal boundary layers}.  \jt{Physics of Fluids}  \bvol{12}~(10),  \pg{2555--2568}.

\bibitem[Lee \& Moser(2015)]{lee2015direct}
{\sc \au{Lee, Myoungkyu} \& \au{Moser, Robert~D.}} \yr{2015}  \at{Direct numerical simulation of turbulent channel flow up to $re_{\tau} \approx 5200$}.  \jt{Journal of Fluid Mechanics}  \bvol{774},  \pg{395--415}.

\bibitem[Lees(1956)]{lees1956laminar}
{\sc \au{Lees, Lester}} \yr{1956}  \at{Laminar heat transfer over blunt-nosed bodies at hypersonic flight speeds}.  \jt{Journal of Jet Propulsion}  \bvol{26}~(4),  \pg{259--269}.

\bibitem[Lele(1994)]{lele1994compressibility}
{\sc \au{Lele, Sanjiva~K}} \yr{1994}  \at{Compressibility effects on turbulence}.  \jt{Annual review of fluid mechanics}  \bvol{26}~(1),  \pg{211--254}.

\bibitem[Lozano-Dur{\'a}n {\em et~al.\/}(2012)Lozano-Dur{\'a}n, Flores \& Jim{\'e}nez]{lozano2012three}
{\sc \au{Lozano-Dur{\'a}n, Adri{\'a}n}, \au{Flores, Oscar} \& \au{Jim{\'e}nez, Javier}} \yr{2012}  \at{The three-dimensional structure of momentum transfer in turbulent channels}.  \jt{Journal of Fluid Mechanics}  \bvol{694},  \pg{100--130}.

\bibitem[Lozano-Dur{\'a}n {\em et~al.\/}(2020)Lozano-Dur{\'a}n, Giometto, Park \& Moin]{lozano2020non}
{\sc \au{Lozano-Dur{\'a}n, Adri{\'a}n}, \au{Giometto, Marco~G}, \au{Park, George~Ilhwan} \& \au{Moin, Parviz}} \yr{2020}  \at{Non-equilibrium three-dimensional boundary layers at moderate reynolds numbers}.  \jt{Journal of Fluid Mechanics}  \bvol{883},  \pg{A20}.

\bibitem[Marusic {\em et~al.\/}(2021)Marusic, Chandran, Rouhi, Fu, Wine, Holloway, Chung \& Smits]{marusic2021energy}
{\sc \au{Marusic, Ivan}, \au{Chandran, Dileep}, \au{Rouhi, Amirreza}, \au{Fu, Matt~K}, \au{Wine, David}, \au{Holloway, Brian}, \au{Chung, Daniel} \& \au{Smits, Alexander~J}} \yr{2021}  \at{An energy-efficient pathway to turbulent drag reduction}.  \jt{Nature communications}  \bvol{12}~(1),  \pg{5805}.

\bibitem[Millikan(1938)]{millikan1938critical}
{\sc \au{Millikan, Clark~B}} \yr{1938}  \at{A critical discussion of turbulent flows in channels and circular tubes}.  \jt{Proc. Fifth Intern. Congr. Appl. Mech., Cambridge}  \pg{pp. 386--392}.

\bibitem[Modesti \& Pirozzoli(2016)]{modesti2016Reynolds}
{\sc \au{Modesti, Davide} \& \au{Pirozzoli, Sergio}} \yr{2016}  \at{Reynolds and mach number effects in compressible turbulent channel flow}.  \jt{International Journal of Heat and Fluid Flow}  \bvol{59},  \pg{33--49}.

\bibitem[Moin {\em et~al.\/}(1990)Moin, Shih, Driver \& Mansour]{moin1990direct}
{\sc \au{Moin, P}, \au{Shih, T-H}, \au{Driver, D} \& \au{Mansour, NN}} \yr{1990}  \at{Direct numerical simulation of a three-dimensional turbulent boundary layer}.  \jt{Physics of Fluids A: Fluid Dynamics}  \bvol{2}~(10),  \pg{1846--1853}.

\bibitem[Morkovin(1962)]{morkovin1962effects}
{\sc \au{Morkovin, Mark~V.}} \yr{1962}  \at{Effects of compressibility on turbulent flows}.  \jt{M{\'e}canique de la Turbulence}  \bvol{367}~(380),  \pg{26}.

\bibitem[Nagano \& Tagawa(1988)]{nagano1988statistical}
{\sc \au{Nagano, Y} \& \au{Tagawa, M}} \yr{1988}  \at{Statistical characteristics of wall turbulence with a passive scalar}.  \jt{Journal of Fluid Mechanics}  \bvol{196},  \pg{157--185}.

\bibitem[Ni {\em et~al.\/}(2016)Ni, Lu, Ribault \& Fang]{ni2016direct}
{\sc \au{Ni, Weidan}, \au{Lu, Lipeng}, \au{Ribault, Catherine~Le} \& \au{Fang, Jian}} \yr{2016}  \at{Direct numerical simulation of supersonic turbulent boundary layer with spanwise wall oscillation}.  \jt{Energies}  \bvol{9}~(3),  \pg{154}.

\bibitem[Perry \& Hoffmann(1976)]{perry1976experimental}
{\sc \au{Perry, AE} \& \au{Hoffmann, PH}} \yr{1976}  \at{An experimental study of turbulent convective heat transfer from a flat plate}.  \jt{Journal of Fluid Mechanics}  \bvol{77}~(2),  \pg{355--368}.

\bibitem[Pirozzoli \& Bernardini(2011)]{pirozzoli2011turbulence}
{\sc \au{Pirozzoli, Sergio} \& \au{Bernardini, Matteo}} \yr{2011}  \at{Turbulence in supersonic boundary layers at moderate reynolds number}.  \jt{Journal of Fluid Mechanics}  \bvol{688},  \pg{120--168}.

\bibitem[Pirozzoli {\em et~al.\/}(2016)Pirozzoli, Bernardini \& Orlandi]{pirozzoli2016passive}
{\sc \au{Pirozzoli, Sergio}, \au{Bernardini, Matteo} \& \au{Orlandi, Paolo}} \yr{2016}  \at{Passive scalars in turbulent channel flow at high reynolds number}.  \jt{Journal of Fluid Mechanics}  \bvol{788},  \pg{614--639}.

\bibitem[Pope(2000)]{pope2000turbulent}
{\sc \au{Pope, Stephen~B.}} \yr{2000} {\em Turbulent Flows\/}.  \publ{Cambridge University Press}.

\bibitem[Quadrio \& Sibilla(2000)]{quadrio2000numerical}
{\sc \au{Quadrio, Maurizio} \& \au{Sibilla, Stefano}} \yr{2000}  \at{Numerical simulation of turbulent flow in a pipe oscillating around its axis}.  \jt{Journal of Fluid Mechanics}  \bvol{424},  \pg{217--241}.

\bibitem[Reynolds(1874)]{Reynolds1874Analogy}
{\sc \au{Reynolds, Osborne}} \yr{1874}  \at{On the extent and action of the heating surface of steam boilers}.  \jt{Proceedings of the Literary and Phlosophical Society of Manchester}  \bvol{14},  \pg{7--12}.

\bibitem[Ricco {\em et~al.\/}(2021)Ricco, Skote \& Leschziner]{ricco2021review}
{\sc \au{Ricco, Pierre}, \au{Skote, Martin} \& \au{Leschziner, Michael~A}} \yr{2021}  \at{A review of turbulent skin-friction drag reduction by near-wall transverse forcing}.  \jt{Progress in Aerospace Sciences}  \bvol{123},  \pg{100713}.

\bibitem[Rouhi {\em et~al.\/}(2023)Rouhi, Fu, Chandran, Zampiron, Smits \& Marusic]{rouhi2023turbulent}
{\sc \au{Rouhi, Amirreza}, \au{Fu, Matt~K}, \au{Chandran, Dileep}, \au{Zampiron, Andrea}, \au{Smits, Alexander~J} \& \au{Marusic, Ivan}} \yr{2023}  \at{Turbulent drag reduction by spanwise wall forcing. part 1. large-eddy simulations}.  \jt{Journal of Fluid Mechanics}  \bvol{968},  \pg{A6}.

\bibitem[Schlichting \& Gersten(2016)]{schlichting2016boundary}
{\sc \au{Schlichting, Hermann} \& \au{Gersten, Klaus}} \yr{2016} {\em Boundary-layer theory\/}.  \publ{springer}.

\bibitem[Subbareddy {\em et~al.\/}(2014)Subbareddy, Bartkowicz \& Candler]{subbareddy2014direct}
{\sc \au{Subbareddy, Pramod~K}, \au{Bartkowicz, Matthew~D} \& \au{Candler, Graham~V}} \yr{2014}  \at{Direct numerical simulation of high-speed transition due to an isolated roughness element}.  \jt{Journal of Fluid Mechanics}  \bvol{748},  \pg{848--878}.

\bibitem[Trettel \& Larsson(2016)]{trettel2016mean}
{\sc \au{Trettel, Andrew} \& \au{Larsson, Johan}} \yr{2016}  \at{Mean velocity scaling for compressible wall turbulence with heat transfer}.  \jt{Physics of Fluids}  \bvol{28}~(2).

\bibitem[Van~Driest(1951)]{van1951turbulent}
{\sc \au{Van~Driest, Edward~R}} \yr{1951}  \at{Turbulent boundary layer in compressible fluids}.  \jt{Journal of the Aeronautical Sciences}  \bvol{18}~(3),  \pg{145--160}.

\bibitem[{von K\'{a}rm\'{a}n}(1934)]{von1934turbulence}
{\sc \au{{von K\'{a}rm\'{a}n}, Theodore}} \yr{1934}  \at{Turbulence and skin friction}.  \jt{Journal of the Aeronautical Sciences}  \bvol{1}~(1),  \pg{1--20}.

\bibitem[Vos \& Farokhi(2015)]{vos2015introduction}
{\sc \au{Vos, Roelof} \& \au{Farokhi, Saeed}} \yr{2015} {\em Introduction to transonic aerodynamics\/}, ,  \vol{vol. 110}.  \publ{Springer}.

\bibitem[Wallace(2016)]{wallace2016quadrant}
{\sc \au{Wallace, James~M}} \yr{2016}  \at{Quadrant analysis in turbulence research: history and evolution}.  \jt{Annual Review of Fluid Mechanics}  \bvol{48}~(1),  \pg{131--158}.

\bibitem[Wallace {\em et~al.\/}(1972)Wallace, Eckelmann \& Brodkey]{wallace1972wall}
{\sc \au{Wallace, James~M}, \au{Eckelmann, Helmut} \& \au{Brodkey, Robert~S}} \yr{1972}  \at{The wall region in turbulent shear flow}.  \jt{Journal of fluid mechanics}  \bvol{54}~(1),  \pg{39--48}.

\bibitem[Walz(1962)]{walz1962compressible}
{\sc \au{Walz, Alfred}} \yr{1962} {\em Compressible turbulent boundary layers\/}.  \publ{CNRS}.

\bibitem[Yao \& Hussain(2019)]{yao2019supersonic}
{\sc \au{Yao, Jie} \& \au{Hussain, Fazle}} \yr{2019}  \at{Supersonic turbulent boundary layer drag control using spanwise wall oscillation}.  \jt{Journal of Fluid Mechanics}  \bvol{880},  \pg{388--429}.

\bibitem[Yu {\em et~al.\/}(2019)Yu, Xu \& Pirozzoli]{yu2019genuine}
{\sc \au{Yu, Ming}, \au{Xu, Chun-Xiao} \& \au{Pirozzoli, Sergio}} \yr{2019}  \at{Genuine compressibility effects in wall-bounded turbulence}.  \jt{Physical Review Fluids}  \bvol{4}~(12),  \pg{123402}.

\bibitem[Zhang {\em et~al.\/}(2012)Zhang, Bi, Hussain, Li \& She]{zhang2012Mach}
{\sc \au{Zhang, You-Sheng}, \au{Bi, Wei-Tao}, \au{Hussain, Fazle}, \au{Li, Xin-Liang} \& \au{She, Zhen-Su}} \yr{2012}  \at{Mach-number-invariant mean-velocity profile of compressible turbulent boundary layers}.  \jt{Physical Review Letters}  \bvol{109}~(5),  \pg{054502}.

\bibitem[Zhang {\em et~al.\/}(2014)Zhang, Bi, Hussain \& She]{zhang2014generalized}
{\sc \au{Zhang, You-Sheng}, \au{Bi, Wei-Tao}, \au{Hussain, Fazle} \& \au{She, Zhen-Su}} \yr{2014}  \at{A generalized reynolds analogy for compressible wall-bounded turbulent flows}.  \jt{Journal of Fluid Mechanics}  \bvol{739},  \pg{392--420}.

\end{thebibliography}
\end{document}